\documentclass[%
reprint,
nofootinbib,
amsmath,amssymb,
aps,
]{revtex4-2}

\usepackage{times}

\usepackage[paperwidth=199.8mm,paperheight=297mm,centering,hmargin=15mm,vmargin=2cm]{geometry}
\usepackage[titletoc,title]{appendix}
\usepackage{hyperref}
\hypersetup{colorlinks=true,citecolor=blue,linkcolor=blue,filecolor=blue,urlcolor=blue,breaklinks=true}
\usepackage[dvipsnames]{xcolor}
\usepackage{color}
\usepackage[marginal]{footmisc}

\usepackage{url}

\makeatletter
\g@addto@macro{\UrlBreaks}{\UrlOrds}
\makeatother

\usepackage{mathtools}
\usepackage{amsmath}
\usepackage{amssymb}
\usepackage[shortlabels]{enumitem}
\usepackage{graphicx,epic,eepic,epsfig,latexsym,verbatim}
\usepackage{dsfont}

\usepackage{framed}
\definecolor{shadecolor}{rgb}{0.9,0.9,0.9}
\usepackage{float}
\usepackage{tikz}
\usepackage{tcolorbox}
\usepackage{relsize}

\usepackage{theorem}

\def\squareforqed{\hbox{\rlap{$\sqcap$}$\sqcup$}}
\def\qed{\ifmmode\squareforqed\else{\unskip\nobreak\hfil
\penalty50\hskip1em\null\nobreak\hfil\squareforqed
\parfillskip=0pt\finalhyphendemerits=0\endgraf}\fi}
\def\endenv{\ifmmode\;\else{\unskip\nobreak\hfil
\penalty50\hskip1em\null\nobreak\hfil\;
\parfillskip=0pt\finalhyphendemerits=0\endgraf}\fi}

\newcounter{remark}

\newcounter{example}

\mathchardef\ordinarycolon\mathcode`\:
\mathcode`\:=\string"8000
\def\vcentcolon{\mathrel{\mathop\ordinarycolon}}
\begingroup \catcode`\:=\active
  \lowercase{\endgroup
  \let :\vcentcolon
  }

\usepackage{colortbl}
\usepackage{pifont}
\definecolor{Gray}{gray}{0.92}
\definecolor{Gray2}{gray}{0.75}
\definecolor{maroon}{cmyk}{0,0.87,0.68,0.32}
\usepackage{booktabs}
\usepackage{makecell}
\usepackage{diagbox}
\usepackage{multirow}

\newcommand{\nc}{\newcommand}
\nc{\rnc}{\renewcommand}
\nc{\bra}[1]{\langle#1|}
\nc{\ket}[1]{|#1\rangle}
\nc{\<}{\langle}
\rnc{\>}{\rangle}
\nc{\ketbra}[2]{|#1\rangle\!\langle#2|}
\nc{\braket}[2]{\langle#1|#2\rangle}
\nc{\braandket}[3]{\langle #1|#2|#3\rangle}
\nc{\proj}[1]{| #1\rangle\!\langle #1 |}
\nc{\avg}[1]{\langle#1\rangle}
\nc{\Rank}{\operatorname{Rank}}
\nc{\smfrac}[2]{\mbox{$\frac{#1}{#2}$}}
\nc{\tr}{\operatorname{Tr}}
\nc{\ox}{\otimes}

\nc{\cA}{{\cal A}}
\nc{\cB}{{\cal B}}
\nc{\cC}{{\cal C}}
\nc{\cD}{{\cal D}}
\nc{\cE}{{\cal E}}
\nc{\cF}{{\cal F}}
\nc{\cG}{{\cal G}}
\nc{\cH}{{\cal H}}
\nc{\cI}{{\cal I}}
\nc{\cJ}{{\cal J}}
\nc{\cK}{{\cal K}}
\nc{\cL}{{\cal L}}
\nc{\cM}{{\cal M}}
\nc{\cN}{{\cal N}}
\nc{\cO}{{\cal O}}
\nc{\cP}{{\cal P}}
\nc{\cQ}{{\cal Q}}
\nc{\cR}{{\cal R}}
\nc{\cS}{{\cal S}}
\nc{\cT}{{\cal T}}
\nc{\cV}{{\cal V}}
\nc{\cX}{{\cal X}}
\nc{\cY}{{\cal Y}}
\nc{\cZ}{{\cal Z}}
\nc{\cW}{{\cal W}}

\nc{\RR}{{{\mathbb R}}}
\nc{\CC}{{{\mathbb C}}}
\nc{\FF}{{{\mathbb F}}}
\nc{\NN}{{{\mathbb N}}}
\nc{\ZZ}{{{\mathbb Z}}}
\nc{\PP}{{{\mathbb P}}}
\nc{\QQ}{{{\mathbb Q}}}
\nc{\UU}{{{\mathbb U}}}
\nc{\EE}{{{\mathbb E}}}
\nc{\id}{{\operatorname{id}}}

\nc{\1}{{\mathds{1}}}
\nc{\supp}{{\operatorname{supp}}}

\makeatletter
\def\grd@save@target#1{%
  \def\grd@target{#1}}
\def\grd@save@start#1{%
  \def\grd@start{#1}}
\tikzset{
  grid with coordinates/.style={
    to path={%
      \pgfextra{%
        \edef\grd@@target{(\tikztotarget)}%
        \tikz@scan@one@point\grd@save@target\grd@@target\relax
        \edef\grd@@start{(\tikztostart)}%
        \tikz@scan@one@point\grd@save@start\grd@@start\relax
        \draw[minor help lines,magenta] (\tikztostart) grid (\tikztotarget);
        \draw[major help lines] (\tikztostart) grid (\tikztotarget);
        \grd@start
        \pgfmathsetmacro{\grd@xa}{\the\pgf@x/1cm}
        \pgfmathsetmacro{\grd@ya}{\the\pgf@y/1cm}
        \grd@target
        \pgfmathsetmacro{\grd@xb}{\the\pgf@x/1cm}
        \pgfmathsetmacro{\grd@yb}{\the\pgf@y/1cm}
        \pgfmathsetmacro{\grd@xc}{\grd@xa + \pgfkeysvalueof{/tikz/grid with coordinates/major step}}
        \pgfmathsetmacro{\grd@yc}{\grd@ya + \pgfkeysvalueof{/tikz/grid with coordinates/major step}}
        \foreach \x in {\grd@xa,\grd@xc,...,\grd@xb}
        \node[anchor=north] at (\x,\grd@ya) {\pgfmathprintnumber{\x}};
        \foreach \y in {\grd@ya,\grd@yc,...,\grd@yb}
        \node[anchor=east] at (\grd@xa,\y) {\pgfmathprintnumber{\y}};
      }
    }
  },
  minor help lines/.style={
    help lines,
    step=\pgfkeysvalueof{/tikz/grid with coordinates/minor step}
  },
  major help lines/.style={
    help lines,
    line width=\pgfkeysvalueof{/tikz/grid with coordinates/major line width},
    step=\pgfkeysvalueof{/tikz/grid with coordinates/major step}
  },
  grid with coordinates/.cd,
  minor step/.initial=.2,
  major step/.initial=1,
  major line width/.initial=2pt,
}
\makeatother

\makeatletter
\newcommand*\rel@kern[1]{\kern#1\dimexpr\macc@kerna}
\newcommand*\widebar[1]{%
  \begingroup
  \def\mathaccent##1##2{%
    \rel@kern{0.8}%
    \overline{\rel@kern{-0.8}\macc@nucleus\rel@kern{0.2}}%
    \rel@kern{-0.2}%
  }%
  \macc@depth\@ne
  \let\math@bgroup\@empty \let\math@egroup\macc@set@skewchar
  \mathsurround\z@ \frozen@everymath{\mathgroup\macc@group\relax}%
  \macc@set@skewchar\relax
  \let\mathaccentV\macc@nested@a
  \macc@nested@a\relax111{#1}%
  \endgroup
}
\makeatother

\newcommand{\CZ}{\text{CZ}}

\usepackage{tikz}
\usetikzlibrary{quantikz}

\definecolor{light-gray}{gray}{0.95}
\newcommand{\code}[1]{\colorbox{light-gray}{\texttt{\small \text{#1}}}}

\usepackage[ruled]{algorithm2e}

\usepackage{listings}
\usepackage{xcolor}
\usepackage{setspace}

\definecolor{backcolor}{rgb}{0.95, 0.95, 0.95}
\definecolor{codegreen}{rgb}{0, 0.6, 0}
\definecolor{codegray}{rgb}{0.5,0.5,0.5}
\definecolor{codepurple}{rgb}{0.58,0,0.82}

\lstdefinestyle{python}{
    backgroundcolor=\color{backcolor},
    commentstyle=\color{codegreen},
    keywordstyle=\color{magenta},
    numberstyle=\tiny\color{codegray},
    stringstyle=\color{codepurple},
    basicstyle=\ttfamily\footnotesize,
    breakatwhitespace=false,         
    breaklines=true,                 
    captionpos=b,                    
    keepspaces=true,                 
    numbers=left,                    
    numbersep=5pt,                  
    showspaces=false,                
    showstringspaces=false,
    showtabs=false,                  
    tabsize=2
}
\makeatletter
\def\l@subsection#1#2{}
\def\l@subsubsection#1#2{}
\makeatother

\begin{document}

\title{\Large \textbf{Quantum NETwork: from theory to practice}}

\author{Kun Fang}
\email{fangkun02@baidu.com}
\affiliation{Institute for Quantum Computing, Baidu Research, Beijing 100193, China}

\author{Jingtian Zhao}
\affiliation{Institute for Quantum Computing, Baidu Research, Beijing 100193, China}

\author{Xiufan Li}
\affiliation{Institute for Quantum Computing, Baidu Research, Beijing 100193, China}

\author{Yifei Li}
\affiliation{Institute for Quantum Computing, Baidu Research, Beijing 100193, China}

\author{Runyao Duan}
\email{duanrunyao@baidu.com}
\affiliation{Institute for Quantum Computing, Baidu Research, Beijing 100193, China}

\date{\today}

\begin{abstract}

The quantum internet is envisioned as the ultimate stage of the quantum revolution, which surpasses its classical counterpart in various aspects, such as the efficiency of data transmission, the security of network services, and the capability of information processing. Given its disruptive impact on the national security and the digital economy, a global race to build scalable quantum networks has already begun. With the joint effort of national governments, industrial participants and research institutes, the development of quantum networks has advanced rapidly in recent years, bringing the first primitive quantum networks within reach. In this work, we aim to provide an up-to-date review of the field of quantum networks from both theoretical and experimental perspectives, contributing to a better understanding of the building blocks required for the establishment of a global quantum internet. We also introduce a newly developed quantum network toolkit to facilitate the exploration and evaluation of innovative ideas. Particularly, it provides dual quantum computing engines, supporting simulations in both the quantum circuit and measurement-based models. It also includes a compilation scheme for mapping quantum network protocols onto quantum circuits, enabling their emulations on real-world quantum hardware devices. We showcase the power of this toolkit with several featured demonstrations, including a simulation of the Micius quantum satellite experiment, a testing of a four-layer quantum network architecture with resource management, and a quantum emulation of the CHSH game. We hope this work can give a better understanding of the state-of-the-art development of quantum networks and provide the necessary tools to make further contributions along the way.

\end{abstract}
\maketitle

\section{Introduction}

The internet has become a truly global phenomenon that drives the growth of the digital economy across international borders and alters our way of life in ways that nothing else has. We can attend international conferences online, read news from around the world, and even compute on the cloud. The idea of quantum internet aims to expand these possibilities further with the integration of quantum technologies.
With the proper use of quantum features, such as interference, superposition, and entanglement, quantum internet can greatly improve the capability of information processing, the security of information transmission, the precision of information synchronization, and even accomplish tasks impossible in classical networks~\cite{kimble2008quantum, wehner2018quantum, caleffi2018quantum, castelvecchi2018quantum, kozlowski2019towards, singh2021quantum, wei2022towards, awschalom2021development,Gyongyosi2022, rohde_2021,pirandola2016physics, ramakrishnan2022quantum}. 
Particularly, it is anticipated that the creation of a fully-fledged quantum internet will have a wide range of fascinating applications. The most well-known is quantum key distribution, a cryptographic paradigm whose security is ensured by the laws of physics~\cite{bennett1984quantum,ekert1991quantum}. It can also support other security-related applications, such as blind quantum computation, which enables users to delegate a computational task to a quantum computer without disclosing the computation details to the cloud server~\cite{broadbent2009universal}. Given long-distance quantum entanglement, a quantum internet would also enable more precise time synchronization that would help to improve the current navigation systems~\cite{Jozsa2000} and facilitate new fundamental tests for a deeper understanding of the universe~\cite{gottesman2012longer}. Distributed quantum computation is also promising in the long run. As the computational power grows exponentially with the number of quantum computers connected, we may exploit such power to handle computationally demanding problems, such as weather forecasts, protein folding, and so on, that are beyond the capabilities of classical computers or individual quantum computers~\cite{cirac1999distributed,crepeau2002secure,cuomo2020towards}.

Given the potential impact that a quantum internet could have on the economy and society, national governments, industrial participants and research institutes are currently conducting extensive research and development on quantum networks worldwide. It is clear that quantum internet research has reached a national level. In 2016, China launched the world's first quantum communications satellite~\cite{Lu2022,liao2017satellite,castelvecchi2017china} and established the world's first trunk line for secure quantum communications between Beijing and Shanghai~\cite{chen2021integrated,courtland2016china}, contributing to the initial steps to make global quantum internet a reality. China has also included quantum internet technology in its 14th Five-Year Plan for National Informatization as one of the strategic core technologies for future development~\cite{china-national-informatization}. Since 2019, all 27 EU Member States have signed the European Quantum Communication Infrastructure Declaration, demonstrating their support for the initiative and aiming to boost the European capabilities in quantum technology, cybersecurity, and industrial competitiveness~\cite{EuroQCI}. In 2020, the National Quantum Coordination Office in the White House published A Strategic Vision for America's Quantum Networks, outlining five-year and twenty-year objectives for the development of quantum internet~\cite{USAVision}. Besides, countries such as Australia, Japan, Russia, etc., have all actively made their moves for the quantum internet. 
The potential of quantum internet has also attracted numerous industrial participants. Among these are the traditional tech giants such as Baidu and Amazon, the world's biggest telecommunication companies such as AT\&T, BT, and NTT, manufacturing companies such as Toshiba and Cisco, and a series of start-ups including Qunnect, evolutionQ, Aliro Quantum, Xanadu, and QuantumCTek, etc. A majority of these companies are concentrating their efforts on developing quantum cryptography applications that are ready to implement with current quantum technologies, such as quantum random number generation and quantum key distribution. Part of them also plan for the long run and study use cases in a more advanced stage of the quantum internet. A more detailed list of global efforts in quantum networks is summarized in Appendix~\ref{sec: Global efforts in quantum network}.

The state-of-the-art quantum networks are mostly aimed at proof-of-concept experiments for quantum physics and the preliminary demonstration of quantum cryptography tasks. The scalability of such networks is largely based on trusted repeaters, where no end-to-end quantum communication actually happens. Compared to a classical network, we are still at a primitive stage where we can merely start to entangle different quantum devices at a small distance in the laboratory~\cite{pompili2021realization}. Fully functional quantum networks capable of transmitting and controlling entangled quantum states in the field are still under development, the success of which requires sustained and concerted efforts in physics, computer science, and engineering. Recently, extensive works have been done to connect multiple quantum devices and improve their efficiency and scalability. There is also a Quantum Internet Research Group established in the well-known Internet Engineering Task Force by leading experts, focusing on the design of quantum network architectures and core topics such as routing, resource allocation, connection establishment, interoperability, security, and API design~\cite{pirker2019quantum,irtf-qirg-principles-11,van2021quantum}. 

The structure of a quantum internet can be much more complex than its classical counterpart. In addition to the usual network performance metrics such as bandwidth, delay, throughput, and packet loss rate, quantum factors such as fidelity, success probability, and coherence time should also be considered. The interplay of tons of parameters makes the designing and testing of quantum network protocols beyond the scope of conventional mathematical frameworks. To address this, various quantum network simulators have been released, among which are the general simulators such as NetSquid~\cite{netsquid}, \text{SeQUeNCe}~\cite{wu2021sequence}, QuISP~\cite{satoh2021quisp}, and SQUANCH~\cite{bartlett2018distributed}, as well as specialized simulators such as qkdX~\cite{mailloux2015model} and QuNet~\cite{leone2021qunet}. As quantum networks are still under exploration, there exist numerous unknowns but also ample opportunities for innovation. Quantum network simulators can serve as a sandbox for idea verification and help to significantly reduce the possible trial-and-error costs.

\begin{figure}[H]
    \centering
    \includegraphics[width=0.45\textwidth]{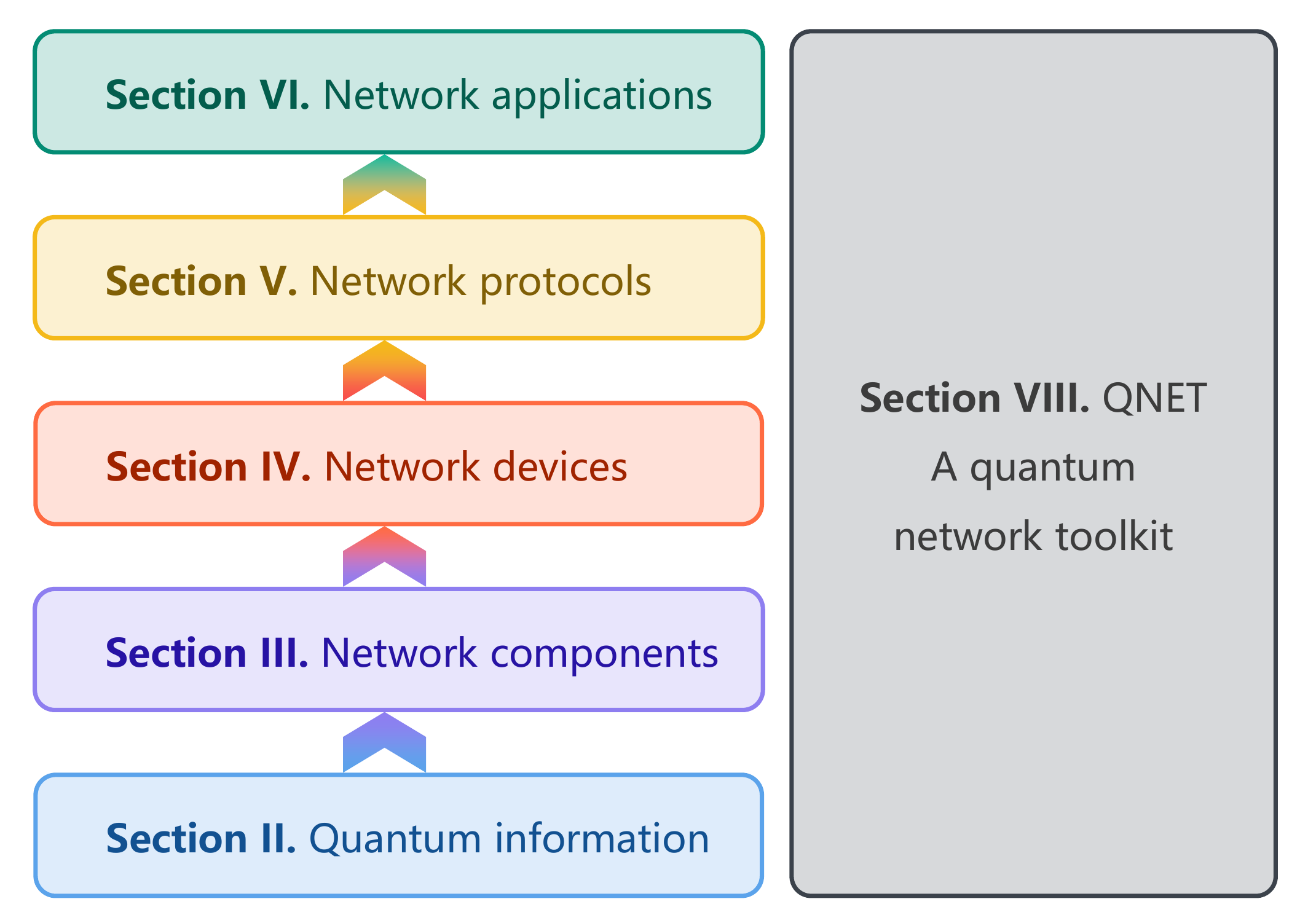}
    \caption{The structure of key contents in this work. The left part is a review of quantum networks, from the underlying principles all the way to the applications. The right part is the introduction of our software practice in quantum network simulation.}
    \label{fig: toc}
\end{figure}

This work aims at providing an up-to-date review of the field of quantum networks from both theoretical and experimental perspectives. It also introduces a newly developed toolkit, QNET~\footnote{The name of QNET is taken from the abbreviation of Quantum NETwork.}, for quantum network simulation. We hope this work can provide a better understanding of the state-of-the-art development of quantum networks and provide the necessary tools to make further contributions along the way. It should be useful to people unfamiliar with the field and can also serve as a reference for those within it.
The remainder of the content is organized as follows (see also the structure in Figure~\ref{fig: toc}). In Section~\ref{sec: preliminaries}, we give a gentle introduction to the basic concepts of quantum information science, laying the theoretical foundations for the understanding of quantum networks. In Section~\ref{sec: basic components}, we introduce the key components that make up a quantum network, including different categories of quantum nodes and the communication links connecting them. Quantum networks necessarily operate on a set of infrastructures comprising hardware, software, and applications. In Section~\ref{sec: devices}, we introduce the hardware part of a quantum network by some essential devices. In Section~\ref{sec: protocols}, we go to the software part by discussing quantum network architectures and relevant protocols. In Section~\ref{sec: applications}, we discuss some representative applications in quantum cryptography, quantum computing, and quantum sensing. In Section~\ref{sec: challenges}, we give a brief summary of the challenges of developing a fully functional quantum network. In Section~\ref{sec: QNET}, we introduce the key techniques in QNET, its basic usages and several featured demonstrations. Finally, a conclusion is given in Section~\ref{sec: conclusions}, and some technical details in the main text are delegated to the appendices.

\section{Quantum information in a nutshell}
\label{sec: preliminaries}

In this section, we discuss some prerequisites of quantum information that are necessary for understanding the working mechanism of quantum networks. Readers who are already familiar with these concepts may skip this part and jump to the later sections. 

\paragraph{Quantum bit and superposition}  

Central to the superiority of a quantum internet is the ability to process quantum bits (qubits). A qubit is a two-level quantum system that can be physically implemented by different approaches, such as the polarization of a photon~\cite{slussarenko2019photonic}, the current of a Josephson junction~\cite{krantz2019quantum}, and the spin of an electron~\cite{bruzewicz2019trapped}, etc. Every qubit is associated with a state, which represents the current status of the system. But different from the binary choice of a classical bit, a qubit can possess a state beyond $0$ or $1$. This phenomenon is called quantum superposition, one of the fundamental principles of quantum mechanics. More specifically, besides the state of $0$ or $1$ (or $\ket{0}$ or $\ket{1}$ in the Dirac notation), a qubit can possess a state that is the linear combination of these isolated states, such as $\alpha\ket{0} + \beta\ket{1}$ for any complex numbers $\alpha$ and $\beta$ with $|\alpha|^2 + |\beta|^2 = 1$. Simply speaking, any unit vector in a two-dimensional complex vector space represents a state of qubit. We can also map these states to points on the unit sphere, known as the Bloch sphere. In this perspective, a classical bit could only be at one of the two poles, where $\ket{0}$ and $\ket{1}$ are located; but a qubit can appear at any point on the sphere. This difference can become even more significant as the number of qubits increases. Generally, while $n$ classical bits can only encode one of the $2^n$ possible states at a time, $n$ qubits can coherently encode all $2^n$ states through superposition. This remarkable phenomenon enables the possibility of parallel information processing and eventually underpins the speedup of quantum computation~\cite{hillery2016coherence,hayashi2021finite,Diaz2018usingreusing,fang2018,regula2018,ahnefeld2022role}.

\paragraph{Quantum entanglement}

Quantum entanglement is another striking feature of quantum mechanics, revealing the quantum nature in the multipartite scenario. Two features of entanglement make it an ideal resource for quantum networks. First, entangled states can be strongly correlated such that the measurement outcomes of two qubits can always be matched, even though their outcomes are completely random. For example, if two parties share a Bell state $(\ket{00}+\ket{11})/\sqrt{2}$, by measuring each of the two qubits independently, one obtains a random distribution of $0$ and $1$ with equal probability. However, if the outcomes of the two independent measurements are compared, one observes that every time the measurement of a qubit yielded $0$, so did the measurement of the other qubit, and the same happened with outcome $1$. Remarkably, such coordination is independent of the distance between the separated parties, making entanglement an ideal resource for data synchronization (e.g., in entanglement-based quantum key distribution~\cite{ekert1991quantum}, clock synchronization~\cite{Jozsa2000}) and transmission (e.g., in the quantum teleportation protocol~\cite{bennett1993teleporting}). Second, due to the inherent property of monogamy, entanglement cannot be shared with a third party. This underpins the security of entanglement-based or device-independent types of quantum key distribution and many other quantum cryptography applications~\cite{Acin2007}. Despite being spooky and counter-intuitive, the existence of quantum entanglement has been rigorously confirmed by plenty of experiments~\cite{bell1964einstein, freedman1972experimental, aspect1982experimental, ou1988violation, hensen2015loophole, Giustina2015, Shalm2015, handsteiner2017,yin2017satellite}. Notably, a Nobel Prize in Physics was awarded in 2022 to Alain Aspect, John F. Clauser and Anton Zeilinger for their pioneering work in this direction~\cite{nobel-prize-2022}.

\paragraph{Quantum measurement}

Quantum measurement is a fundamental operation that helps to extract information from a quantum system~\cite{nielsen2010quantum,xiao2019complementary}. A quantum measurement, according to the postulate of quantum mechanics, will collapse the measured quantum state and return a random outcome, the probability of which depends on the measurement operation and the state itself. For instance, if the outcome of measuring a superposed quantum state $\alpha \ket{0} + \beta \ket{1}$ (in the computational basis) is zero, then the qubit collapses into the corresponding state~$\ket{0}$ and any further measurement will give zero as the outcome with certainty, regardless of the parameters $\alpha,\beta$ in the superposed state. As a result, although the superposition principle permits a qubit to retain more than one bit of information, only one bit can be retrieved at a time by measurement. The disruptive nature of quantum measurement makes it difficult to extract information from a quantum state without causing disturbance. This, however, happens to make qubit an ideal information carrier for the defense against attackers. Besides information extraction, quantum measurement can also be leveraged to steer quantum computation in the measurement-based model~\cite{briegel2009measurement}, which lays the foundation for blind quantum computation, one of the most important applications in a quantum internet~\cite{broadbent2009universal}.

\paragraph{No-cloning theorem}

Unlike classical information that can be readily duplicated, the no-cloning theorem states that it is impossible to create an identical copy of an arbitrary and unknown quantum state~\cite{wootters1982single}. That is, there exists no unitary transformation $U$ such that $U\ket{\varphi}\ket{0} = \ket{\varphi}\ket{\varphi}$ for any $\ket{\varphi}$, where $\ket{0}$ is a blank quantum state. This results in both advantages and limitations. On the one hand, the no-cloning theorem guarantees that information embedded in a quantum state cannot be easily stolen by a malicious party because the intercepted qubits cannot be copied, and extracting the information by quantum measurement will inevitably disturb the state and be detected. This underpins the security of many quantum network applications, such as quantum key distribution~\cite{bennett1984quantum}. On the other hand, no-cloning makes quantum information processing much more challenging than its classical counterpart. For instance, techniques for classical information processing, such as signal amplification, retransmission, and so on, cannot be directly applied to quantum networks, making it necessary to find new schemes to counteract losses and errors.

\section{Quantum network components}
\label{sec: basic components}

\begin{figure*}[!htb]
    \centering
    \includegraphics[width=\textwidth]{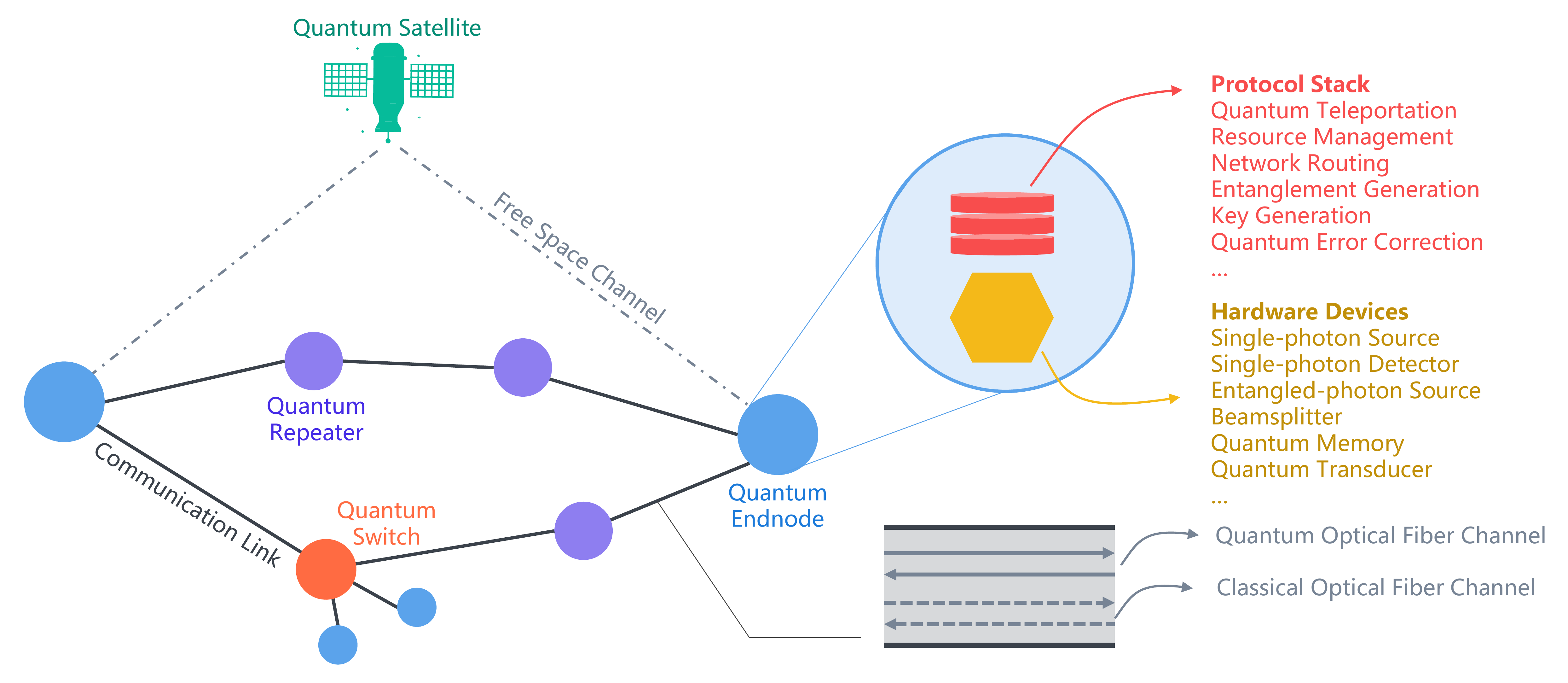}
    \caption{An illustration of a quantum network and its key components. A quantum network comprises a collection of quantum nodes and communication links between them. Typical quantum nodes include quantum endnodes, quantum repeaters, quantum switches/routers, and quantum satellites. Each node is equipped with a protocol stack and relevant hardware devices that work together to provide network services. Communication links can consist of both classical and quantum channels connecting two ends for data transmission.}
    \label{fig: quantum_network}
\end{figure*}

A quantum network will work synergistically with the existing classical internet, but it also demands some extra \emph{quantum} components to support new applications and services. These primarily include different categories of quantum nodes and communication links connecting them. Depending on the role it plays, a quantum node can be a quantum endnode, quantum repeater, quantum switch, quantum router, quantum satellite and so on. As the classical information processing is always indispensable, we assume that every quantum node has the capability of classical computation built-in. On top of this, every quantum node should contain both the software and hardware parts, particularly a protocol stack defining the rules of information exchange and quantum hardware devices responsible for the implementation of specific requests. As for the communication links, they may consist of classical or quantum channels, depending on the type of information to propagate. In the following, we highlight the basic components required to construct a quantum network. An illustrative diagram of these components and their relationships is depicted in Figure~\ref{fig: quantum_network}.

\paragraph{Quantum endnode}

The quantum endnode usually represents a user or client who can send or receive quantum information. It can also be a cloud server that provides quantum computing services to the public. These nodes typically play the role of the source or destination of a service request. For instance, a request is started by a quantum endnode and travels through intermediary nodes before reaching the destination endnode. This request is subsequently handled by the target endnode in accordance with the applicable protocols, and it is then sent back to the initiator. Finally, end-to-end quantum services can be accomplished by repeating these transactions back and forth. A quantum endnode (as well as other nodes) typically contains two types of qubits: communication qubits for generating entanglement with distant nodes and memory qubits for data storage and computation. Depending on the use cases, a quantum endnode can be equipped with very simple devices consisting of barely a photon source or a photon detector. Such an endnode can work with applications like quantum key distribution. To support sophisticated applications such as blind quantum computation, a quantum endnode may require durable quantum memories and fault-tolerant quantum computers.

\paragraph{Quantum repeater}

Because of the environmental noise and channel attenuation, information can readily be contaminated and get lost over a long distance transmission. In particular, quantum communication requires the transmission of fragile photons that cannot travel indefinitely in fibers or through the air. The longer the distance, the more likely they will be absorbed or scattered. This issue is resolved in classical networks by repeaters that can copy and amplify the received signals for further transmission. As for quantum networks, information carried in qubits cannot be copied or amplified due to the no-cloning theorem~\cite{wootters1982single}, making classical repeaters not applicable for quantum communication.
To overcome the barrier of long-distance qubit transmission, quantum repeaters were introduced~\cite{briegel1998quantum}. These nodes are deployed in the middle of endnodes to divide the whole transmission line into smaller segments, thereby reducing the distance of direct transmission to an acceptable range.
Depending on the underlying mechanism, there are at least two types of quantum repeaters in principle~\cite{duan2001long,muralidharan2016optimal,zwerger2018long,brand2020efficient,li2020efficient,rozpkedek2020quantum,rozpkedek2018parameter}. The first type starts with the generation of long-range entanglement through a combination of protocols, such as herald entanglement generation, entanglement swapping and entanglement distillation. Then quantum information is transferred via quantum teleportation with the established entanglement. (More discussions on these protocols will be given in Section~\ref{sec: protocols}.) The second type relies on the techniques of quantum error correction that directly encode the quantum information in a block of physical qubits and send it toward the next hop. For the time being, the practical use of quantum repeaters is still hindered by the challenges of simultaneously realizing and integrating all the key capabilities~\cite{sangouard2011quantum}. In particular, the first type of quantum repeater requires entanglement generation for the elementary links in a heralded way; that is, one has to know when the entanglement has been successfully established. It also requires durable quantum memories to temporarily store the established short-range entanglement so as to perform entanglement swapping when two adjacent entangled pairs are both ready. The second type of quantum repeater can be much more efficient in time but requires joint manipulation of numerous qubits and higher initial fidelity of the transmitted states (refer to the threshold theorem~\cite{gottesman2009introduction}). Besides what has been mentioned, a hybrid quantum repeater that combines different approaches can also be considered~\cite{muralidharan2016optimal,munro2015inside}. Although practical quantum repeaters do not yet exist, several lab experiments involving networks of two or three nodes have made some progresses~\cite{pompili2021realization,lago2021telecom,liu2021heralded,pu2021experimental}.
It is also worth noting that trusted repeaters are frequently used as a temporary solution at the current stage, for example, in the Beijing-Shanghai Trunk Line, the Tokyo QKD network, and many other networks in the field, where no end-to-end quantum communication actually happens~\cite{cao2022evolution}. Photons that arrive at the trusted repeaters are measured, and fresh photons in the right state are sent toward the next repeater station in the chain. While trusted repeaters do help to extend the applicable distances for specific applications, they are essentially classical and will compromise the security of the network.

\paragraph{Quantum satellite} 

Due to the photon loss, the effective distance of photon transmission in optical fibers or terrestrial free space is limited to a few hundred kilometers. Quantum repeaters on the ground may extend the communication distance to a few thousand kilometers. To further expand the coverage of quantum networks and realize a truly global quantum internet, quantum satellites and space-based links should be exploited~\cite{sidhu2021advances,belenchia2022quantum}. In this case, most of the photons' propagation paths are in vacuum, and thus the photon loss and decoherence can be significantly reduced. In 2016, China launched the first quantum communications satellite, successfully demonstrating its integration with ground-based networks and the distribution of secure keys and quantum entanglement at space scales~\cite{Lu2022}. In 2022, the first micro-nano quantum satellite has also been launched by China, which is smaller but more efficient in its performance~\cite{china-micro-nano-satellite}. Stimulated by these outstanding results, other countries such as Austria~\cite{neumann2018q3sat}, Germany~\cite{haber2018qube}, French~\cite{kerstel2018nanobob}, Singapore~\cite{dalibot2020preliminary}, Canada~\cite{cananda-satellite} and the United Kingdom~\cite{uk-satellite}, have all making efforts in building and launching their quantum satellite missions, especially nanosatellites that helps to reduce cost and development time. Despite significant progress, there remains plenty of room for improvement in both science and engineering aspects~\cite{Lu2022}. First, the propagation of quantum signals through free space is prone to atmospheric turbulence and weather conditions, significantly limiting the practical application of quantum satellites. New technologies to release the constraints of working conditions for quantum satellites are desired. Second, compared to the classical satellite, quantum satellites require adapting the existing optical tracking techniques to the needs of quantum communication and building devices that can operate in a satellite without the need for heavy maintenance. Third, a constellation combining low-earth-orbit and high-earth-orbit quantum satellites is necessary for constructing a global-scale quantum network. While more satellites with varying orbital altitudes should be launched, the cost of manufacture is also an important consideration to take into account.

\paragraph{Quantum switch/router}

In a classical network, information can be transmitted when two users are directly connected by a communication link. However, as the number of users grows, it becomes impractical to connect every pair of users via a physical link. Switches are then created as central nodes to facilitate information exchange by connecting various network users. A quantum switch would play a similar role by linking several users to it and attempting to establish multipartite quantum entanglement, such as Bell states or general Greenberger–Horne–Zeilinger (GHZ) states~\cite{greenberger1989going}, among them. The entanglement switching mainly involves two steps. First, bipartite entanglement between the user and the switch is created in parallel. Second, local operations and measurements are carried out on the bipartite entangled states to produce multipartite entanglement for a set of users. In this sense, a quantum switch can be considered a multipartite generalization of a quantum repeater, but it should also include routing logic to determine the best route to switch and the capability of resource management to handle multiple requests. Because of the decoherence and limited size of quantum memories, prioritizing entanglement switching affects the switch's performance, such as the fidelity of the distributed entanglement, the latency of the entanglement requests, and the throughput of the switch. Some analytical studies on the interplay of these factors have been given in~\cite{dai2021entanglement,nain2020analysis,vardoyan2019capacity,vardoyan2019stochastic}.

\paragraph{Communication link}

A link represents the connection between two nodes in the network. It is usually referred to as a topological concept and should contain specific classical or quantum channels for the actual information transmission. Classical channels in a quantum network are the same as those in a classical network and are responsible for transmitting classical messages between the nodes they connect to. Quantum channels, however, are responsible for transmitting quantum information. Given that quantum teleportation allows the transmission of quantum states by consuming shared entanglement, a link may also refer to the nonlocal correlation of quantum entanglement between different nodes.

\section{Quantum network devices}
\label{sec: devices}

In this section, we investigate the hardware part of a quantum network, which is responsible for the physical implementation of all kinds of quantum services and applications. We will put our focus on the discussion of \emph{quantum} devices, particularly, sources for photon emission and entanglement generation, detectors for information readout, quantum transducers for signal conversion, quantum memories for data storage, and quantum channels for photon transmission. Before proceeding to their details, we remark that devices responsible for classical computation and communication also play a significant part in quantum networks and that their performance can constitute a bottleneck for the overall efficiency of the network~\cite{coordinated2021}. However, these are beyond the scope of this work. More in-depth discussions on the hardware aspect of quantum networks can refer to~\cite{ramakrishnan2022quantum,wei2022towards} and references therein.

\paragraph{Single-photon source}
Different from coherent or thermal light sources, a single-photon source is a device that emits single photons. Such a device is needed in quantum key distribution and other quantum cryptography applications because multi-photon pulses would compromise security by allowing eavesdroppers to collect information, e.g., by the photon-number-splitting attack~\cite{lutkenhaus2002quantum}.
An ideal single-photon source should meet the following criteria: a single photon can be emitted with certainty and on demand; vacuum or multiple-photon emission occurs with zero probability; successively emitted photons are indistinguishable; and the repetition rate is arbitrarily large. Apparently, deviations from these ideal features are always present in reality. These deviations lead to the performance metrics, most importantly, the frequency that determines the repetition rate of the light pulse emission, the wavelength and bandwidth that define the intrinsic attributes of the released photons, and the mean photon number that represents the average number of photons per signal pulse.
In terms of physical realizations, two distinct technologies have been explored~\cite{eisaman2011invited,migdall2013single}. The first one is deterministic sources, also referred to as on-demand sources, that can be realized in systems such as semiconductor quantum dots~\cite{senellart2017high, arakawa2020progress, strauf2007high, shields2007semiconductor}, mesoscopic quantum wells~\cite{kim1999single}, single molecules~\cite{chu2017single, de1996single, kako2006gallium}, single atoms~\cite{kuhn2002deterministic}, single ions~\cite{blinov2004observation, maurer2004single}, and color centers~\cite{kurtsiefer2000stable, babinec2010diamond, alleaume2004experimental}. The other is probabilistic sources, which can be realized by weak lasers~\cite{pittman2007investigation}, atomic cascades~\cite{grangier1986experimental}, and parametric down-conversion~\cite{hong1986experimental, fasel2004high}. These sources emit photons in pairs so that the detection of one photon heralds another photon. The distinction between deterministic and probabilistic sources can be blurred in practice. Deterministic sources can become more probabilistic with the decrease of extraction loss and coupling efficiency, and probabilistic sources can become more deterministic by multiplexing~\cite{meyer2020single}.
Although perfect sources are not achievable in reality, significant advancements have been made towards ever-improving approximations~\cite{wang2019boson, thomas2021race}. The state-of-the-art source can generate a single photon on demand with a probability of $57\%$ at a repetition rate of $1$ GHz and with average two-photon interference visibility of $97.5\%$~\cite{tomm2021bright}. Another recent experiment successfully yields single-photon source characteristics such as a single-photon purity of $99.8\%$ and indistinguishability of $92.6\%$ by tailoring solid-state single-photon sources with stimulated emissions~\cite{wei2022tailoring}.

\paragraph{Single-photon detector}
A photon detector is a highly sensitive device that converts incoming photons into electrical signals and is capable of detecting and specifying the arrival times of individual photons~\cite{hadfield2009single,migdall2013single}. An ideal detector should meet the following criteria: the probability that the detector successfully detects a photon when it arrives is one hundred percent; the rate at which the detector is triggered in the absence of any incident photons is zero; the time after a photon-detection event during which the detector is incapable of detecting a photon is zero; and the variation from event to event in the delay between the input of the optical signal and the output of the electrical signal is zero. Deviations from these ideals negatively impact experiments in various ways, and thereby the performance of a detector is usually characterized by a set of metrics, including detection efficiency, dark count rate, dead time, and timing jitter.
Besides photon detection, an ideal photon detector should be able to distinguish the number of photons in an incident pulse, which is referred to as the photon-number resolution. 
While non-photon-number-resolving detectors can be realized by the quantum dot~\cite{komiyama2000single}, single-photon avalanche photodiode~\cite{cova2004evolution},  superconducting nanowire~\cite{rosfjord2006nanowire} and up-conversion technique~\cite{takesue2006}, the more demanding photon-number resolving detectors can be realized by the superconducting-tunne-junction~\cite{peacock1996single}, transition-edge sensor~\cite{Rosenberg2005}, charge-integration technique~\cite{fujiwara2007direct}, quantum-dot optically gated field-effect transistor~\cite{gansen2007photon} and so on.
The state-of-the-art detectors are made by the cryogenic superconducting nanowire single-photon detectors, whose efficiency can be as high as $98\%$~\cite{reddy2020superconducting} and timing jitter below $3$ps~\cite{korzh2020demonstration}.

\paragraph{Entangled photon source} 
An entangled photon source is a device that produces the entangled states, required for reliable quantum communication and the majority of quantum network applications. An ideal entangled photon source should satisfy the following criteria: at a given time, only one pair of entangled photons is emitted; the emitted two-photon state is exactly the standard Bell state; individual photons emitted in different trials should be indistinguishable from each other; and the photons can be extracted and collected from the source with certainty~\cite{lu2014pushbutton}. There are at least two approaches that are frequently used in realizing entangled photon sources in practice. The first approach is by using the technique of spontaneous parametric down-conversion (SPDC)~\cite{Burnham1970,pan2012multiphoton,anwar2021entangled}, which is a nonlinear optical process in which a strong pump field induces the spontaneous emission of two lower-energy photons (referred to as signal and idler photons). Because SPDC is fundamentally a probabilistic process, the correlations of entangled photon pairs can be revealed only after post-selection, and the conversion efficiency is typically as low as $10^{-6}$~\cite{bock2016highly}. Nevertheless, probabilistic entangled pair generation via SPDC remains the most widely used approach, owing mainly to its relative simplicity and versatility. The other approach for making an entangled photon source is using the radiative cascades in single quantum emitters, such as atoms~\cite{Aspect1981} and quantum dots~\cite{young2006improved,Akopian2006}. A state-of-the-art entangled photon source based on quantum dots has been reported in~\cite{wang2019ondemand} that achieves an entanglement fidelity of $90\%$, pair generation rate of $59\%$, pair extraction efficiency of $62\%$, and photon indistinguishability of $90\%$, simultaneously.

\paragraph{Quantum transducer}
Transducers are devices that can convert a signal from one physical platform to another. Examples are microphones that convert air pressure into an electrical signal and antennae that convert propagating electromagnetic waves into conducted electrical signals. Quantum transducers are designed to play a similar role by faithfully converting quantum information between different platforms. Extensive efforts have been made in the physics community to investigate schemes and technologies enabling such an interface~\cite{hammerer2010quantum, stannigel2010optomechanical, awschalom2021development,lauk2020perspectives}. Of particular importance is the microwave-to-optical transducer~\cite{bochmann2013nanomechanical,andrews2014bidirectional,jiang2020efficient}. On the one hand, superconducting quantum computers, which manipulate quantum information via microwave signals, are one of the most developed approaches to implementing quantum computers. On the other hand, there is broad consensus in the field that optical photons should be used as the information carrier for long-distance communication due to their advantages, such as low interaction with the environment, ease of control with standard optical devices, and high-speed low-loss transmissivity. As a result, a scalable quantum network that connects disparate quantum computers requires a quantum transducer that reliably converts quantum signals between microwave modes and optical modes (used in quantum computing and quantum communication, respectively). However, because the energy (frequency) gap between these two domains exceeds five orders of magnitude, such a microwave-to-optical conversion can be incredibly challenging. The quality of an ideal quantum transducer can be evaluated by its high conversion efficiency, broad bandwidth, low added noise, and high quantum capacity~\cite{wang2022quantum}. In terms of conversion efficiency, a state-of-the-art experiment is reported in~\cite{higginbotham2018harnessing} that achieves a conversion efficiency of $47\%$. For a more comprehensive summary of quantum transducers, please refer to the recent review paper~\cite{han2021microwaveoptical} and references therein.

\paragraph{Quantum memory}
Quantum memory is a quantum analog of classical memory, used to store and retrieve quantum states. It can be of significant use in multiple contexts. One of the simple but potentially important applications is to enhance the efficiency of a single-photon source. Given a probabilistic single-photon source such as parametric down-conversion, one can realize a deterministic source of single photons by detecting a heralding photon in a pair while storing the other one in memory and releasing it on demand. Quantum memories are also indispensable for developing long-distance quantum communication via quantum repeaters. As photons from different entangled sources do not always arrive at the repeater stations simultaneously, quantum memories can be used as a synchronization tool by storing the first arriving qubits and waiting for the later ones, allowing for the effective implementation of entanglement swapping. Moreover, quantum memories can also serve as an essential component in advanced quantum computers to store the intermediate computational results and synchronize data in a distributed computational task. As for physical implementations, quantum memories can be realized through a variety of approaches, such as atomic ensembles~\cite{julsgaard2001experimental, chou2005measurement}, cold neutral atoms~\cite{choi2008mapping}, doped crystals~\cite{usmani2012heralded}, and particularly in photonic systems~\cite{lvovsky2009optical,simon2010quantum,heshami2016quantum,bussieres2013prospective,novikova2012electromagnetically,ma2017optical,vernaz2018highly, wang2019efficient,ma2020optical,zhu2022ondemand}. The performance of a quantum memory can be evaluated by metrics such as the probability that the stored information can be successfully retrieved, the fidelity of the retrieved information compared to the original one, the storage time that information can be faithfully retained, and the bandwidth that determines the achievable repetition rates. A state-of-the-art quantum memory that uses silicon vacancies in diamond has been reported in~\cite{Stas2022} that achieves a lifetime exceeding $2$ seconds at temperatures up to $4.3$ kelvin.

\paragraph{Quantum channel}
The propagation of quantum information in a network from one end to the other relies on the quantum channels connecting them. These channels can be either optical fiber channels or free-space channels, each of which has advantages and limitations. The former can leverage the well-established network of telecommunication fibers and has the advantage that the environment inside the fiber is relatively stable, thus preserving the coherence of photons. Nevertheless, the distance of direct transmission of quantum states in fibers is significantly restricted because the probability of photon loss grows exponentially with the channel distance~\cite{sangouard2011quantum,inagaki2013entanglement,yin2016measurement}. As an example, it would take nearly three hundred years to successfully deliver one photon over a thousand kilometers using a typical standard telecom optical fiber with a loss of about $0.2$ decibel per kilometer~\cite{sangouard2011quantum}.
In contrast, free-space optical communication is less developed but does not have to account for polarization scrambling induced by optical fibers and can typically allow for higher transmission rates. Particularly in satellite communication, most of the photon loss happens lower than $10$km in the atmosphere, leaving most of the photon path as vacuum with negligible loss of photons and decoherence. Once photons leave the atmosphere, only the dispersion effect is dominated~\cite{villoresi2008experimental}. Thus, the satellite free-space optical channel keeps the quantum information in photons intact and increases the system fidelity and channel capacity, making it an ideal approach to establishing long-distance connections~\cite{hughes2017quantum}. The drawback of free-space communication is that it is more susceptible to environmental disturbances on the photons (e.g. weather changes), as well as the challenge of stabilizing the free-space coupling using pointing and tracking methods. However, many of these challenges have been overcome in a series of free-space quantum communication demonstrations~\cite{buttler1998practical,ma2012quantum,steinlechner2017distribution,valivarthi2016quantum,sun2016quantum,pugh2017airborne,wang2013direct,liu2020drone,ren2017ground,liao2018satellite}. It is also worth noting that the quality and efficiency of the transmission of quantum information depend not only on the communication channel but also on the coding scheme we use. The ultimate capability of a quantum channel to transmit quantum information is characterized by its channel capacity, which is a fundamental research topic in quantum Shannon theory. Unlike the classical theory, there is currently no efficient algorithm for calculating quantum channel capacity in general. Plenty of efforts have been made in this direction. In particular, the quantum capacity of a lossy channel was determined in~\cite{pirandola2017fundamental}. For general quantum channel models, the state-of-the-art results are given by~\cite{pirandola2016capacities, wang2019converse, wang2018semidefinite, fang2021geometric, fang2021ultimate, fawzi2021defining, fawzi2022hierarchy,wang2017semidefinite,fang2019quantum,Fang2020}.

\section{Quantum network protocols}
\label{sec: protocols}

Designing a quantum network capable of providing both efficient and stable quantum services to a massive number of users would require the development of a quantum-updated version of the network architecture. In this architecture, the network functionalities should be organized as a stack of layers, each containing specific protocols and interfaces that all hardware and software must comply with. Since quantum networks are still in the early stages of development, there is currently no consensus on a standard architecture yet. Different architectures have been proposed in~\cite{dahlberg2019link,kozlowski2020designing, van2021quantum, illiano2022quantum,diadamo2022packet,pirker2018modular,pirker2019quantum}, and the fundamental principles have been outlined in~\cite{irtf-qirg-principles-11}.

\begin{figure}[!ht]
    \centering
    \includegraphics[width=7.6cm]{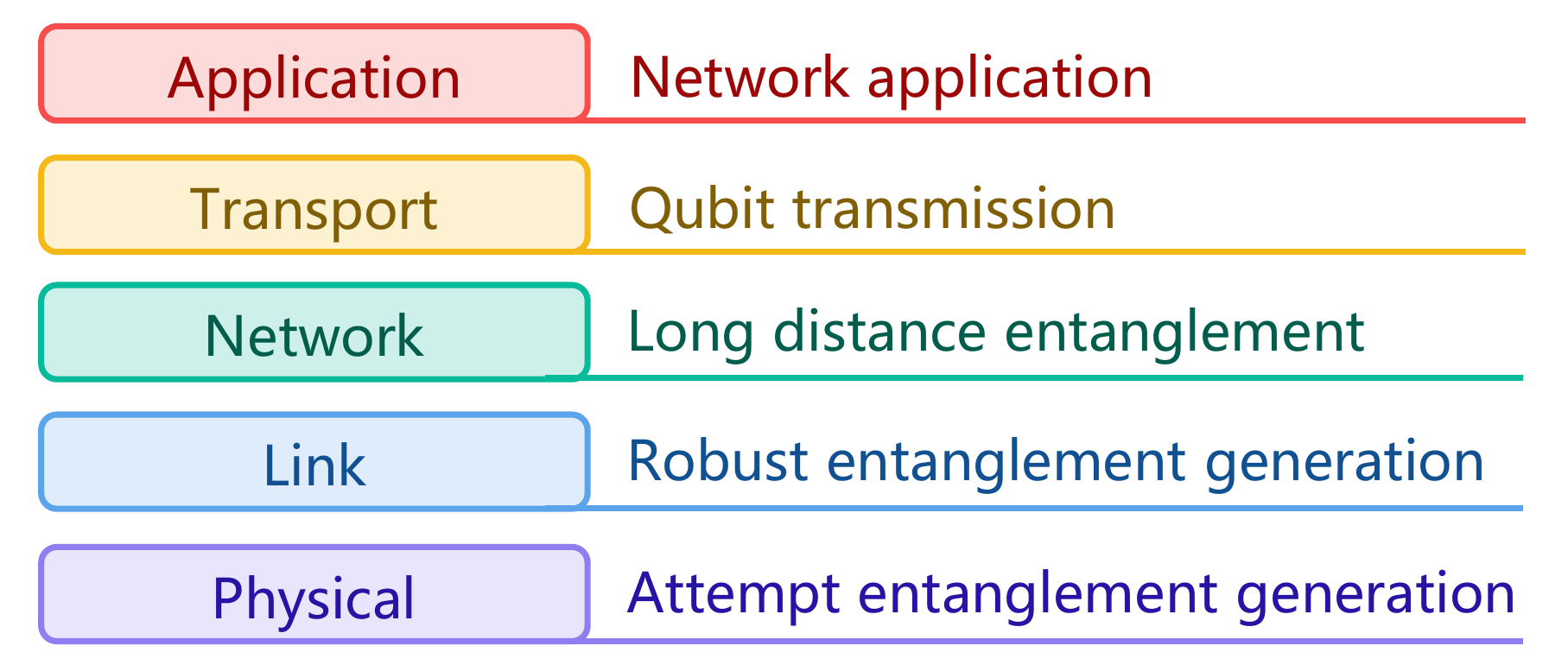}
    \caption{An illustration of the TCP/IP architecture used in classical networks and its quantum generalization proposed in~\cite{dahlberg2019link}.}
    \label{fig: tcp-ip-stack}
\end{figure}

\begin{figure}[!ht]
    \centering
    \includegraphics[width=8.2cm]{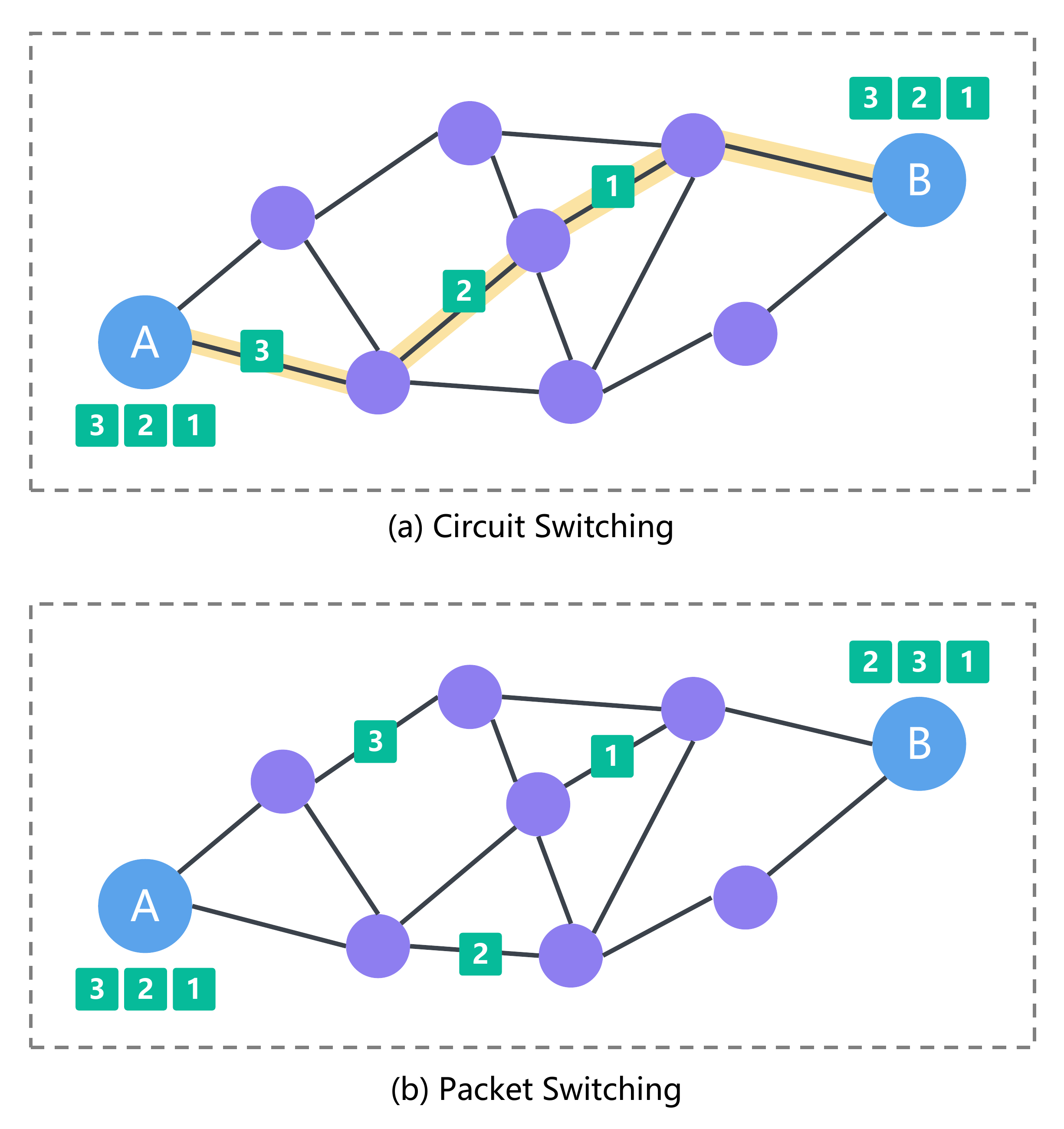}
    \caption{An illustration of the circuit switching and packet switching used in classical networks. Suppose node A wants to send information to node B. In circuit switching, a dedicated circuit path is established before communication occurs. Once set up, all communication takes place over this circuit. In packet switching, no circuit is set up prior to the data transmission. A message is
    divided into many parts, each of which is routed from the source to the destination as individual packets.}
    \label{fig: circuit_packet}
\end{figure}

One possible quantum network architecture is similar to the idea of circuit switching in the classical network, where a dedicated communication path is established before the actual data transfer happens~\cite{dahlberg2019link} (see also Figure~\ref{fig: tcp-ip-stack}). More specifically, the lowest element will be the physical layer, which includes all classical and quantum hardware devices and is responsible for the attempt at entanglement generation. The link layer runs on top of the physical layer, aiming to produce robust entanglement between neighboring stations. The network layer will be responsible for creating long-distance entanglement between nodes that are not directly connected, by means of entanglement swapping over short-distance entanglement established from the link layer. The transport layer will be responsible for reliable qubit transmission using quantum teleportation. Finally, the application layer will enable services and applications to use both classical and quantum network capabilities.
Besides circuit switching, a different approach widely used in today's internet is called packet switching, in which a message is divided into many parts which are sent independently over whatever route is optimum for each packet, and then get reassembled at the destination (see Figure~\ref{fig: circuit_packet}). However, the incapability of quantum cloning and the difficulty of performing quantum error correction make it hard to apply this idea to quantum networks. Nevertheless, a conceptual design for a packet switching quantum network has been recently proposed in~\cite{diadamo2022packet}. Quantum network protocols and architectures based on quantum graph states and the measurement-based quantum computation model have also been considered in~\cite{zwerger2012measurement,zwerger2016measurement,pirker2018modular,pirker2019quantum}.

How future quantum networks will be deployed is yet to be known. But given the nature of quantum information processing and our experience with classical networks, it is conceivable that several building blocks would be necessary. In the following, we introduce some of the most important quantum protocols that could serve as critical components for future quantum network stacks.

\paragraph{Quantum teleportation}
In classical packet switching networks, data is duplicated before transmission; if the data is lost, retransmission is done with no harm beyond a performance penalty (refer to the Transmission Control Protocol~\cite{cerf1974protocol}). With only one copy of quantum data and the restrictions of the no-cloning theorem, we must treat that data with extreme caution. Given the risk of data loss in direct transmission, quantum teleportation provides a technique for reliable transmission of quantum information without physical movement of the object itself~\cite{bennett1993teleporting}. With local operations and shared entanglement resources, quantum teleportation allows transmitting unknown quantum states between two disparate quantum nodes regardless of their physical distance, thereby providing a fascinating strategy for the remote transmission of quantum information in the network. Suppose that the data qubit to teleport is $\ket{\psi}_{A'}$ held by Alice and the shared entanglement between Alice and Bob is given by $\ket{\Phi}_{AB}$. The teleportation starts with Alice's implementation of a Bell state measurement on her data qubit and half of the entangled pair in her possession. Then Alice sends the measurement outcomes, i.e., two random bits of classical information, to Bob. Upon receiving the classical message, Bob performs a corresponding correction that locally restores his state to $\ket{\psi}$. Notably, the information that completely characterizes the data qubit is locally destroyed on Alice's side and reconstructed on Bob's side, ensuring that the no-cloning theorem is not violated. Moreover, as the exchange of classical information from Alice to Bob is necessary, quantum teleportation does not violate the speed of light either. Owing to the noise and errors of direct transmission, quantum teleportation will serve as one of the main approaches to transmitting qubits across a quantum network. Experimental determinations of quantum teleportation have been made in different physical systems~\cite{boschi1998experimental,bouwmeester1997experimental,pirandola2015advances} such as photons~\cite{ursin2004quantum}, atoms~\cite{riebe2004deterministic}, electrons~\cite{Leuenberger2005}. Utilizing the Micius satellite for space-based quantum communication, the cutting-edge experiment sets a record for quantum teleportation at a distance of $1400$ kilometers~\cite{ren2017ground}.

\paragraph{Entanglement swapping}
Quantum teleportation provides a promising approach for remote state transmission by consuming pre-shared entanglement resources. But the preparation of long-range entanglement remains to be solved. Since the distance of direct transmission of entanglement is severely limited by the channel loss, entanglement swapping was introduced to extend the distance by converting multiple neighboring entangled pairs into one long-distance entangled pair~\cite{bennett1993teleporting,Zukowski1993,briegel1998quantum}. Technically, entanglement swapping can be understood as a variant of quantum teleportation. For instance, given two independent entangled pairs, say, entangled state $\ket{\Phi}_{AB_1}$ between Alice and Bob and entangled state $\ket{\Phi}_{B_2C}$ between Bob and Charlie, a new entangled pair between Alice and Charlie can be created by performing a Bell state measurement on Bob's two qubits, such that qubit $B_1$ is teleported to Charlie using the entangled link $\ket{\Phi}_{B_2C}$ and its entanglement with qubit $A$ is maintained. After a suitable correction by Charlie upon receiving Bob's measurement outcomes, an entangled pair between Alice and Charlie can be established. Notably, entanglement swapping can be recursively applied in practice to extend the effective distance of entanglement as demanded. Entanglement swapping has received significant attention since its initial proposal in 1998 as a subroutine of a quantum repeater~\cite{briegel1998quantum}, and has been experimentally demonstrated in different platforms under both lab and field conditions up to one hundred kilometers~\cite{pan1998experimental,kaltenbaek2009high,jennewein2001experimental,de2005long,wu2013high,halder2007entangling,xue2012polarization,takesue2009entanglement,sun2017field,pompili2021realization,huang2022,Liu2022alloptical}.

\paragraph{Entanglement routing}
Since two endnodes in a network may not be directly connected by a communication channel, data traveling between them would have to pass through a series of intermediary stations. Routing is the procedure used to determine the best route for data transmission across a network. Particularly, given that Bell pairs can be prepared between two adjacent nodes, entanglement routing considers choosing a specific network path to establish long-distance entanglement from the adjacent Bell pairs. Because quantum information suffers from noise and decoherence, some quantum factors, such as fidelity, coherence time, and success probability, should be considered when designing entanglement routing protocols, making them much more involved than their classical counterpart. Routing protocols based on the idea of classical techniques such as the Dijkstra’s algorithm are proposed, which should be easy to implement in the near term~\cite{van2013path,caleffi2017optimal,schoute2016shortcuts}. The essential strategy employed in these studies is first to create a path using a path selection algorithm with cost metrics tailored to quantum networks. Then entanglement can be distributed along the path. Most of these are centralized algorithms that work under the assumption that each node has information about the entire network, whereas global knowledge about the nodes or the properties of the entangled links may not always be available in practice. Decentralized routing schemes that use only local knowledge about the nodes and their neighbors have been considered in~\cite{gyongyosi2018decentralized,gyongyosi2017entanglement,chakraborty2019distributed}. More efforts in considering the scalability and practical limitations of entanglement routing can be found in~\cite{perseguers2013distribution,hahn2019quantum,pant2019routing,shi2020concurrent}.

\paragraph{Entanglement distillation} 
Quantum entanglement is a critical resource in quantum networks for teleporting quantum information~\cite{bennett1993teleporting}, reducing communication complexity~\cite{bennett1992communication}, and performing quantum cryptography tasks~\cite{bennett1984quantum,ekert1991quantum}. The reliability of quantum entanglement, however, may be undermined by environmental noise and faulty controls. To counteract such detrimental effects, entanglement distillation, also referred to as entanglement purification, was introduced, aiming to extract high-quality entangled states better suited for application from many copies of the noisy ones~\cite{bennett1996mixed}. More specifically, entanglement distillation usually starts with sharing many copies of entangled pairs between different parties and then performing unitary transformations on their individual quantum systems. After that, quantum measurements are performed on a portion of the entangled pairs, and the outcomes are shared through classical communication. Unitary transformations can again be performed on the rest of the entangled pairs, conditioned on the measurement outcomes. These processes can be applied recursively as needed. Note that entangled pairs being measured will be destroyed in the process, and when the test fails, the tested state must also be discarded. Consequently, there are three parameters of particular relevance in entanglement distillation: the rate of transformation of primitively entangled pairs into the final pairs, $r$, the success probability of distillation, $p$ and the fidelity of the final pairs, $F$. The interplay of these parameters has been extensively studied in quantum information theory~\cite{bennett1996purification,Deutsch1996quantum,buscemi2010distilling,brandao2011one,datta2014second,rozpkedek2018optimizing,Fang2022,regula2019one,fang2019non,zhao2021practical,fang2021sum,regula2021fundamental}. Notably, an "uncertainty" type of relation between the success probability and the fidelity has been recently discovered in~\cite{Fang2020nogo}, indicating that ideal distillation where $p=1,F=1$ is generically impossible. Experimental advances of entanglement distillation can be found in~\cite{kalb2017entanglement,abdelkhalek2016efficient,ecker2021experimental}.

\paragraph{Quantum error correction} 
Quantum error correction is a set of techniques used to protect quantum information against noise and errors in both computation and communication. Its main idea is to utilize the redundancy of physical qubits to protect the information of logical qubits. More specifically, quantum error correction maps $k$ qubits into $n$ qubits, where $n > k$. The $k$ qubits are the logical qubits that we want to protect from error, and the extra $n - k$ qubits help to encode the $k$ logical qubits redundantly so that the encoded information is not easily contaminated. Quantum error correction is more sophisticated than its classical counterpart due to the no-cloning theorem, the disturbance of quantum measurement, and the involved quantum error models. In general, error correction codes cannot protect against all possible errors, and they need to be designed to match the most likely types of noise. One of the most famous quantum error correction codes was given by Shor in 1995, which encodes the information of one logical qubit onto a highly entangled state of nine physical qubits, capable of correcting both bit flip and phase flip errors. Since then, quantum error correction has been an extremely active field of research~\cite{Gottesman1997,gaitan2008quantum,lidar_brun_2013,terhal2015quantum,suter2016colloquium}. Although the theory of quantum error correction has been worked out in great detail, the experimental difficulties of implementing it remain formidable.
Several proof-of-concept implementations have been recently demonstrated in~\cite{ofek2016extending,linke2017fault,Wootton2018repetition,harper2019fault,livingston2022experimental}. 

\section{Quantum network applications}
\label{sec: applications}

It is believed that the quantum internet in the future will significantly improve the capabilities of the existing internet in areas such as cryptography, computing, and sensing. Although it is difficult to foresee all its use cases, several representative applications have been identified. In this section, we discuss several quantum internet applications that are of particular interest. 

\subsection{Applications in quantum cryptography}

\subsubsection{Quantum key distribution}

Public key cryptosystems, such as RSA, are commonly employed for secure data transmission over the internet~\cite{rivest1978method}. However, the security of such systems relies on computational assumptions, such as factoring large integers is hard for an eavesdropper. Unfortunately, these assumptions can be threatened by advances in hardware and unanticipated algorithms for computation. Shor's quantum algorithm for efficient factorization represents a milestone in this regard~\cite{shor1994algorithms}. As primitive small-scale quantum computers have already been constructed and their computational power is rapidly growing every year (refer to the quantum Moore's law), the implementation of Shor's algorithm would be expected to happen in the near future, putting the conventional key distribution method and, consequently, the security of relevant applications in great danger. A countermeasure to this is quantum key distribution (QKD), which encodes secret keys into quantum information carriers for transmission~\cite{bennett1984quantum}. Thanks to the no-cloning theorem and the disruptive nature of quantum measurement, any eavesdropping on the distributed quantum states will unavoidably be detected. It has been proven that quantum key distribution is information-theoretically secure, meaning that the probability of any information leakage can be made arbitrarily small. Such a level of security is solely based on the fundamental laws of quantum physics and is unaffected by future advancements in computing power or algorithms.

While the application of QKD in large-scale quantum networks and high-security scenarios has a promising future (see~\cite{travagnin2019quantum} for a review of the national deployment of QKD networks), there are still numerous theoretical and experimental hurdles to overcome before its widespread commercial use can be realized. In particular, the unconditional security of QKD is based on ideally perfect single-photon sources and detectors. However, ideal devices do not exist in practice, which introduces security risks in the practical implementation of QKD protocols~\cite{zhang2018large}. Designing secure protocols with imperfect devices, such as decoy-state protocols~\cite{hwang2003quantum, wang2005beating, lo2005decoy} and device-independent protocols~\cite{lo2012measurement} is of particular importance. The security proof under different attacks and with imperfect devices should also be continuously strengthened~\cite{xu2020secure,pirandola2020advances,Ian2022}. Another challenge is the implementation of long-range key distribution with high key rates. One solution to fix the issue is to split a long-distance channel into multiple short-distance channels and connect them with quantum repeaters to expand the effective distance~\cite{sangouard2011quantum}. However, quantum repeaters impose stringent demands on quantum memories, which only provide limited performance at the moment and thus cannot meet the requirements of quantum repeaters yet. An alternative scheme is to replace quantum repeaters with trusted repeaters, but this will sacrifice the security at the middle stations. Another solution to extending the distance of key distribution is given by the technique of twin-field quantum key distribution~\cite{lucamarini2018overcoming, wang2022twin}, but the distance is still limited to a few hundreds of kilometers.

\subsubsection{Blind quantum computing}

Blind quantum computing (BQC), also known as secure delegated quantum computing, is a cloud-based quantum computing scheme that enables users to conduct calculations privately without disclosing any information to the server (see a review paper~\cite{fitzsimons2017private}). At the very least, there are two ways in which BQC benefits. First, most users will not own large devices (classical or quantum) due to the high cost and the requirement for specialized facilities to host them, making cloud computing the primary delivery mechanism for future quantum computing services. Second, as computational activities may involve sensitive information that the user does not want a third party to know, the blindness of BQC assures that no one in the network, even the service provider, cannot see what algorithm you are running or what data you are trying to process. This is advantageous for governments, non-governmental organizations, and corporations that want to solve computational problems that are politically or commercially sensitive. For example, a manufacturer can simulate an original material design without worrying that the service provider will figure out its secret.

The ideal scenario for implementing BQC would be a verifiable protocol between a client without any quantum capabilities (purely classical client) and a single quantum server. However, it has been shown that one-round perfectly secure BQC in this setting is impossible, i.e., the protocol cannot satisfy both the correctness of the result and the perfect blindness of the client's privacy~\cite{morimae2014impossibility}. There is also negative evidence that BQC is impossible with a purely classical client and a single quantum server~\cite{aaronson2017complexity}, even though the problem itself remains open. With these in mind, two relaxed scenarios are considered: a client with minor quantum capability (quantum state preparation or quantum measurement) remotely accesses one quantum server~\cite{broadbent2009universal,mantri2017universality,morimae2013blind} or a completely classical client remotely accesses multiple non-communicating quantum servers~\cite{reichardt2013classical,mckague2013interactive}. The latter is very similar to the source-replacement technique used in entanglement-based QKD, that is, to prepare a quantum state by measuring part of the entangled states. But the non-communicating assumption might be difficult to guarantee. Some proof-of-concept demonstrations of BQC have been given in~\cite{barz2012demonstration,huang2017experimental,greganti2016demonstration}. However, the lack of practical quantum memories and quantum computers makes it hard to implement large-scale BQC algorithms at present. The frequent classical communication between the client and service may also hamper the performance of BQC in practice, necessitating further improvements on both classical and quantum infrastructures.

\subsection{Applications in quantum computing}
\subsubsection{Quantum leader election}

Leader election is a primitive subroutine in distributed computing to avoid data inconsistencies and operation conflicts. It is intimately connected to several fundamental problems in distributed computing, such as the construction of spanning trees and the computation of functions on distributed inputs~\cite{Lynch1996}. The objective of the leader election problem is to select a single leader from a collection of distributed parties. For instance, when there is a writing request, we hope that only one processor in the distributed system will perform this task. If each party possesses a unique identifier, the problem can be handled deterministically by designating the party with the largest identifier as the leader. In the case of anonymous networks, the leader election problem can be handled probabilistically by tossing a coin~\cite{tani2012exact}. However, this approach cannot solve the problem exactly, i.e., in bounded time and with zero error, due to the chance that all parties would receive the same result from independent random bit generators. In fact, it has been proven that there exist an infinite number of graphs for which the problem cannot be exactly solved even with an unbounded (but finite) amount of communication~\cite{Angluin1980,itai1990symmetry,yamashita1996computing}. Quantum leader election aims to complete the leader election task with the help of quantum resources. It has been shown that the leader election problem can be exactly solved in $O(n^2)$ rounds with $O(n^4)$ bits of communication for the number $n$ of parties, assuming quantum computation and communication are permitted~\cite{tani2012exact}. This result was further improved in~\cite{kobayashi2014simpler} which achieves a round complexity linear in the number of parties, demonstrating significant superiority in the computational power of distributed quantum computing over its classical counterpart. Experimental demonstration of quantum leader election has been conducted in linear optics~\cite{okubo2008experimental}.

\subsubsection{Distributed quantum computing}

Distributed quantum computing usually refers to the act of performing a quantum computational task on two or more distinct quantum processors and combining the results to produce a complete output~\cite{cirac1999distributed,van2007communication,buhrman2003distributed,van2016path,cuomo2020towards,beals2013efficient}. Consider a quantum computational task that is divisible into several subroutines. Each of them can be executed on a distinct quantum processor, which may be located at different nodes in a computational network and may exchange data with a central processor. Each processor operates on a partial input with a fixed size, which may be uncorrelated, correlated, or even entangled. When the computation is finished, the central processor collects partial outputs from the other processors and integrates them into a final solution. In a sense, distributed quantum computing effectively combines multiple quantum computers into a single, more powerful system, thereby considerably improving the capability of information processing. When classical computers are unified by means of cloud-based distributed architectures, the net computational capacity is equal to the sum of the components. Quantum computers, however, have fundamentally distinct scaling characteristics: the power of a quantum computer increases exponentially with the number of quantum bits it processes. Consequently, upon integrating quantum devices via a quantum internet, we are left with something significantly more powerful than the sum of its parts. Distributed quantum computing is also applicable at the nanoscale, specifically in designing quantum chips. As the architecture of a single quantum processor approaches its limits, combining multiple processors with classical and quantum communication links could provide a promising solution~\cite{van2016path}. In IBM's recent development roadmap, it is considered that getting beyond single-chip processors is the key to overcoming the scalability problem~\cite{bravyi2022futurea,piveteau2022circuit}. Nevertheless, the distributed architecture would also introduce new challenges, particularly, the compilation of a quantum algorithm into a distributed manner and the realization of multi-qubit gates across different quantum processing units. Some frameworks for compiling distributed quantum programs have been recently proposed in~\cite{diadamo2021distributed,parekh2022quantum,haner2021distributed,ferrari2021compiler}, where non-local operations are replaced with local operations on shared entanglement~\cite{eisert2003optimal,gottesman1999demonstrating}. Advances in the implementation of nonlocal gates across different modules have been reported in~\cite{chou2018deterministic,wan2019quantum,daiss2021quantumlogic}. From another perspective, distributed quantum computing may also refer to computational tasks that are inherently distributed~\cite{cleve1998quantum,buhrman1998quantum,anshu2017exponential,Brukner2004,wei2019experimental}. One representative example is
quantum fingerprinting, where a third-party needs to check the consistency of the other two parties' input. Utilizing quantum superposition, the amount of communication can be exponentially smaller than the comparable classical case~\cite{buhrman2001quantum,zhong2021efficienta}.

\subsection{Applications in quantum sensing}

\subsubsection{Clock synchronization}

A wide range of technologies, such as data transfer networks, financial trading, airport traffic control, telecommunication networks, global positioning systems, and long baseline interferometry, rely heavily on access to a globally agreed-upon standard time, making clock synchronization a critical task on the internet. Conventionally, there are two basic techniques for synchronizing clocks. The usual procedure is Einstein Synchronization, which requires an operational line-of-sight exchange of light pulses between two observers~\cite{einstein1905elektrodynamik}. A less popular alternative is Eddington’s Slow Clock Transport~\cite{eddington1923mathematical}, in which the two clocks A and B are first synchronized locally and then transported adiabatically to their separate destinations. One obvious drawback of these methods is that either light or matter is physically exchanged, making them vulnerable to the quality of the intervening medium (e.g., the atmosphere). In 2000, Jozsa et al. proposed a clock synchronization scheme based on quantum entanglement that leverages pre-shared entanglement between two clocks located at different spatial locations~\cite{Jozsa2000}. The intrinsic correlation in quantum entanglement makes it a reliable method for synchronizing clocks regardless of the intervening medium or distance, distinguishing it from the conventional methods. The original two-partite quantum clock synchronization protocol has been improved and generalized to the multi-partite scenarios ~\cite{yurtsever2002lorentz,krvco2002quantum,ben2011optimized,komar2014quantum}. Experimental verifications of these protocols have been reported in~\cite{quan2016demonstration,giovannetti2001quantum,valencia2004distant,kong2017implementation}. In particular, experimental tests with quantum clock synchronization have shown timing precision up to $51$ picosecond~\cite{lee2019symmetrical}. It is believed that a highly accurate and secure global time distribution network in the future quantum internet would enable many terrestrial applications for scientific exploration, commercial utility, and national defense~\cite{troupe2022quantum}.

\subsubsection{Longer-baseline telescopes}
For a better understanding of the universe, we need to observe smaller and finer details of astronomical objects. This can be achieved by enlarging the primary mirror of a telescope and thereby increasing its resolution. However, the diameter of mirrors we can manufacture is limited. An alternative approach is using the interferometry technique, which combines telescope arrays into a virtual instrument with a diameter equivalent to the size of the separation between the individual telescopes. By interfering the light gathered at sites across the array, a synthetic aperture with resolution proportional to the length of the array can be realized~\cite{lawson2000principles}. Nonetheless, due to the photon loss during transmission, the baseline lengths of the interferometers are constrained, which subsequently restricts the resolution of the telescopes. One potential method to circumvent the limit is to adopt quantum teleportation-based interference, which was first suggested by Gottesman et al.~\cite{gottesman2012longer}. Utilizing post-selected quantum teleportation, one can measure the interference visibility (which is a function of the baseline and, according to the van Cittert-Zernike theorem, is the Fourier transform of the intensity distribution of the light source~\cite{zernike1938concept}) of all baselines of telescope arrays and image the light source. By introducing quantum repeaters, the photon loss during transmission can be effectively reduced, thus extending the baseline of interferometers. The scheme proposed by Gottesman et al., however, faces several technical challenges. For instance, it demands an extremely high rate of entanglement distribution as well as detectors with high efficiency and time resolution, which seems not feasible at the current stage. To provide a more practicable solution, new schemes have been proposed to reduce the technical requirements~\cite{khabiboulline2019optical, khabiboulline2019quantum}.

\section{Challenges}
\label{sec: challenges}

Despite the promising future of the quantum internet, there are still plenty of problems to overcome before a large-scale quantum network can be realized. These problems can come from various aspects, the most important of which are hardware, software, applications, and ecosystems. In this section, we will have a brief discussion on some of the most important issues.

\paragraph{Hardware} 
The functionality of future quantum network services and applications depends on the coordinated development of both quantum and classical hardware. Generally, the present quantum network components have mostly been tested in a lab setting and have not yet been put to use in a full network configuration. However, it can be inferred from classical networks that to create wide-area, operational quantum networks, more capable devices satisfying suitable requirements for reliability, scalability, and maintenance are needed. These include high-rate quantum entanglement sources; robust quantum memory buffers; highly efficient photon detectors; high-speed, low-loss quantum switches; and transducers from optical and telecommunications regimes to quantum computer-relevant domains and vice versa. In addition, the development of classical communication is also important for quantum networks, as classical control information is indispensable for the functionality of quantum network protocols. For instance, the implementation of entanglement swapping and quantum teleportation requires the classical message to do suitable correction; quantum key distribution requires the classical message to do key sifting and parameter estimation; and blind quantum computing requires the classical message to control the measurement angles of computation, etc. Besides the hardware performance, another thing to consider is the cost. The fact is that current quantum-related devices are extremely expensive and apparently beyond the reach of most users. To make quantum network services widely accessible to the public, their operating and maintenance costs need to be significantly reduced.

\paragraph{Software}

The nature of quantum information determines that fundamental operations like information readout and duplication in quantum networks are fundamentally different from those in classical networks. Consequently, a one-to-one mapping from classical network architecture (e.g., TCP/IP or OSI) to quantum networks appears implausible. Instead, a significant paradigm shift is required for the design of a quantum network architecture that exploits the peculiarities of quantum mechanics. This leaves us with tons of challenges in building quantum network architectures for efficient and robust services, developing quantum operating systems for resource and error management, and designing software interfaces for network users~\cite{cacciapuoti2019quantum}. Challenges also include entanglement generation over multiple users, as well as the integration of the existing networks with quantum network technologies for optimal control and management. Despite the fact that several architecture principles have been proposed, it is not yet possible to evaluate, compare, and determine exactly which one is superior in terms of network performance, readiness for implementation, and scalability for large networks. A systematic benchmarking framework with relevant simulation and evaluation tools is in high demand. 

\paragraph{Applications}

Quantum networks are proven to provide advantages in various aspects, such as computation, communication, and sensing. However, most of the existing applications are hard to realize on current devices with limited performance. Based on the experience with classical networks, the development of network applications and network infrastructure are strongly intertwined. On the one hand, the development of network infrastructure facilitates the creation of suitable applications, while on the other hand, the pursuit of better applications will stimulate the development of network infrastructure. Instead of waiting until the quantum network infrastructure is fully built, it is more important to design applications that work at different stages. This demands the collaboration of theorists and experimentalists, as well as innovations from both network designers and users. The currently known applications are more on the theoretical side to demonstrate the superiority of quantum networks. Yet these are just the tip of the iceberg, and there are tons of real-world application scenarios and use cases in all walks of life that remain to be explored.

\paragraph{Ecosystem}

Several prototype quantum networks have been demonstrated in the lab environment. But the transition from the demonstration to the operational infrastructure requires the collaborative efforts of all communities. Particularly, it requires scientific researchers to lay the foundation of quantum network principles; governments and business investors to finance the construction of network infrastructures; and users and customers to actually make use of the services and propose new demands on applications. It is anticipated that existing technologies will be significantly disrupted by the future quantum internet. But from another perspective, the introduction of new things inevitably results in the emergence of new challenges. When large-scale quantum networks become a reality, new issues about security, privacy, ethics, and legislation will arise. For instance, many quantum network applications are proven to be unconditionally secure in principle. But they may not be hack-proof in practice~\cite{satoh2021attacking}. With this in mind, an open and sustainable quantum network ecosystem is demanded along with the development of quantum network technologies.

\section{QNET: a quantum network toolkit}
\label{sec: QNET}

\begin{figure*}[ht]
\centering
\includegraphics[width=\textwidth]{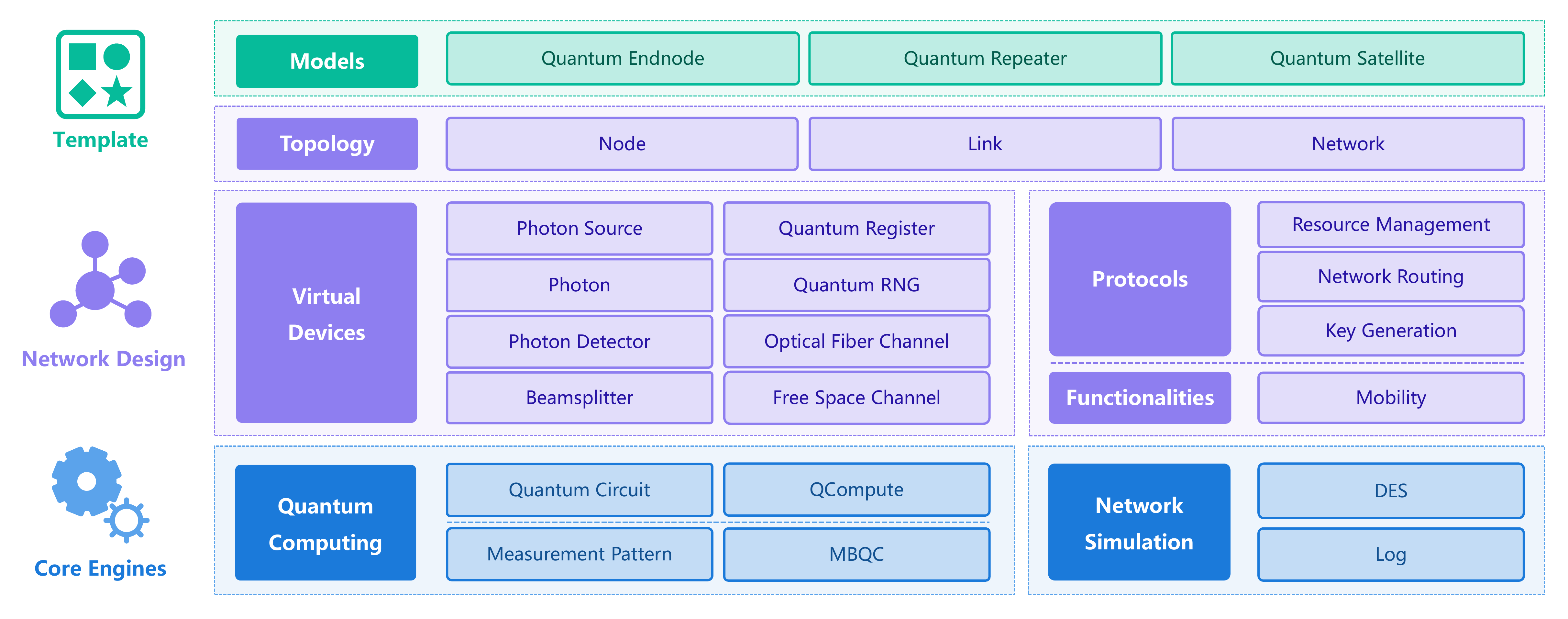}
\caption{Overview of QNET's key modules (based on version 1.2.0). The bottom layer provides core engines for both quantum computation and network simulation. The middle layer offers tools such as virtual devices and protocols for designing a quantum network with customized topology. The top layer provides the users with various easy-to-use model templates for convenience. A UML class diagram laying out the structure of QNET with more details is given in Appendix~\ref{sec: UML class diagram}. Note that QNET is under active development, and the latest version can be accessed at~\cite{QNET}.}
\label{fig: architecture}
\end{figure*}

\begin{figure}[H]
\centering
\includegraphics[width=8cm]{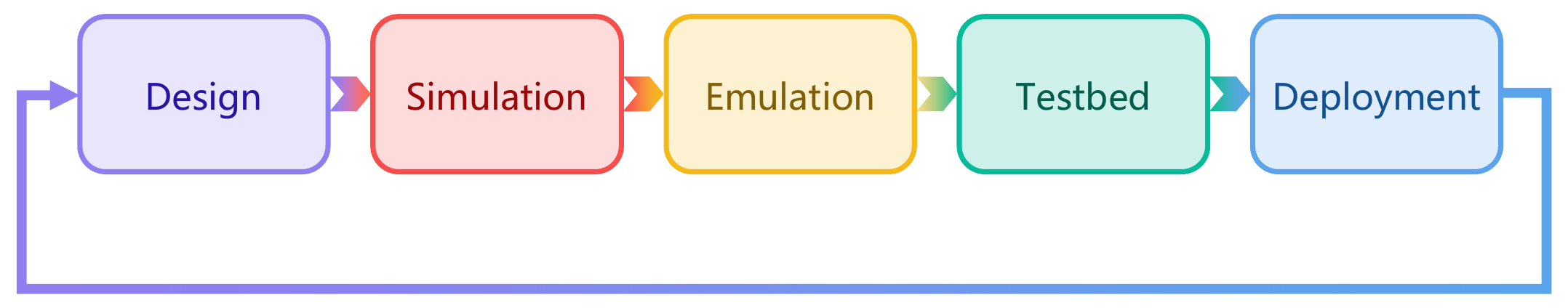}
\caption{The life cycle of a network protocol. A network protocol is born from an initial design and then goes through simulation and emulation steps. After that, the protocol will be evaluated in a small testbed and finally go through the deployment and standardization processes. Sometimes it may also go back to the design step for further calibrations and improvements.}
\label{fig: protocol life cycle}
\end{figure}

A network protocol, whether classical or quantum, usually goes through several stages from its initial design to the final deployment~\cite{wehrle2010modeling}. As depicted in Figure~\ref{fig: protocol life cycle}, once a network protocol is born, it will first undergo simulation-level testing to ensure that it performs as expected. The process then advances to the emulation phase, which entails a more comprehensive simulation of both functionality and execution logic. The emulation of a protocol should simulate the working environment of the protocol as precisely as possible and incorporate related hardware devices, either virtually or physically. The protocol will subsequently be implemented on a small testbed to check whether it works properly in practice. If everything goes as expected, the protocol will be deployed and standardized on a large-scale network. Of course, a protocol may also return to the design phase for more calibrations and improvements.

As quantum networks are still under development, a simulation toolbox would be essential for verifying new ideas and reducing possible trial-and-error costs. Here we introduce QNET, a quantum network toolkit~\cite{QNET}, which aims to stimulate the design of quantum network protocols, accelerate the testing of quantum network architectures, and facilitate the deployment of quantum internet standards. QNET is an open-source project implemented in Python. An overview of its key modules is presented in Figure~\ref{fig: architecture}. The current version (v1.2.0) of QNET mainly contains three layers of functional modules. The bottom layer provides core engines for both quantum computation and network simulation. The middle layer offers tools such as virtual devices and protocols for designing quantum networks with customized topologies. The top layer provides various frequently-used model templates to help speed up the workflow of the research and development on quantum networks. QNET, in particular, contains a fully-featured discrete-event simulation framework for precise and efficient tracking of quantum network status during simulation. It also features dual quantum computing engines that support simulations in both the quantum circuit model and the measurement-based quantum computation model. Furthermore, QNET provides detailed modeling of the underlying physical devices. Its simulation is fine-grained down to the photon level, enabling verification of real-world quantum network experiments. QNET also provides a solution for compiling quantum network protocols into standard quantum circuits, supporting the simulation of network protocols on real-world quantum hardware devices. With its modular design, QNET can serve as a sandbox for testing different quantum network architectures.

In the remainder of this section, we will go through QNET in more detail. We will first discuss the key techniques for its implementation, followed by a quick-start guide to help to understand its basic usage. Lastly, we will use several featured demonstrations to show how QNET can be used in practice.

\subsection{Key techniques}
\subsubsection{Discrete-event simulation}
\label{sec: des}

\begin{figure}[H]
    \centering
    \includegraphics[width=8cm]{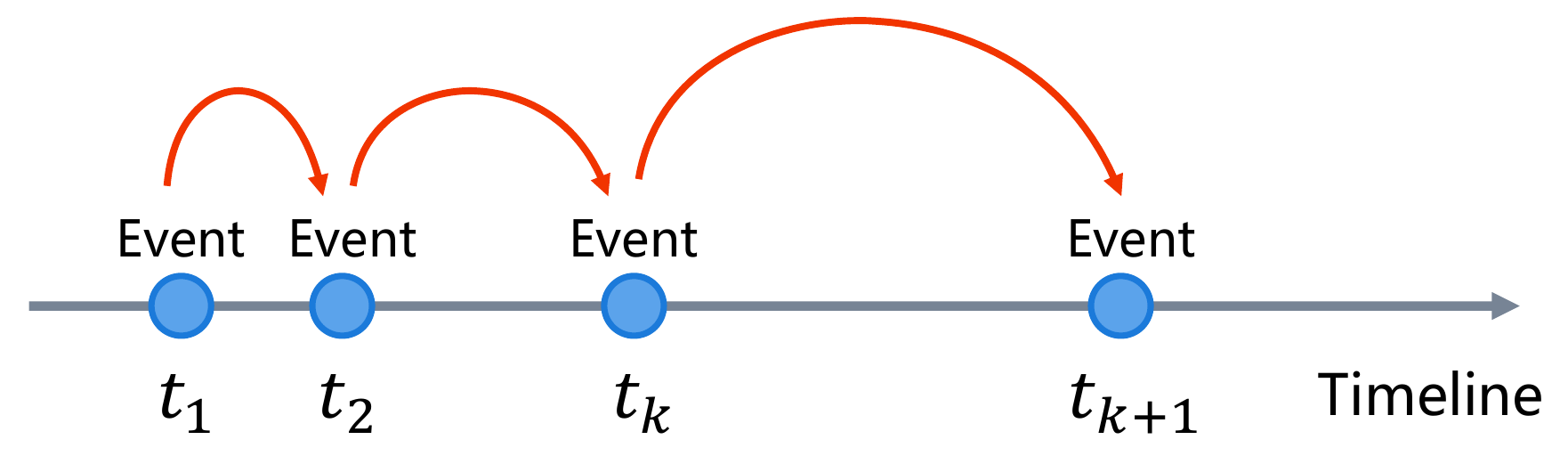}
    \caption{A schematic diagram of discrete-event simulation. Every change in the network status is modeled as an event on the timeline. The simulation proceeds from event to event and holds still until the next event happens. The simulation terminates when all events are executed or a termination condition is reached.}
    \label{fig: des}
\end{figure}

Discrete-event simulation (DES) is a well-studied modeling technique widely used in problems such as dispatching, resource allocation, capacity planning, and particularly network simulations~\cite{wehrle2010modeling}. A schematic diagram of discrete-event simulation is depicted in Figure~\ref{fig: des}. In this framework, all changes in the network are regarded as a series of events occurring at discrete time points. These events are then scheduled and processed in chronological order along the timeline. In general, discrete-event simulation has the following characteristics: 1) randomness, i.e., state variables of the simulation system contain random factors; 2) dynamicity, i.e., the time evolution dynamically affects the system status, which may create new events in the future as the current event is executed; and 3) discreteness, i.e., the system status only changes when the corresponding event is executed and holds still until the next event happens.

Given that DES is well-suited to network simulation, we develop a fully-featured DES engine in QNET to accurately and efficiently track the system status in quantum network simulation. The design philosophy of our DES engine mostly refers to~\cite{wehrle2010modeling}. In particular, it contains the following key elements: 1) {"DESEnv"}, i.e., the simulation environment with a built-in timeline that drives the simulation forward; 2) {"Entity"}, i.e., the main object of the system to simulate, such as an endnode or a physical device; 3) {"Event"}, i.e., the change of the system status at a particular time, which may include the message transmission and reception among network users; 4) {"EventHandler"}, i.e., the specific action of an entity or protocol, such as emitting a photon from the photon source; 5) {"Scheduler"}, i.e., a scheduler that schedules events onto a timeline, usually included as a component in objects that need to schedule events, such as entities or protocols. 

The DES technique has been adopted by many quantum network simulators. Unlike SeQUeNCe, which delegates event scheduling entirely to the timeline, or \text{NetSquid} that regards protocols as virtual entities and allows them to interact directly with the timeline, QNET uses the design of on-demand event schedulers that can be embedded into any object as needed, leaving the flexibility of event scheduling to the users. By default, all components in a quantum network and the network itself are considered entities, each with a built-in scheduler for event scheduling. As a network protocol typically changes the network status and thus needs to schedule events, we set by default that each protocol has a built-in scheduler as well. 

\begin{figure}[htbp]
\centering
\includegraphics[width=0.48\textwidth]{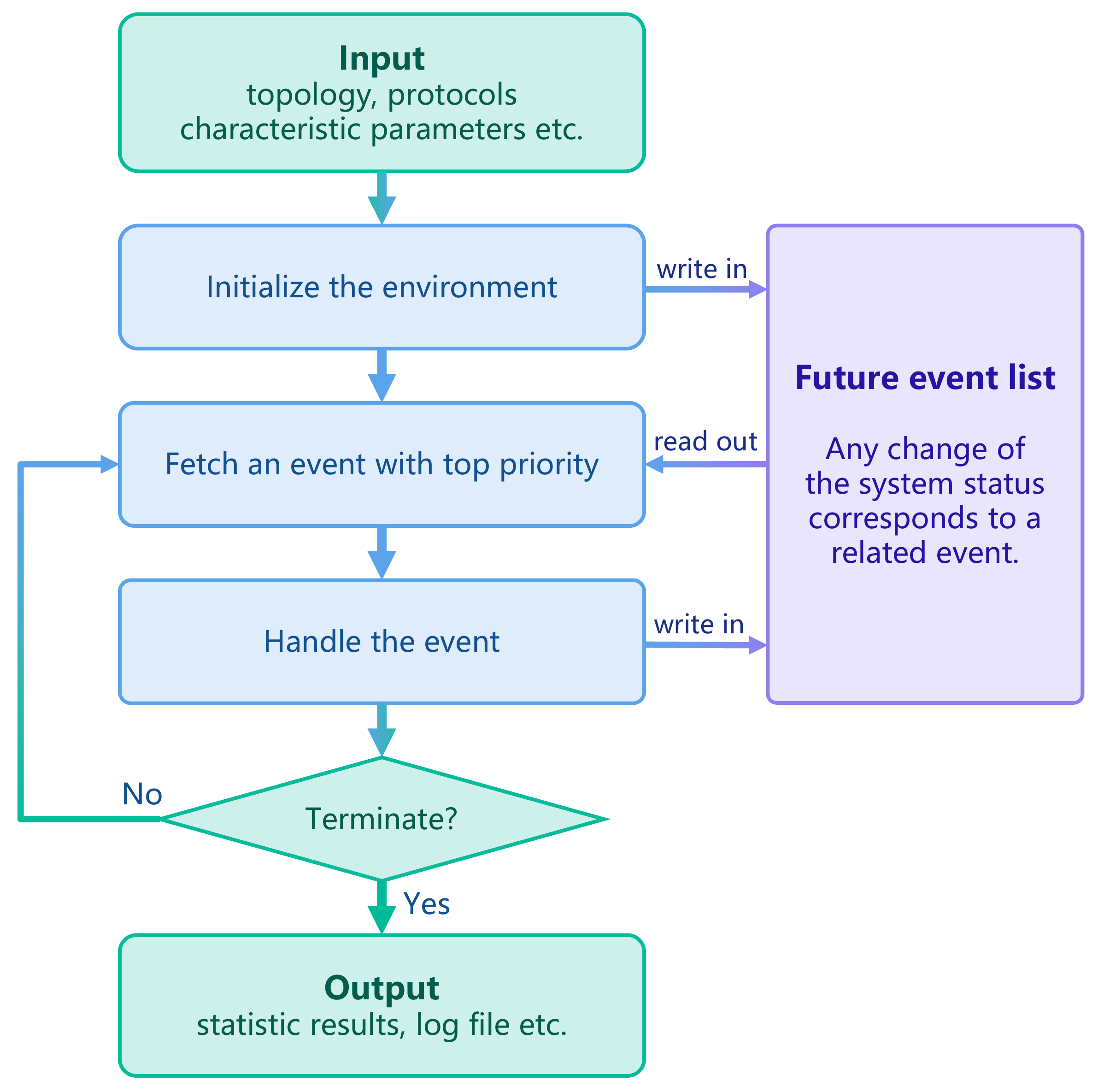}
\caption{A flow chart of the algorithm for the discrete-event simulation. This algorithm takes network topology, protocols, and some characteristic parameters as input and finally outputs the statistical results and log files for network analysis. In the main loop, the algorithm interacts with a future event list that maintains the evolution of the network. At each step, an event with the highest priority is fetched from the future event list and executed according to the given instructions. In the meantime, new events can be scheduled and pushed into the future event list.}
\label{fig: des_algorithm}
\end{figure}

A flow chart of the DES algorithm is depicted in Figure~\ref{fig: des_algorithm}. This algorithm takes network topology, protocols, and some characteristic parameters as input and finally outputs the statistical results and log files for network analysis. In the beginning, the simulation environment is initialized. This will automatically trigger the initialization of all entities in the environment. Meanwhile, relevant events will be written into the future event list, which is maintained as a specialized tree-based data structure (i.e., heap). An event with the top priority is then fetched from the future event list and gets executed according to the given instructions. In the meantime, new events can be scheduled and pushed into the future event list. Once the processing of an event is complete, the next event is fetched from the future event list and gets executed. The main loop of this algorithm will keep running until all events are executed, or a termination condition is reached.

\subsubsection{Dual quantum computing engines}
\label{subsubsec: dual}

\begin{figure}[ht]
\centering
\includegraphics[width=8.2cm]{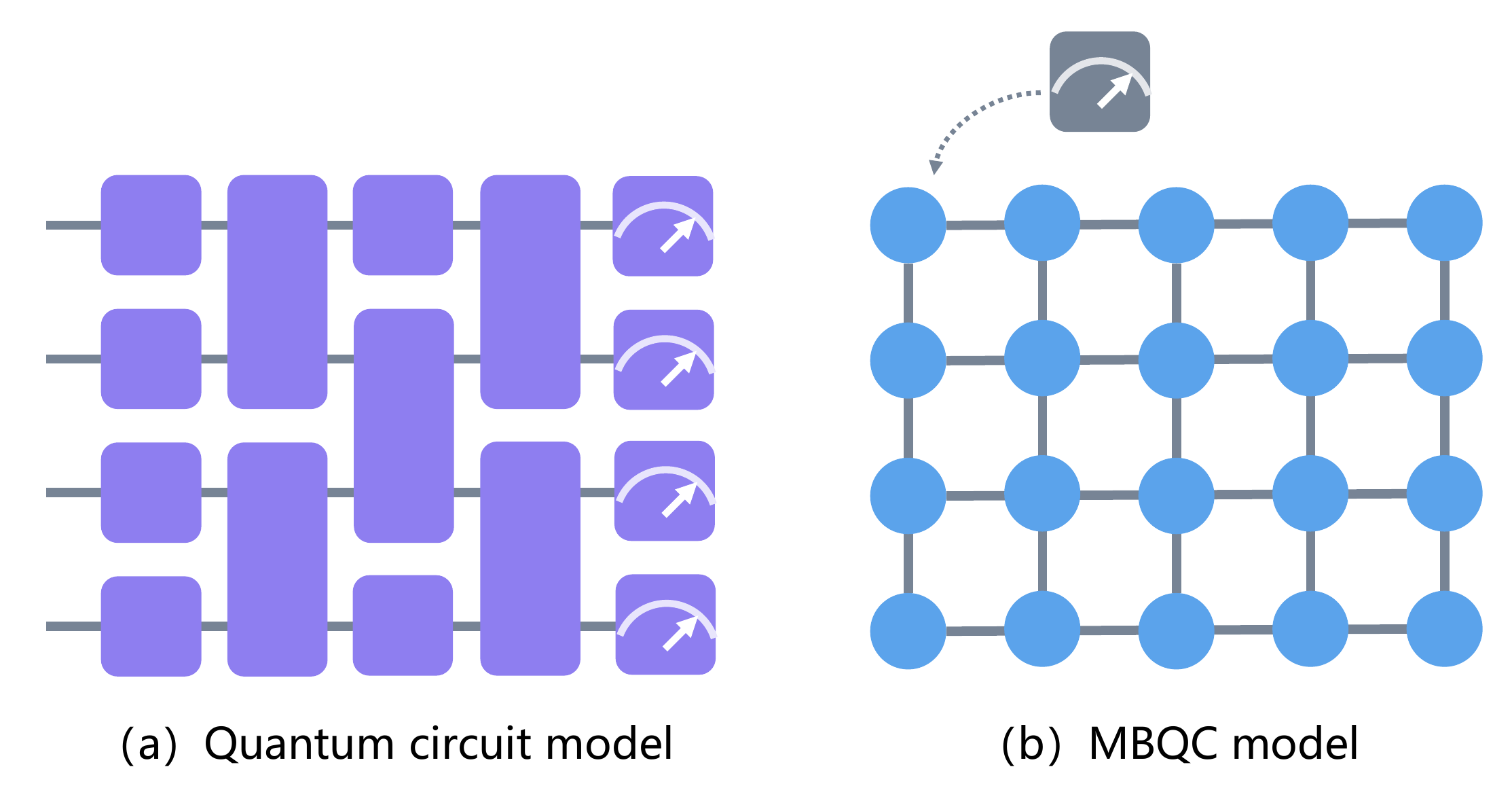}
\caption{Schematic diagrams of two quantum computation models. The diagram on the left represents the quantum circuit model, where computation is accomplished by performing quantum gates, and results are read out by the final quantum measurements. The diagram on the right represents the measurement-based quantum computation model, where the computation is realized by measuring each qubit on a prepared graph state.}
\label{fig: computation_models}
\end{figure}

Quantum computation utilizes the peculiar laws of the quantum world and provides a novel and promising way of processing information. The essence of quantum computation is to evolve the initially prepared quantum state into another expected one and then make measurements to obtain the required classical results. However, the approaches to quantum state evolution vary in different computation models. The widely used quantum circuit model completes the evolution by performing quantum gate operations, which can be regarded as a quantum analog of the classical computing model~\cite{deutsch1989quantum,barenco1995elementary}. Another completely different approach is given by measurement-based quantum computation (MBQC), which is a unique quantum computation model with no classical analog. This model steers the computation by measuring part of the qubits of an entangled state, with those remaining unmeasured undergoing the evolution correspondingly. By controlling measurements, we can complete any desired evolution. More specifically, the computation in MBQC is mainly divided into three steps. The first step is to prepare a resource state, which is a highly entangled many-body quantum state. This state can be prepared offline and independent of specific computational tasks. The second step is to sequentially perform single-qubit measurements on each qubit of the prepared resource state, where subsequent measurements can depend on previous measurement outcomes; that is, measurements can be adaptive. The third step is to perform classical data processing on measurement outcomes to obtain the required computation results. A schematic illustration of MBQC algorithms is shown in Figure~\ref{fig: computation_models}. The grid represents a commonly used quantum resource state (called \emph{cluster state}). Each vertex on the grid represents a qubit, while the entire grid represents a highly entangled quantum state. We measure each qubit one by one in a specific measurement basis to drive the computation.

Quantum network architectures based on different computation models have been proposed~\cite{pirker2019quantum,irtf-qirg-principles-11,van2021quantum,dahlberg2019link,kozlowski2020designing,zwerger2012measurement,zwerger2016measurement,pirker2018modular}, and each architecture admits its own features. For instance, the quantum circuit model has been widely used in current quantum technology. Its working mechanism nicely matches our intuition, which facilitates the migration of classical network architectures to quantum networks. On the other hand, the single-qubit measurements in MBQC are much easier to implement in practice, and non-adaptive quantum measurements can be carried out simultaneously, dramatically reducing the computational depth and consequently the effect of decoherence on fidelity. Moreover, the resource states in MBQC can be independent of specific computational tasks, which can be applied in blind quantum computing (or secure delegated quantum computing in some literature) for privacy preservation~\cite{broadbent2009universal, morimae2014verification,fang5}. To better utilize the characteristics of different computational models, it is likely that a hybrid approach will be adopted in the future quantum internet. Therefore, we develop dual quantum computation engines in QNET to support the simulation of both models. In particular, as far as we are concerned, QNET is the \emph{first} quantum network toolkit that supports the implementation of MBQC algorithms.

As mentioned, MBQC processes the computation in three steps: resource state preparation, single-qubit measurement, and data post-processing. Each resource state corresponds to a graph $G = (V, E)$ where $V$ is the set of vertices and $E$ is the set of edges representing the entangling correlation in the state. Suppose we have a $n$-qubit resource state on systems $V$ whose $m$ qubits are initialized as $\ket{\varphi}$ on systems $I \subseteq V$ and the rest qubits on $V\backslash I$ are initialized as plus states $\ket{+} = (\ket{0}+\ket{1})/\sqrt{2}$. Then the computation can be represented as
\begin{align}\label{eq: MBQC steps}
   \bigg(\prod_{k \in V} M_k\bigg)\bigg(\prod_{(i,j) \in E} \CZ_{ij} \bigg)\bigg(\ket{\varphi}_I \ox \ket{+}^{\ox n-m}\bigg),
\end{align}
where $\CZ_{i,j}$ is a controlled-phase gate on qubits $i$ and $j$, and $M_k$ is a measurement on qubit $k$. After the measurements, we obtain a sequence of outcomes and then perform classical post-processing. To simulate the computation according to Eq.~\eqref{eq: MBQC steps}, it is necessary to store the resource state in a vector with a size of $2^n$. This exponential memory cost can be a problem when $n$ is larger than several dozen, which is usually the case in practice. To simulate the computational process more efficiently, we design a \emph{vertex dynamic classification algorithm} (see Algorithm~\ref{Alg: Vertex dynamic classification}) that divides all the vertices into three categories (active, pending, measured) and dynamically maintains these vertex sets as the computation runs forward~\cite{fang2}. The intuition behind this is that single-qubit measurements only affect the evolution of their neighboring qubits. So we can postpone the activation of irrelevant qubits at the current step. It can also be understood as changing the execution order of entanglement and measurement operations to reduce the number of effective qubits involved in the computation and thereby improve storage and computational efficiency. This technique helps to support the simulation of MBQC algorithms up to thousands of qubits in certain cases. More details of our MBQC simulator and its usage can be found in~\cite{QNET}.

\begin{algorithm}[H]
    \caption{Vertex dynamic classification}
    \label{Alg: Vertex dynamic classification}
	\KwIn{graph $G = (V, E)$ of the resource state, initial input state $\ket{\varphi}$ on systems $I$, measurement operations $M_1,\cdots,M_n$}    
	\KwOut{measurement outcomes $o_1,\cdots,o_n$}    
	\vspace{0.2cm}
	
	1. initialize three empty lists, $V_p, V_a, V_m$;
	
	2. let $V_a = I$, $V_p = V\backslash I$, and the background state $\ket{\psi} = \ket{\varphi}$;
	
	\ForEach{$k \in V$}{
	3. if $k \in V_p$, then append $k$ to $V_a$ and remove it from $V_p$;
	
	4. find all neighbors of $k$ in $G$, denoted as $N(k)$, and compute $W = N(k) \backslash (V_m \cup V_a)$;
	
	5. append $W$ to $V_a$ and remove it from $V_p$;
	
	6. initialize the newly added qubits $W$ as plus states, take tensor product with $\ket{\psi}$, and update the background state $\ket{\psi}$;
	
	7. perform quantum gate $\CZ_{j,k}$ on $\ket{\psi}$ for any $j \in W$ and update the  background state $\ket{\psi}$;
	
	8. perform measurement on qubit $k$ by $M_k$ and record the measurement outcome $o_k$;
	
	9. append $k$ to $V_m$ and remove it from $V_a$.
	
	}
\end{algorithm}

\subsubsection{Compilation of quantum network protocols}
\label{sec: protocol mapping}

As mentioned at the beginning of Section~\ref{sec: QNET}, a network protocol needs to pass through a number of development stages before it can be deployed to the network infrastructures (see the life cycle of a network protocol in Figure~\ref{fig: protocol life cycle}). Emulation, which runs quantum network protocols in the actual quantum environment, is a crucial stage that comes before the testbed. By utilizing the power of quantum computers, we can remove the computational bottleneck of simulating quantum networks with classical computers while also optimizing relevant parameters and checking the protocol's performance in the quantum environment.

\begin{figure}[ht]
\centering
\includegraphics[width=8.2cm]{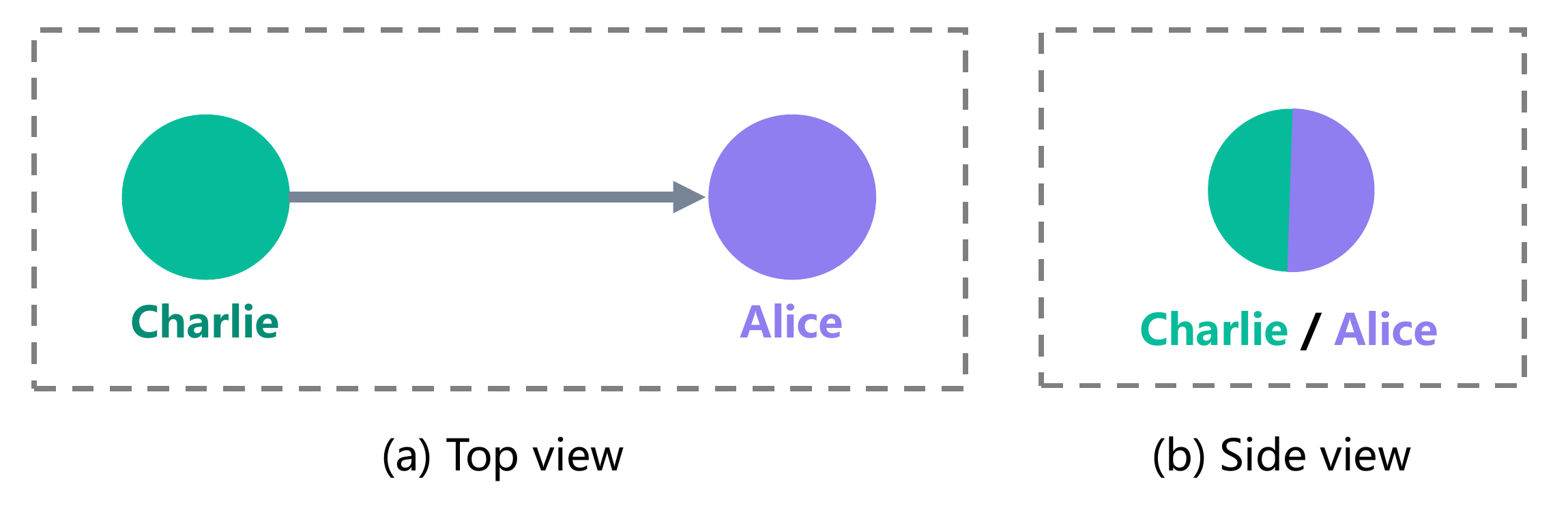}
\caption{An illustration of the qubit transmission in the protocol compilation scheme. The diagram on the left represents the spatial transmission of a qubit from Charlie to Alice through a quantum channel. The diagram on the right represents the perspective of the receiver, Alice. The qubit can be considered to hold its position still, and the transmission corresponds to the transfer of its ownership.}
\label{fig: qubit transmission}
\end{figure}

\begin{figure}[ht]
\centering
\includegraphics[width=8.2cm]{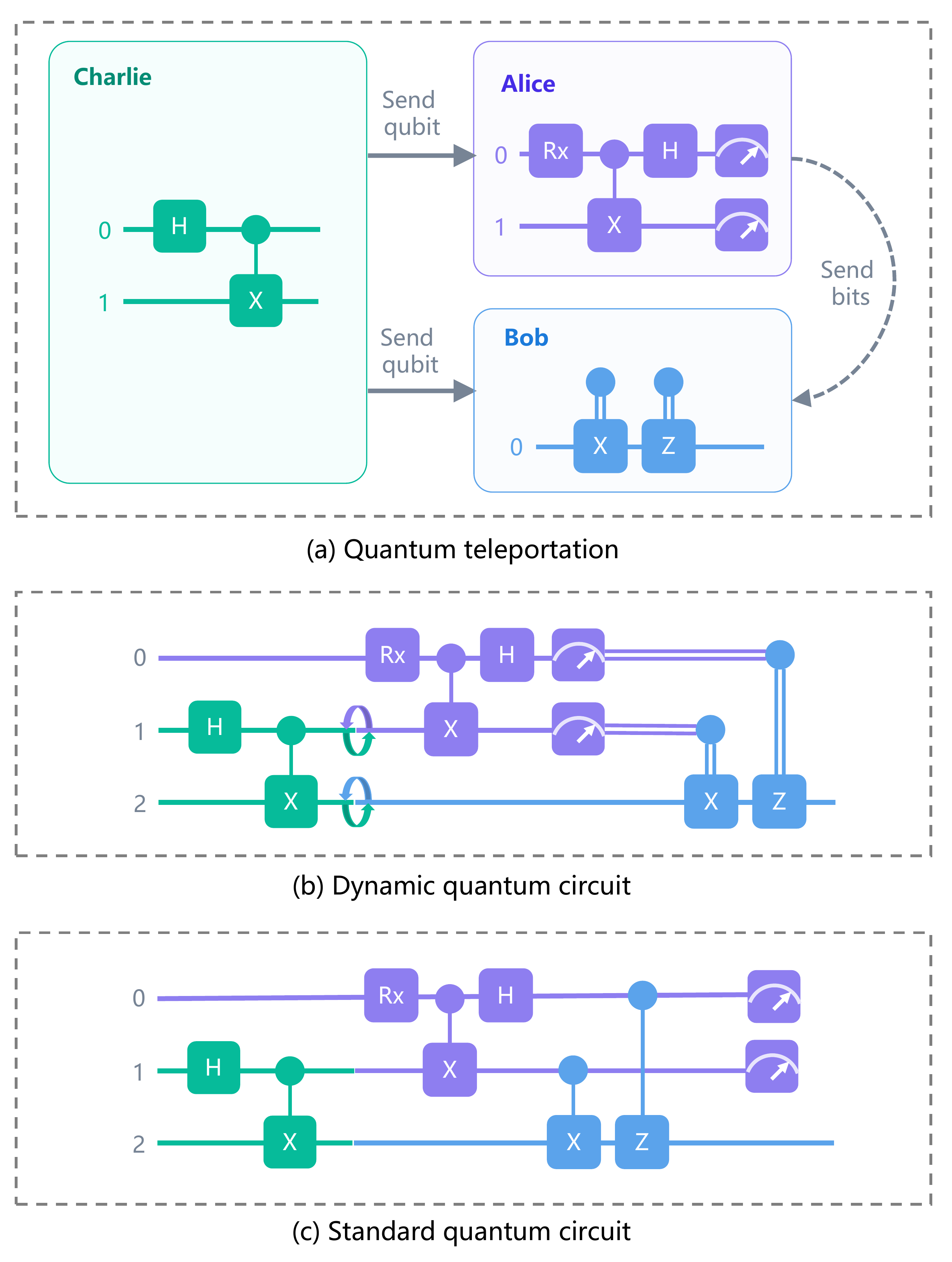}
\caption{An illustrative example of compiling a quantum network protocol into a standard quantum circuit. The first step is realized by replacing the spatial transmission of a qubit with the transfer of its ownership. The second step is done by using the principle of deferred quantum measurement.}
\label{fig: protocol mapping}
\end{figure}

A number of commercially available quantum computers, such as those based on superconducting or trapped-ion qubits, can now support the implementation of standard quantum circuit programs, consisting of initial state preparation, quantum gate evolution, and final quantum measurement. However, a quantum circuit program is not the same as a quantum network protocol, which features both classical and quantum communications between different nodes (not native to quantum circuit programs). This makes it hard to run a quantum network protocol on the current quantum computers. To solve this, we propose a compilation scheme that maps quantum network protocols onto standard quantum circuits in two steps~\cite{fang10}, as shown in Figure~\ref{fig: protocol mapping}. In the first step, we compile the quantum network protocols into dynamic quantum circuits with conditional operations. In the second step, we use the principle of deferred measurement to replace classically controlled operations with quantum-controlled operations~\cite{nielsen2010quantum}. Finally, the resulting circuit is a standard quantum circuit that can be delegated to a quantum computer.
The difficulty of protocol compilation mainly comes from the first step, as we need to emulate quantum communication on a single quantum device. Here, we propose to treat the spatial transmission of a qubit as the transfer of its ownership from the sender to the receiver. That is, Alice sending a qubit to Bob through a quantum channel can be viewed as Alice handing over the ownership of this qubit to Bob. A pictorial description is shown in Figure~\ref{fig: qubit transmission}. In this way, we can simulate the qubit transmission through an identity channel. As for more general noisy quantum channels, they can be simulated by adding suitable gate operations before and after the transmission.

Let us take quantum teleportation as an example. This protocol involves three parties: Alice, Bob and Charlie. Each of them has its own local operations and will interact through classical or quantum communications (see Figure~\ref{fig: protocol mapping}(a)). First, Charlie prepares a Bell state on his two local registers and sends the first qubit to Alice and the second to Bob. Alice prepares a qubit to send (e.g., by a rotation-x gate) and receives the qubit from Charlie, then she takes a Bell state measurement on both qubits and sends the classical outcomes to Bob. Once receiving the qubit from Charlie and classical messages from Alice, Bob performs conditional Pauli operations to restore the qubit that Alice wants to send. By using the idea in Figure~\ref{fig: qubit transmission}, QNET automatically compiles this tripartite network protocol into a dynamic quantum circuit, which contains all information about the local operations as well as classical and quantum communications (see Figure~\ref{fig: protocol mapping}(b)). Then this circuit will be compiled into an equivalent standard circuit, which can be delegated to a quantum computer (see Figure~\ref{fig: protocol mapping}(c)). Notably, all of these compilations are automatically handled in QNET, and users only need to focus on the native descriptions of the network protocols. More technical details about the compilation can be found in Appendix~\ref{sec: protocol mapping supplements}. To the best of our knowledge, QNET is the \emph{first} quantum network simulator that supports the compilation of quantum network protocols to quantum circuits as well as the emulation on quantum hardware devices. The other project capable of interacting with quantum hardware is NetQASM, whose public interface to hardware is still under development at the time of writing this manuscript~\cite{dahlberg2021netqasma}.

\subsection{Quick start guide}

\begin{figure}[ht]
    \centering
    \includegraphics[width=8.2cm]{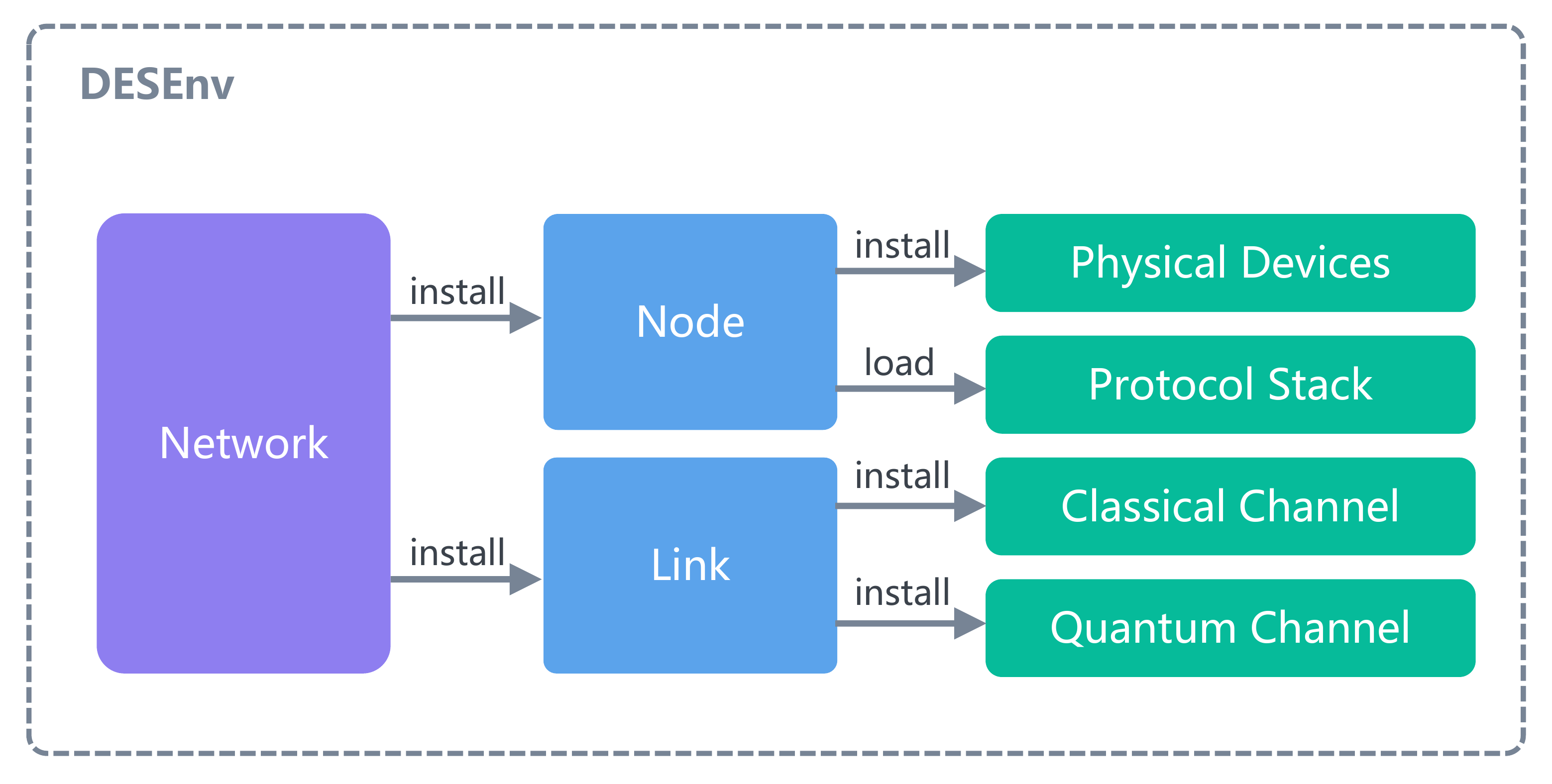}
    \caption{The relationship between different network components. DESEnv is the simulation environment underlying all network components. A network is composed of nodes and links, where each node needs to install the necessary physical devices and load a protocol stack, and each link should contain classical and/or quantum channels.}
    \label{fig: components_relation}
\end{figure}

In this section, we introduce the basic usage of QNET with the corresponding code demonstrations. Note that all the syntax highlighted in the shaded box (e.g., \code{ABC}) corresponds to the Python codes in QNET. As with a classical network, a quantum network comprises nodes and links. Nodes can install specific physical devices and operate in accordance with the loaded protocol stack, whereas network links usually contain classical and quantum channels between nodes. A quantum network and its components all belong to discrete-event simulation entities in QNET, that is, the corresponding classes, including \code{Network}, \code{Node}, \code{Link}, \code{Channel} and other physical devices, all inherit from the class \code{Entity}. Every entity can add its components using the \code{install} method. A network, for example, can install nodes and links; a link can install physical channels, and so on. Moreover, we remark that each entity has a built-in scheduler for event scheduling. As a scheduler needs to find the corresponding simulation environment to schedule events on the timeline, users are required to specify the simulation environment whenever instantiating an entity. Otherwise, it will be attached to the default environment. With these basic settings, we can proceed to introduce each component in further detail and show how to put them together to form a quantum network. The relationship between different network components is summarized in Figure~\ref{fig: components_relation}.

\subsubsection{Simulation environment}

\begin{figure}[ht]
    \centering
    \includegraphics[width=0.48\textwidth]{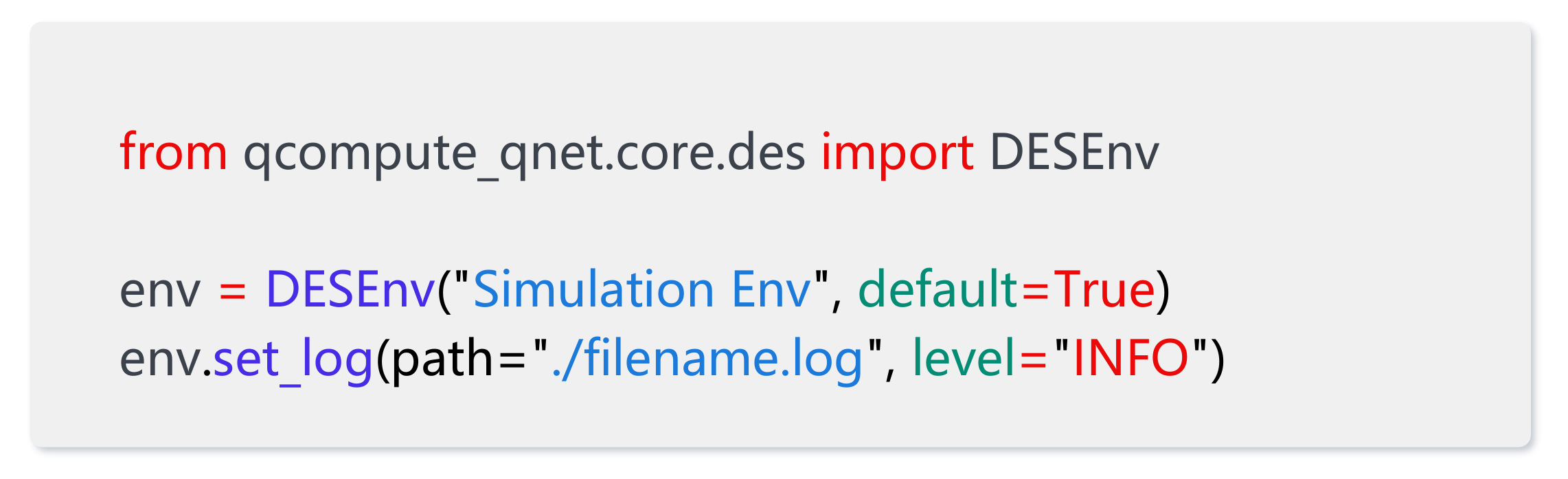}
    \caption{Code example for creating a simulation environment.}
    \label{fig: quickstart-des}
\end{figure}

Here we begin with the simulation environment. Before running a simulation, we need first to create a discrete-event simulation environment \code{DESEnv} that is responsible for driving the simulation of the entire network. Moreover, we can set the simulation environment to default (i.e., set the parameter \code{default=True}) to allow the entities to attach themselves to the default environment automatically. This can be done by calling the method \code{set\_default\_env} as well. In addition, the simulation environment provides a logging function to record critical information during the simulation. We can also specify parameters of the logging function, such as the path of the output file or the level of log records, by \code{set\_log}. A code example is given in Figure~\ref{fig: quickstart-des}.

\begin{figure}[H]
    \centering
    \includegraphics[width=0.48\textwidth]{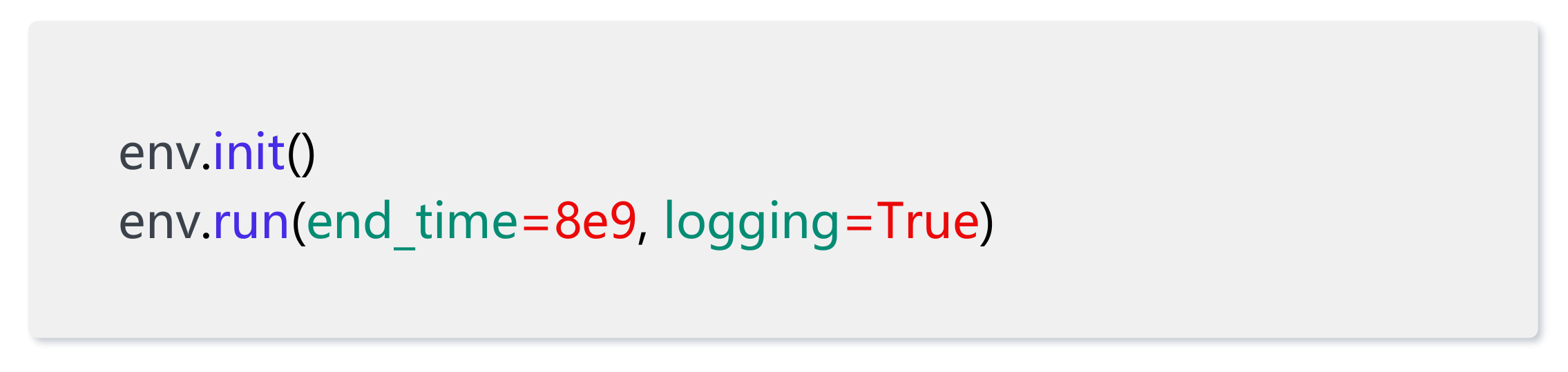}
    \caption{Code example for running the simulation.}
    \label{fig:quick-start-env-init}
\end{figure}

Now we have successfully created a simulation environment and set its configuration. The next step is to construct the quantum network we want to simulate, which will be detailed in subsequent sections. Once this is done, we need to call the \code{init} method to initialize the environment. This step is mandatory, and it will automatically call the \code{init} method of each \code{Entity} for initialization. After this, we can launch the network simulation by calling the \code{run} method of \code{DESEnv}. The duration of the simulation can be specified by passing the parameter \code{end\_time} (the time unit is picosecond by default, and the time we talk about here is not the actual CPU time but a virtual time). If no \code{end\_time} is specified, the \code{run} method will take its default value (which is infinity), meaning that the simulation will keep running until all future events are executed. Moreover, we can set the boolean value of the parameter \code{logging}, which is \code{False} by default, to decide whether to output the log files of the simulation. A code example is given in Figure~\ref{fig:quick-start-env-init}.

\subsubsection{Network}

After the simulation environment, we need to construct a \code{Network} to group nodes and links. The \code{Network} inherits from the class \code{Entity} and will be automatically attached to the default simulation environment once instantiated if no specific environment is assigned. The \code{Network} will be initialized with the environment, which checks relevant network configurations and calculates the routing tables. In this version, we use static routing via the shortest-path algorithm.

\begin{figure}[ht]
    \centering
    \includegraphics[width=0.48\textwidth]{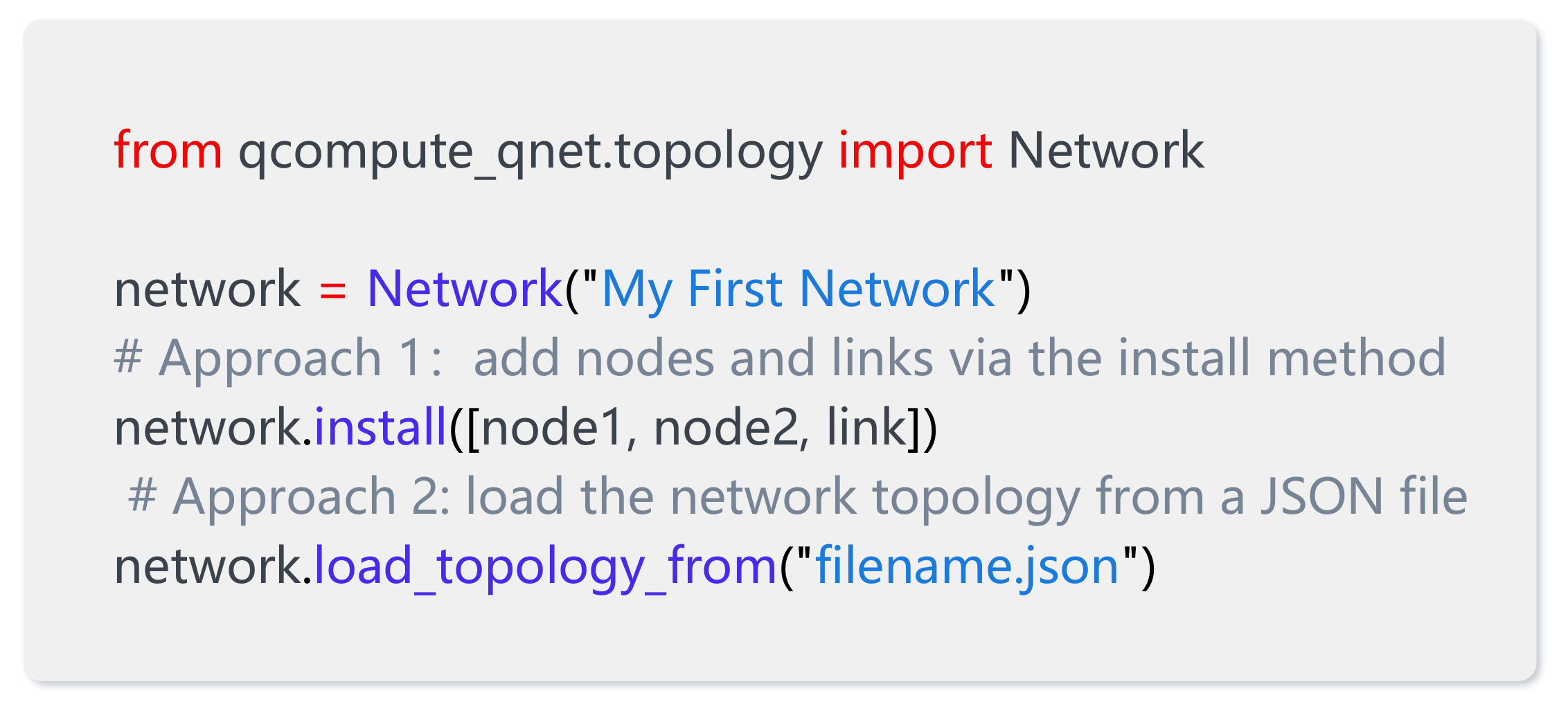}
    \caption{Code example for quantum network configuration.}
    \label{fig: quickstart-network}
\end{figure}

QNET provides two approaches to constructing the quantum network topology. The first approach is to create nodes and links manually and add them via the \code{install} method. The second approach is to load the network topology from a saved JSON file by using the \code{load\_topology\_from} method, which installs nodes and links according to the descriptions therein. For users' convenience, the \code{Network} class allows printing its topology by calling the \code{print\_classical\_topology} or \code{print\_quantum\_topology} method. A code example is given in Figure~\ref{fig: quickstart-network}.

Now we have constructed the framework of a quantum network. The next step is to fill in the nodes and links.

\subsubsection{Node}

\begin{figure}[ht]
    \centering
    \includegraphics[width=0.48\textwidth]{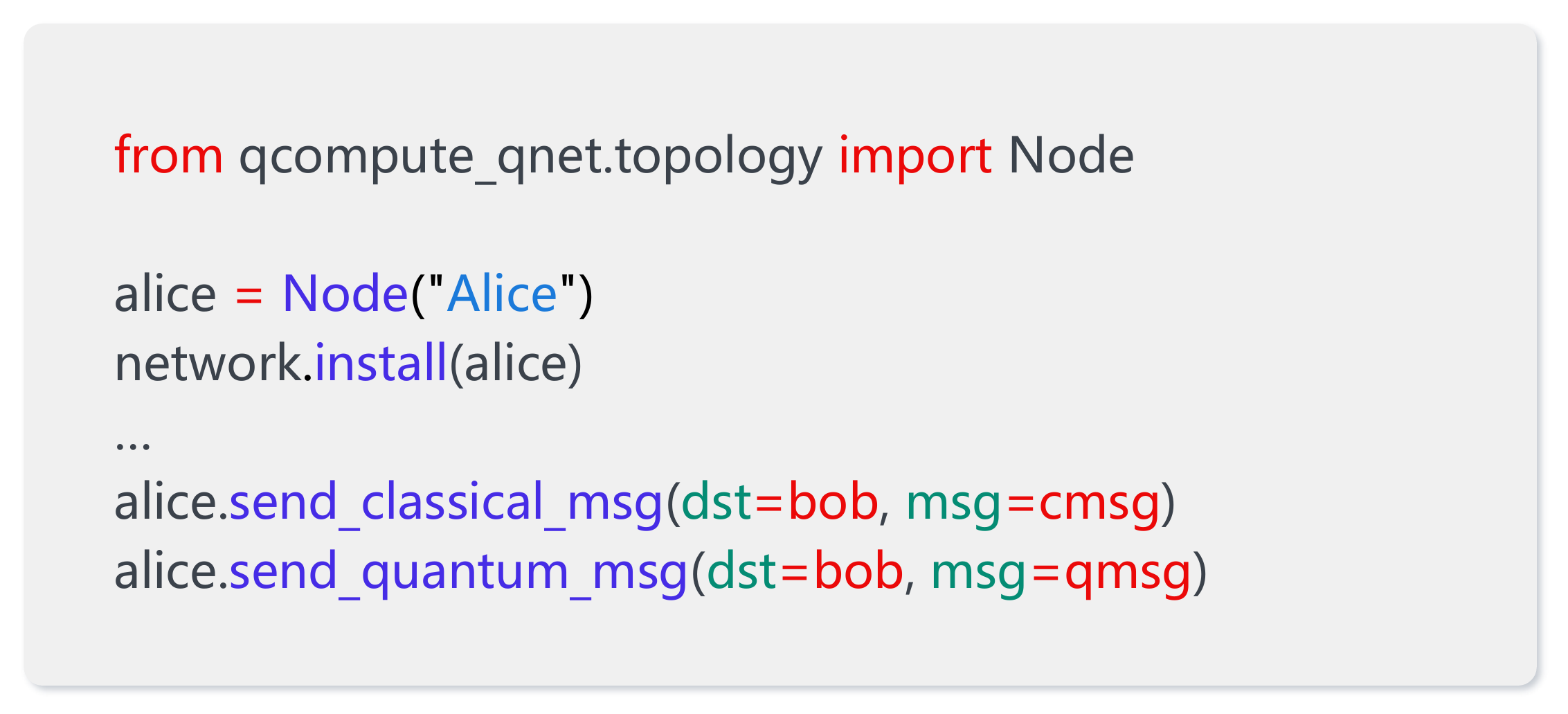}
    \caption{Code example for creating nodes and sending classical/quantum messages.}
    \label{fig:quickstart-node}
\end{figure}

Nodes, such as users and routers in a network, are implemented by \code{Node} in QNET, which are entities capable of information processing and communication. We can call the \code{install} method of \code{Network} to add nodes to it. A node can send classical or quantum messages to the other node by calling the \code{send\_classical\_msg} or \code{send\_quantum\_msg} method with a specified destination node parameter \code{dst}. In discrete-event simulation, calling these methods will actually schedule the corresponding events whose actions are calling methods of \code{receive\_classical\_msg} or \code{receive\_quantum\_msg} in the destination node for message's receiving. A code example is given in Figure~\ref{fig:quickstart-node}.

\begin{figure}[ht]
    \centering
    \includegraphics[width=0.48\textwidth]{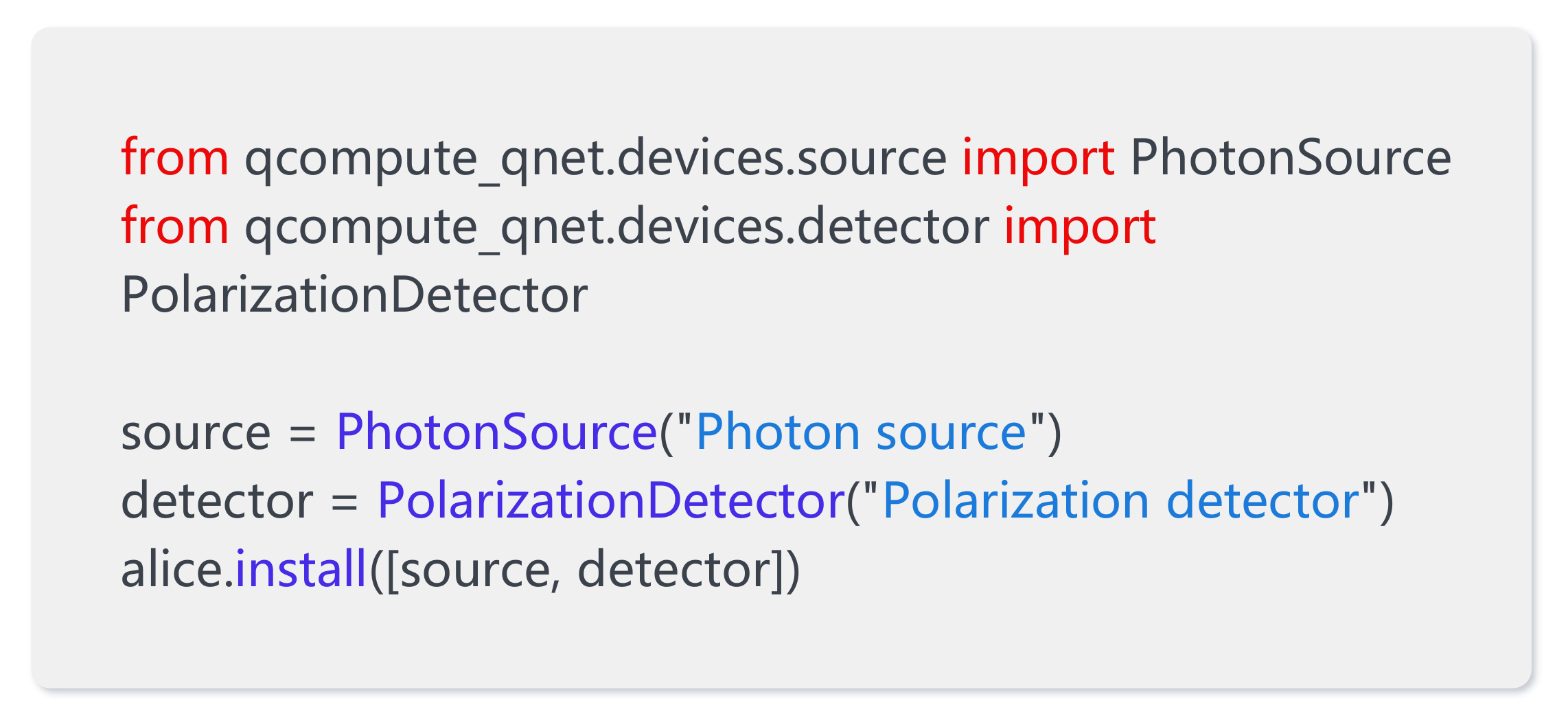}
    \caption{Code example for creating network devices and installing them to a node.}
    \label{fig:quickstart-devices}
\end{figure}

\paragraph{Install physical devices}
After creating nodes, we need to add physical devices to them according to their usage scenarios. For instance, a node for quantum key distribution usually contains at least a photon source (e.g., \code{PhotonSource}) for key encoding or a photon detector (e.g., \code{PolarizationDetector}) for information readout. These physical devices can be added to a \code{Node} by calling the \code{install} method. A code example is given in Figure~\ref{fig:quickstart-devices}.

\paragraph{Load a protocol stack}

\begin{figure}[ht]
    \centering
    \includegraphics[width=0.48\textwidth]{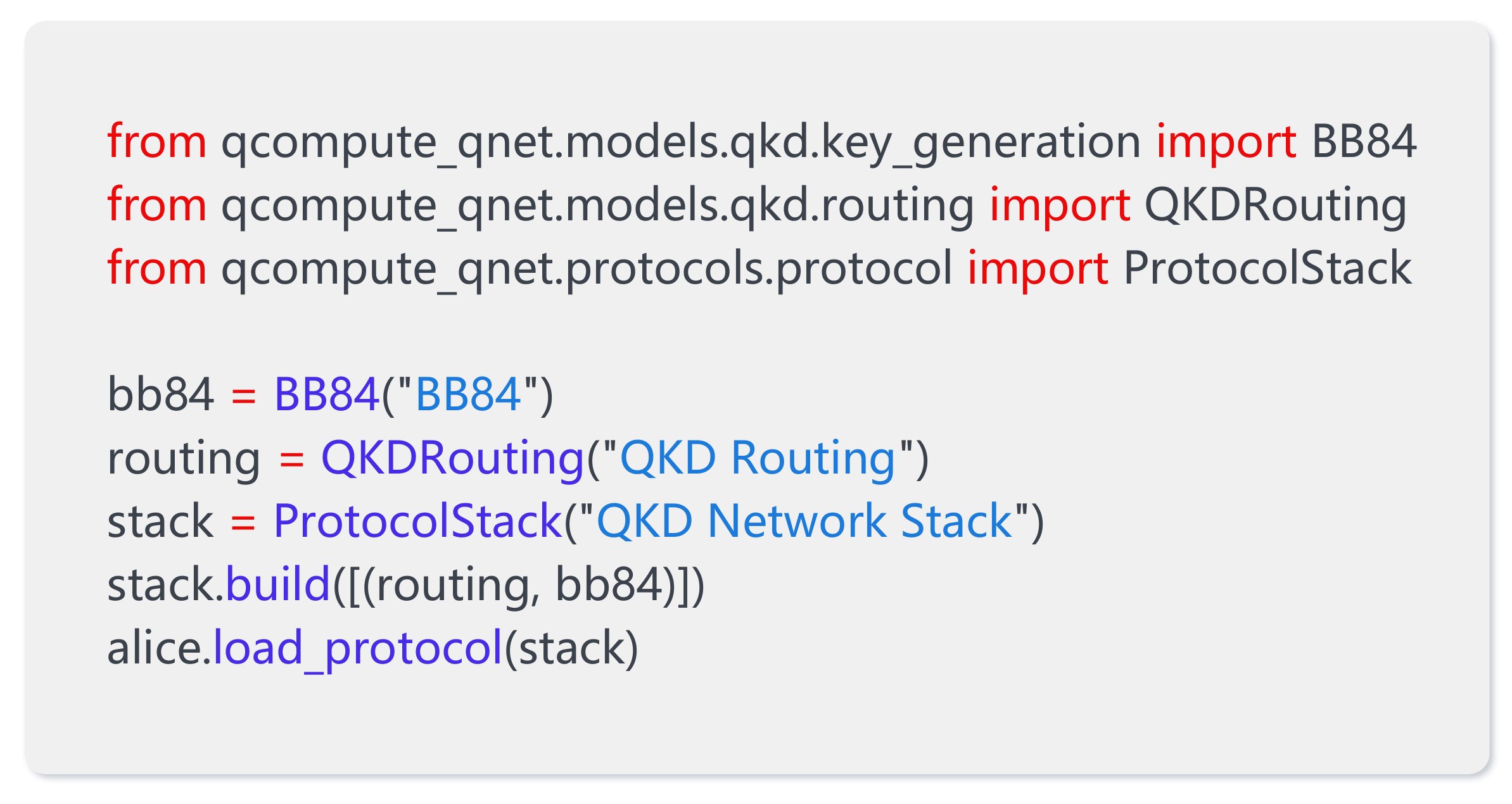}
    \caption{Code example for building a protocol stack and loading it onto a node.}
    \label{fig:quickstart-protocolstack}
\end{figure}

Aside from the physical devices, we need to provide the node with relevant software, i.e., protocol stack (\code{ProtocolStack}), which consists of a collection of protocols (\code{Protocol}) in a layered structure. As with a classical network, each layer of the protocol stack relates to specific functions that determine the rules of classical and quantum communication. A protocol stack can be customized by splicing different protocols. In QNET, the \code{ProtocolStack} class provides a \code{build} method, which allows for building a protocol stack and defining its hierarchical relationships. Once a protocol stack is built, it can be loaded onto a \code{Node} via the \code{load\_protocol} method. A code example is given in Figure~\ref{fig:quickstart-protocolstack}. Note that protocol stacks do not belong to entities and, thereby, should not be loaded via the \code{install} method. 

\paragraph{Frequently-used node templates}

As stated above, nodes can be manually created, followed by the installation of physical devices and the loading of a protocol stack. This gives the most flexibility in designing a specific node. Besides, we also provide some frequently-used node templates, which can be directly used or inherited for further customization.
For example, we provide \code{QKDNode} that can be used for quantum key distribution. In a \code{QKDNode}, a photon source and a photon detector are pre-installed. These physical devices can be accessed through the corresponding attributes. Besides, an empty protocol stack is also loaded by default, leaving the flexibility of setting the key generation protocol (e.g., the \code{BB84} protocol) to the user. Please refer to the API documentation for more node templates and their detailed usage~\cite{QNET}.

Now we know how to create nodes in a network. The next step is to link them together, so they can communicate.

\subsubsection{Link}

\begin{figure}[ht]
    \centering
    \includegraphics[width=0.48\textwidth]{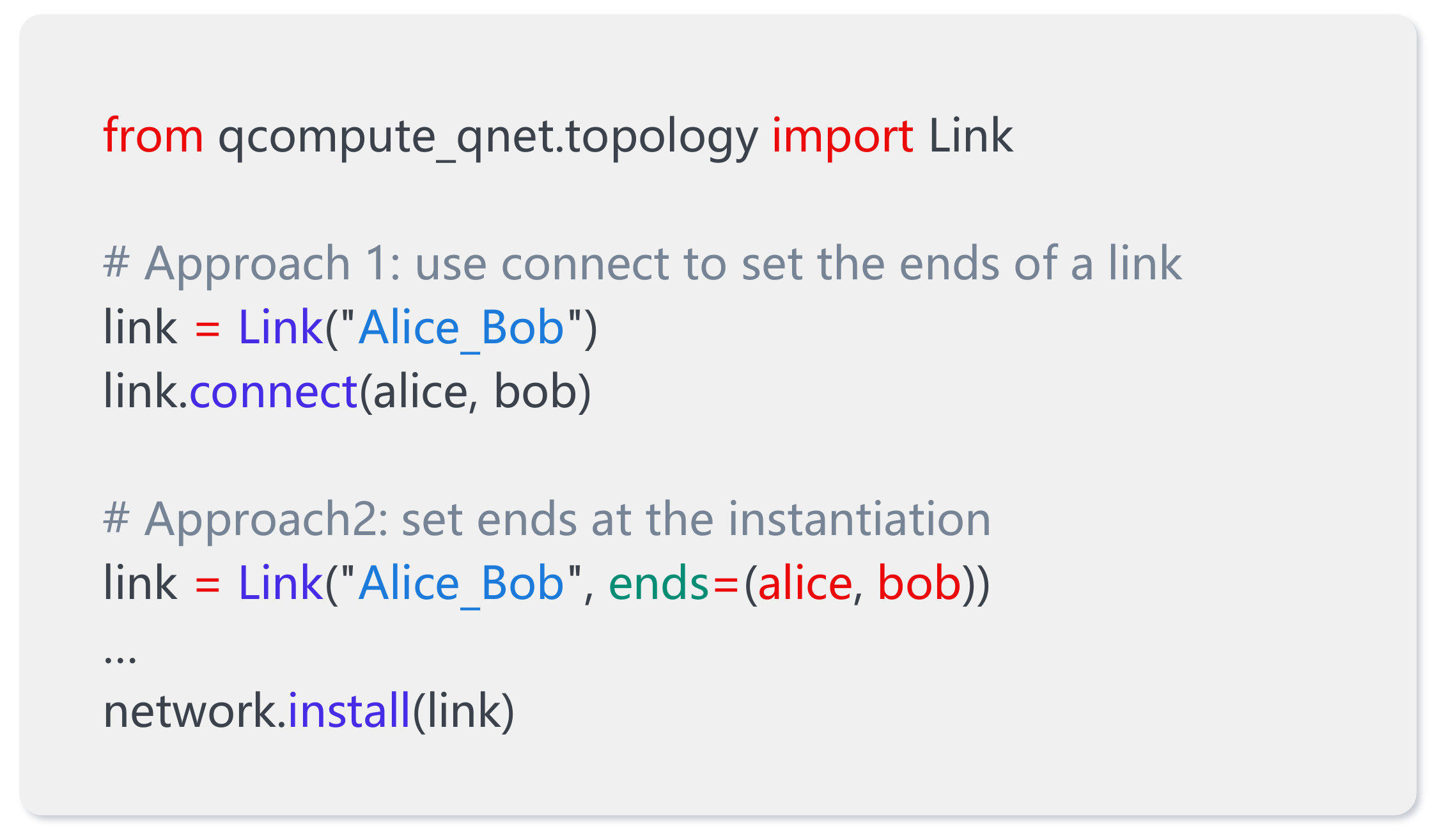}
    \caption{Code example for creating a link between Alice and Bob, and installing it to the network.}
    \label{fig:quickstart-link}
\end{figure}

The link, implemented as \code{Link} in QNET, is the connection between nodes that usually contains specific classical or quantum channels. We can either call the \code{connect} method to set the ends of a link or specify the ends at instantiation. As with the nodes, we can add \code{Link} to a \code{Network} via the \code{install} method. A code example is given in Figure~\ref{fig:quickstart-link}.

\begin{figure}[ht]
    \centering
    \includegraphics[width=0.48\textwidth]{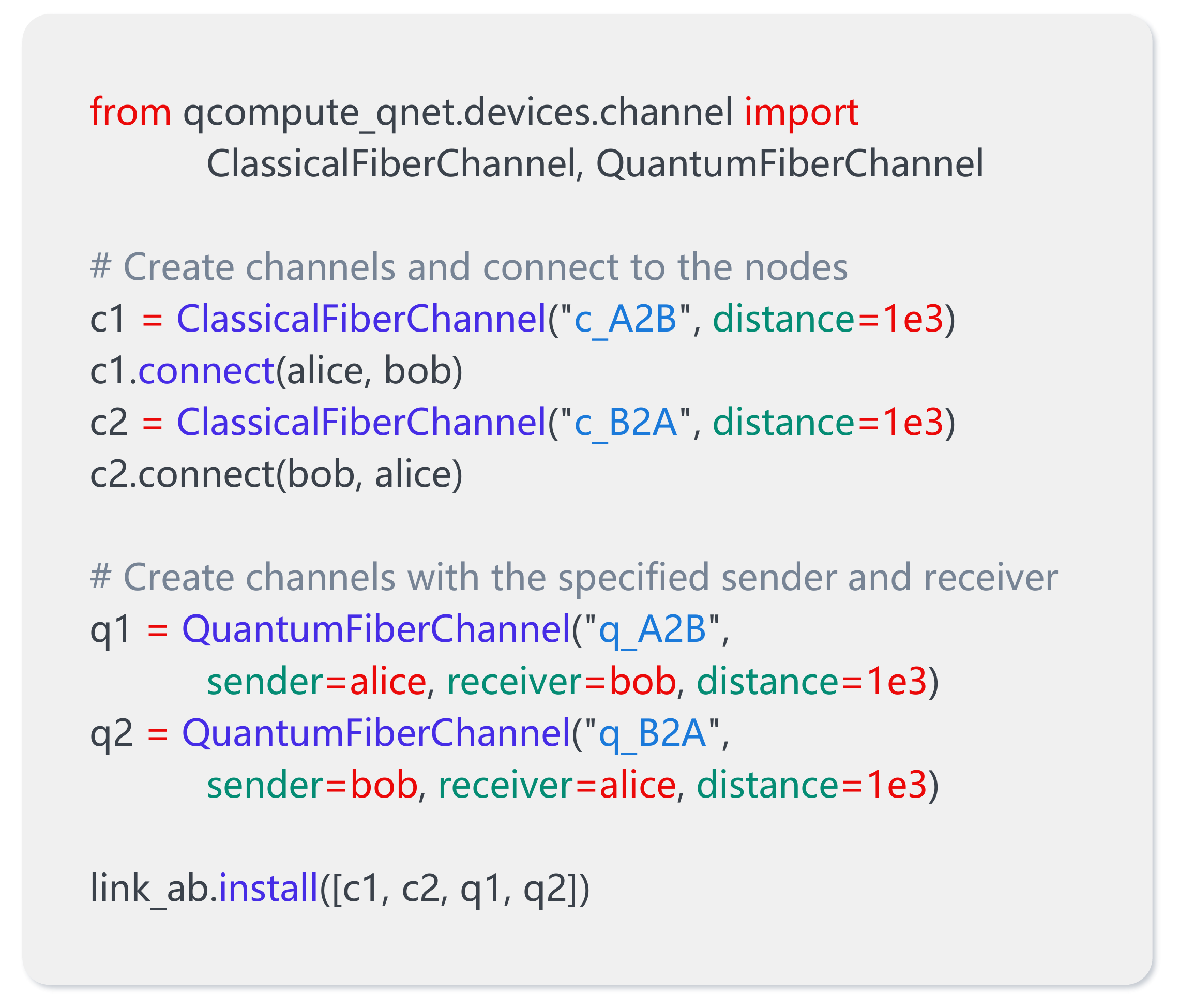}
    \caption{Code example for creating classical and quantum channels between Alice and Bob, and installing it to the link.}
    \label{fig:quickstart-channel}
\end{figure}

As the link is merely a topological concept, we need to further install the classical and quantum channels (\code{Channel}) for actual communications. In QNET, a classical fiber channel or a quantum fiber channel can be created by instantiating \code{ClassicalFiberChannel} (inherited from \code{ClassicalChannel}) or \code{QuantumFiberChannel} (inherited from \code{QuantumChannel}). As with links, we can either call the \code{connect} method to set the sender and receiver of a channel or specify this information at instantiation. Note that both the \code{ClassicalFiberChannel} and \code{QuantumFiberChannel} are both unidirectional channels, so the call of \code{connect} method has to set the sender and receiver in the right order.
In addition to the unidirectional fiber channels, QNET also provides duplex channels (\code{DuplexChannel}) and free-space channels (\code{FreeSpaceChannel}); please refer to the API documentation for more details~\cite{QNET}. Once created, channels can be added to a \code{Link} via the \code{install} method. A code example is given in Figure~\ref{fig:quickstart-channel}.

\subsubsection{A complete example}
\label{subsubsec: quick-start example}

\begin{figure}[ht]
    \centering
    \includegraphics[width=0.48\textwidth]{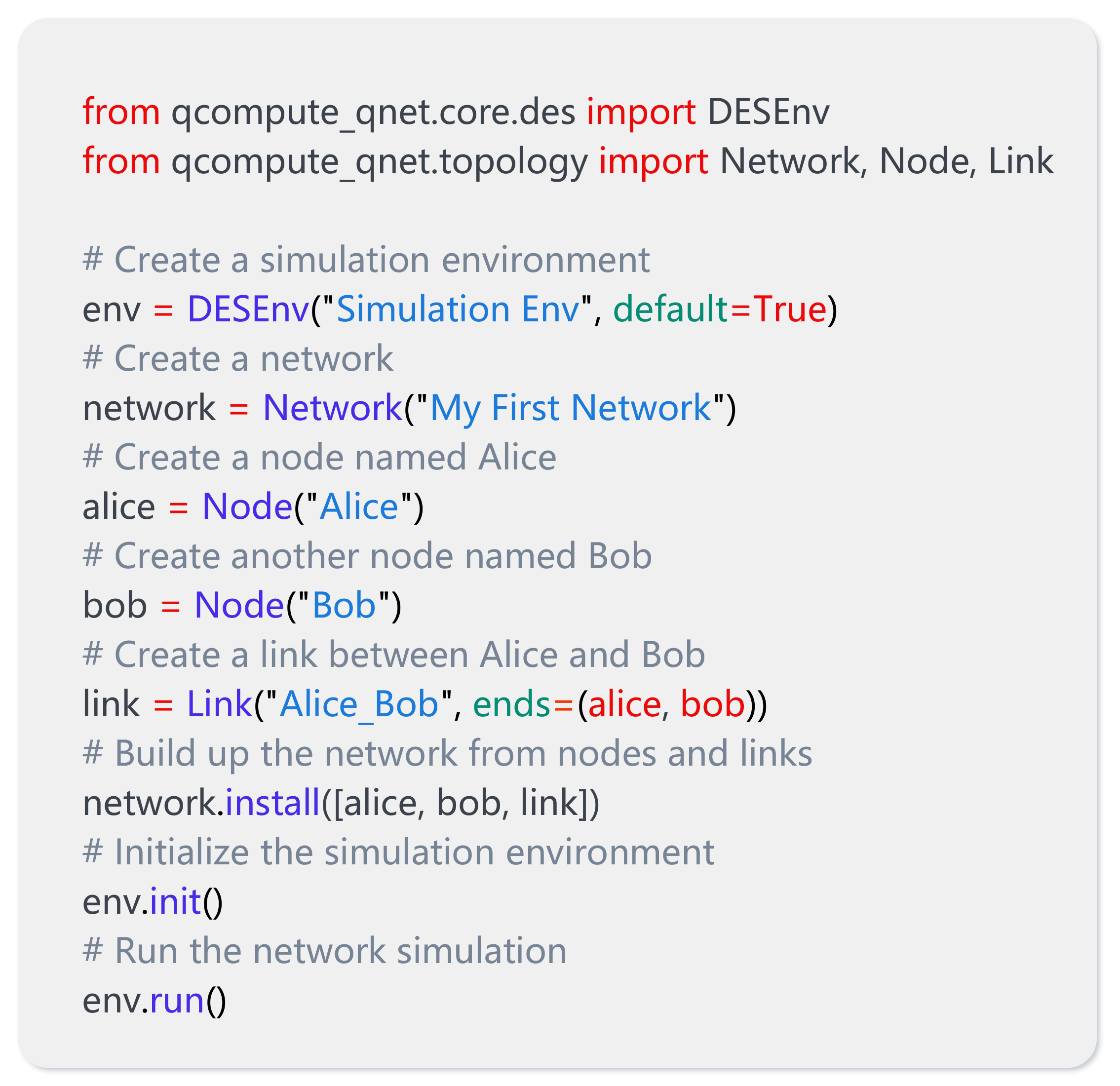}
    \caption{A code example for running a quantum network simulation using QNET. A simulation typically contains the creation of a simulation environment, the construction of a quantum network, and the initialization and running of the environment.}
    \label{fig:quickstart-complete}
\end{figure}

So far, we have covered all the basic components for constructing a quantum network. A vanilla example of running a quantum network simulation using QNET is given in Figure~\ref{fig:quickstart-complete}. Notably, all demonstrations in this section have merely constructed the skeleton of a quantum network; the heart of network simulation lies in the design of network protocols, the development of protocol stacks, and the modeling of physical devices. QNET seeks to provide the building blocks and the simplest set of usage guidelines while leaving users with complete flexibility for innovation. In the next section, we will showcase some featured examples implemented with QNET.

\subsection{Featured demonstrations}
\label{subsec: featured demonstrations}

\subsubsection{Micius quantum satellite experiment}
\label{sec: micius}

Due to ground limitations, many places cannot lay the dedicated fibers for quantum communication. But using quantum satellites for relaying can overcome the effects of many negative factors, such as obstacles and ground curvature, making long-distance quantum communication easier to achieve and aiding in the establishment of a global-scale quantum network. The Micius quantum satellite is the world's first quantum communication satellite, launched in 2016 by China. It has conducted various pioneering experiments, including satellite-to-ground bidirectional entanglement distribution over thousands of kilometers, high-speed satellite-to-ground quantum key distribution, and high-speed ground-to-satellite quantum teleportation. In this example, we simulate Micius' quantum key distribution experiment to showcase the power of QNET in verifying realistic experiments.

\begin{figure}[ht]
\centering
\includegraphics[width=8.5cm]{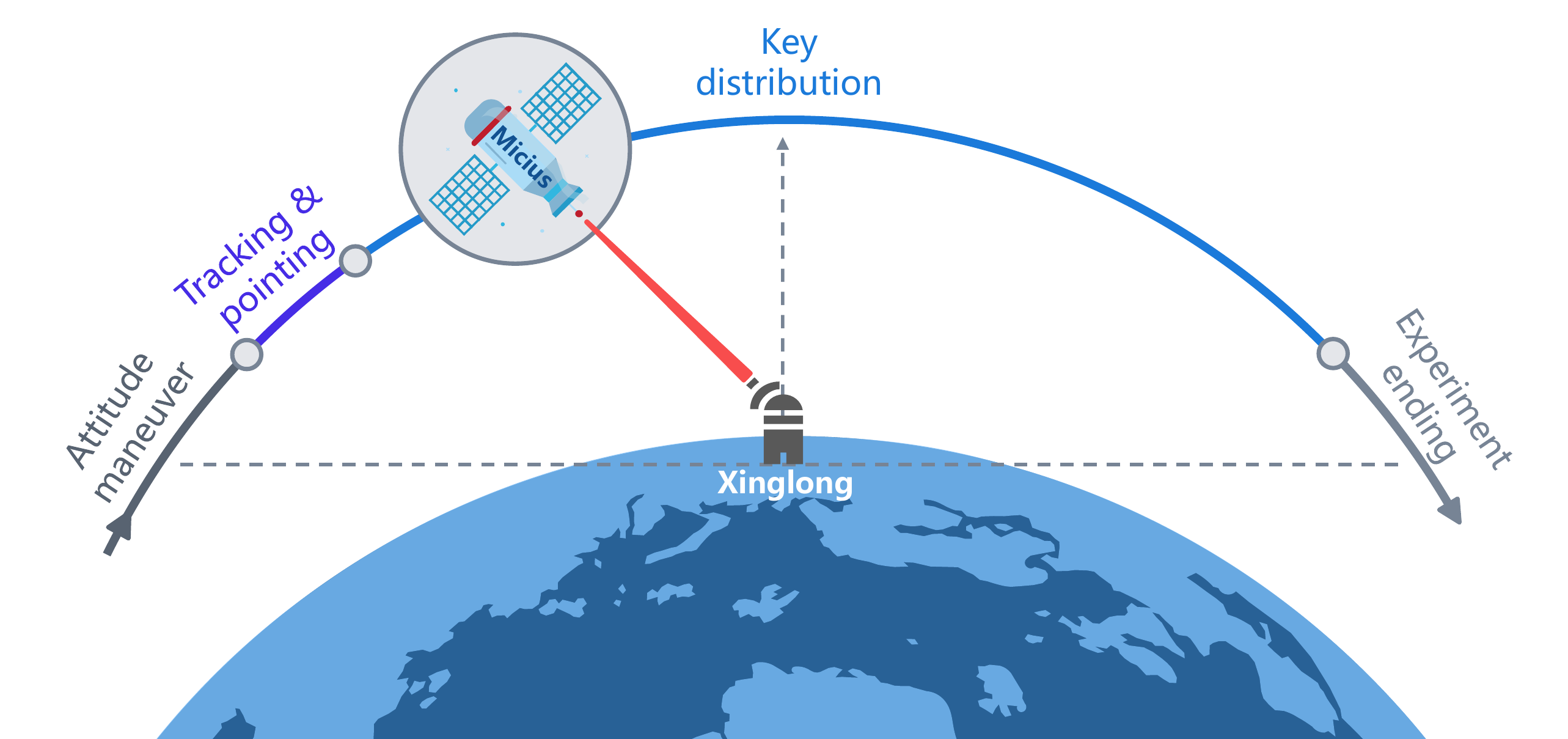}
\caption{A schematic diagram for the satellite-to-ground quantum key distribution experiment~\cite{liao2017satellite}. The experiment uses the down-link strategy, in which the Micius satellite sends quantum signals to the ground station at Xinglong. Before performing the key distribution, the satellite needs to perform the attitude maneuver and tracking \& pointing procedures. The effective time for key distribution is around five minutes.}
\label{fig: micius experiment}
\end{figure}

A schematic diagram for the experimental procedure is presented in Figure~\ref{fig: micius experiment}. Before the key distribution, the satellite needs to go through the processes of attitude maneuver (adjusting the attitude of the satellite and pointing to the ground station) and tracking \& pointing (ensuring the transmitter and receiver are firmly locked throughout the orbit) for preparation. Then, the key distribution experiment adopts the decoy-state BB84 protocol to guarantee the security of the key and the down-link strategy for higher link efficiency (more details can be found in~\cite{QNET}). The entire key distribution process lasts for around five minutes.
The difficulty in carrying out quantum satellite experiments and simulations stems from the fact that the satellite-to-ground distance varies in real-time as the satellite travels. The variation in communication distance affects photon loss and thus the sifted key rate. Generally, the longer the distance, the higher the photon loss rate and the lower the sifted key rate. In QNET, we provide templates for quantum satellites as well as free-space quantum channels. More broadly, we consider mobility to be a functionality that can be loaded onto a node, such as a quantum satellite, to enable its ability to move (see~\cite{QNET} for further details).

\begin{figure}[ht]
\centering
\includegraphics[width=0.48\textwidth]{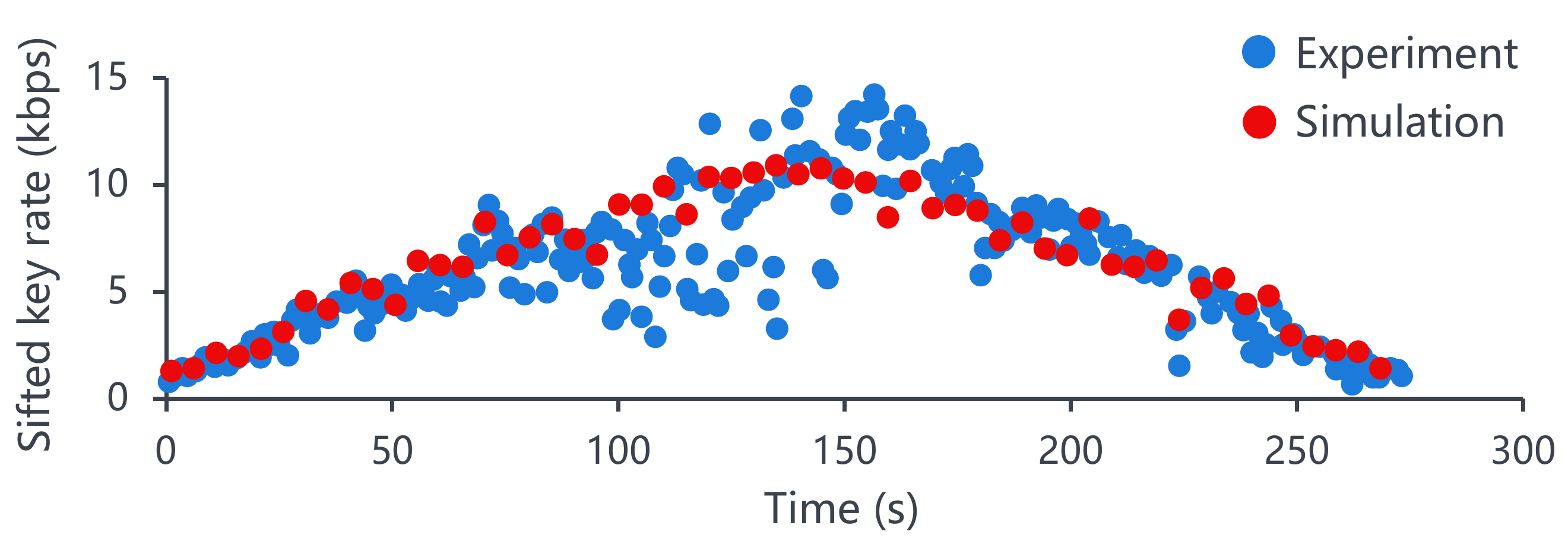}
\caption{Comparison between the simulation results and the experimental data. The horizontal axis is the time of key distribution, and the vertical axis is the sifted key rate.}
\label{fig:micius experiment}
\end{figure}

In the simulation, we use the same parameters as the original experiment in~\cite{liao2017satellite}. The comparison between our simulation results and the experimental data is given in Figure~\ref{fig:micius experiment}. It is clear that the simulation results closely match the experimental data, thus validating the Micius quantum satellite experiment at the software level. This example demonstrates the benefit of using a simulation tool for verifying quantum experiments. Even without real hardware devices, simulation allows us to swiftly examine protocol performance and adjust settings. This can greatly lower the cost of protocol testing and can be necessary for the standardization of quantum network architectures and expensive projects such as quantum satellites.

\subsubsection{Quantum network architecture with resource management}

Most of the existing quantum network infrastructure is still in the experimental testing stage and is not yet ready to serve large volumes of users. To make these facilities and technologies operational in the future, we need a network architecture with efficient resource management. In this example, we propose a quantum network architecture, i.e., a four-layer protocol stack that cooperates with a pooling scheme in the repeater stations, capable of processing large numbers of concurrent requests~\cite{fang12,fang13}. Note that this architecture is presented in the context of quantum key distribution, but it should be readily adapted to other scenarios, such as entanglement distribution.

\begin{figure}[ht]
\centering
\includegraphics[width=7.4cm]{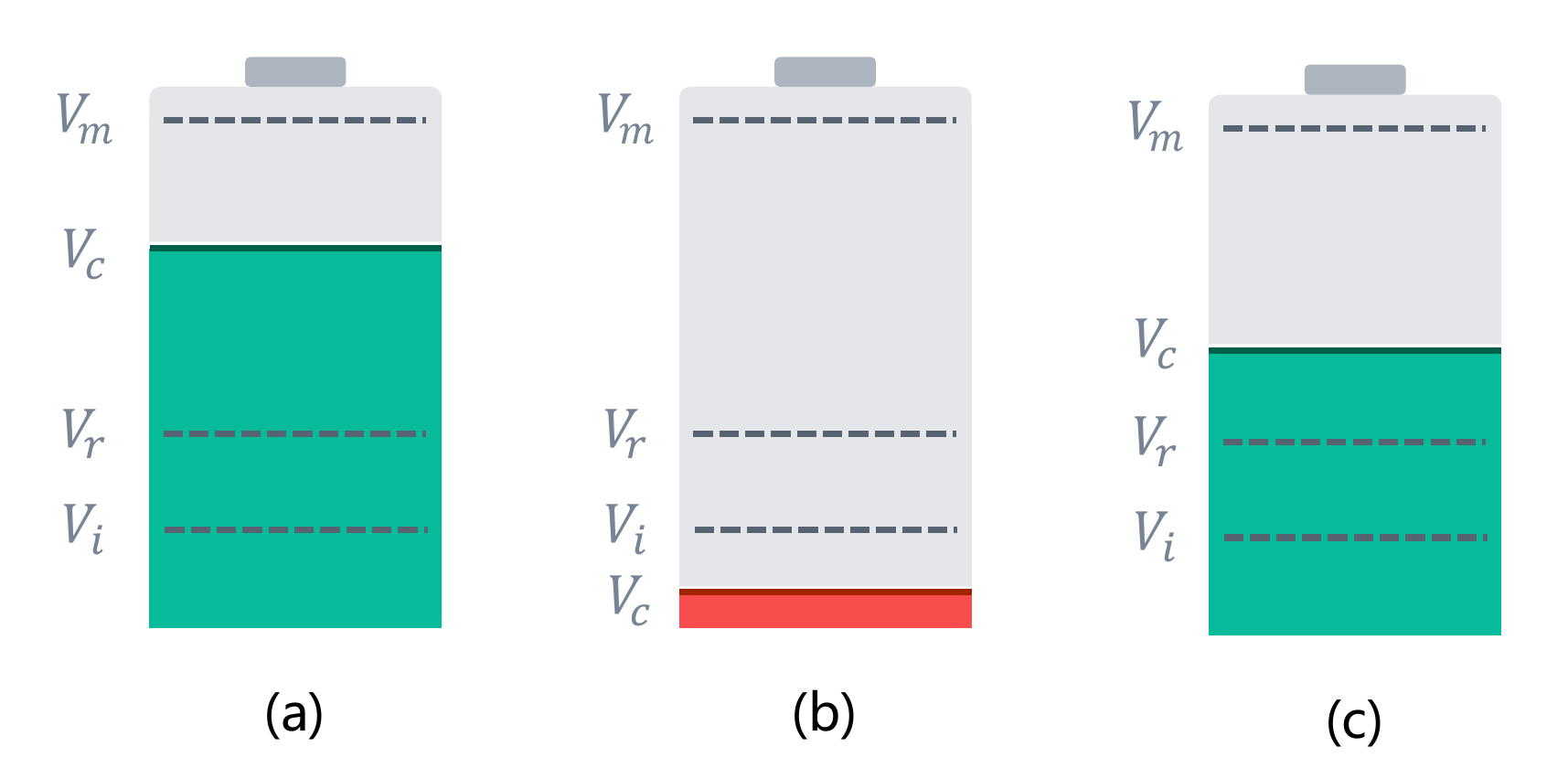}
\caption{An illustration of the key pool structure. Each key pool has a maximum volume $V_m$, a current volume $V_c$, a service interruption volume $V_i$, and a service recovery volume $V_r$.}
\label{fig: battery}
\end{figure}

\paragraph{Key pool} 
Given limited resources, generating keys for a specific request will inevitably block the processing of other requests in the network. To cope with multiple requests, we propose the use of key pool, which is a battery-like queuing structure that stores and manages the keys generated between two directly connected repeater stations. As shown in Figure~\ref{fig: battery}(a), a key pool has a maximum volume (capacity) $V_{m}$ indicating the maximum number of keys it can accommodate, a current volume $V_c$ for the currently available keys it holds, a service interruption volume $V_{i}$ as the threshold for stopping the service, and a service recovery volume $V_r$ for recovering the service. After delivering keys to user requests, a key pool will check its current volume, $V_c$. If it is less than $V_i$ (see Figure~\ref{fig: battery}(b)), the key pool will stop processing new requests and will call the key generation protocol to replenish the pool with new keys until it reaches the service recovery volume $V_r$ (see Figure~\ref{fig: battery}(c)). 

\begin{figure}[ht]
\centering
\includegraphics[width=8.2cm]{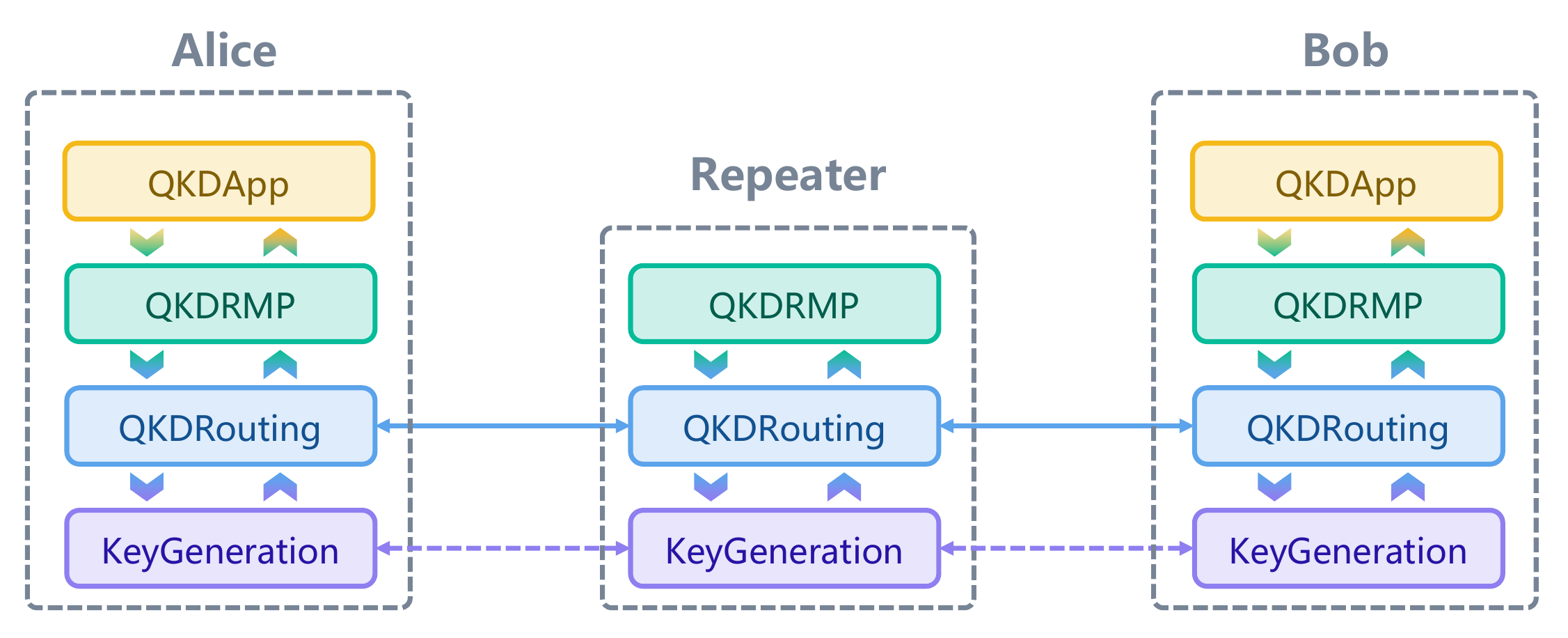}
\caption{A four-layer quantum network architecture. The protocol stack for each endnode has four layers of protocols, and the protocol stack for each repeater has three layers of protocols.}
\label{fig: protocol_stack}
\end{figure}

\paragraph{Network architecture}
The four-layer quantum network architecture we propose is depicted in Figure~\ref{fig: protocol_stack}. Each endnode is equipped with a four-layer protocol stack, with the following protocols running from top to bottom: \text{QKDApp} protocol that generates and receives user requests; \text{QKDRMP} protocol that is responsible for resource management; QKDRouting protocol that determines the path of key distribution; KeyGeneration protocol that performs key generation with directly connected nodes. Each repeater has the same protocol stack as an endnode node, except for the application layer. The protocol on each layer is not restricted and can be specified as needed. For example, the routing protocol can be either static or dynamic, and the key generation protocol can be any protocol from BB84, E91, B92, etc.

The process of an end-to-end key request from Alice to Bob is outlined as follows: 1) Alice initiates a key request and sends it to her connected repeater (the message will be directly sent to Bob if there is a direct channel between Alice and Bob); 2) Repeater stations receive the request and choose to forward or reject this message depending on the resource requirements; 3) Bob receives the request message and locally sets his key generation protocol to be ready to generate keys with the upstream node, then he returns an acceptance message; 4) Repeater stations along the path receive the acceptance message and decide, depending on the node's current status, whether to process the request or append the request to the request queue. When a request is processed, the repeater will generate a ciphertext message after key swapping and send it to Bob; 5) Alice receives the acceptance message and locally sets the key generation protocol to generate keys with the downstream node; 6) Alice and Bob save the generated keys with their downstream and upstream nodes locally; 7) Bob receives the ciphertext message and returns an acknowledgment to the repeater; 8) Upon receiving the ciphertexts from all the intermediate stations on the path, Bob combines all the information to decrypt the ciphertexts, and finally obtains the end-to-end keys with Alice. He then sends a done message to Alice to finalize the key request; 9) The repeater stations directly forward the done message once receiving it; 10) Alice receives the done message and sets the end-to-end keys, then finishes the request. It is worth noting that this end-to-end process can occur concurrently among different network users. Particularly in step 4), the repeater station will add the request to a queue if it cannot process it at the moment of acceptance. This request queue will then collaborate with the key pool for an overall scheduling. A request will wait to be called and get processed until there are enough keys in the pool.

\begin{figure}[ht]
\centering
\includegraphics[width=8.4cm]{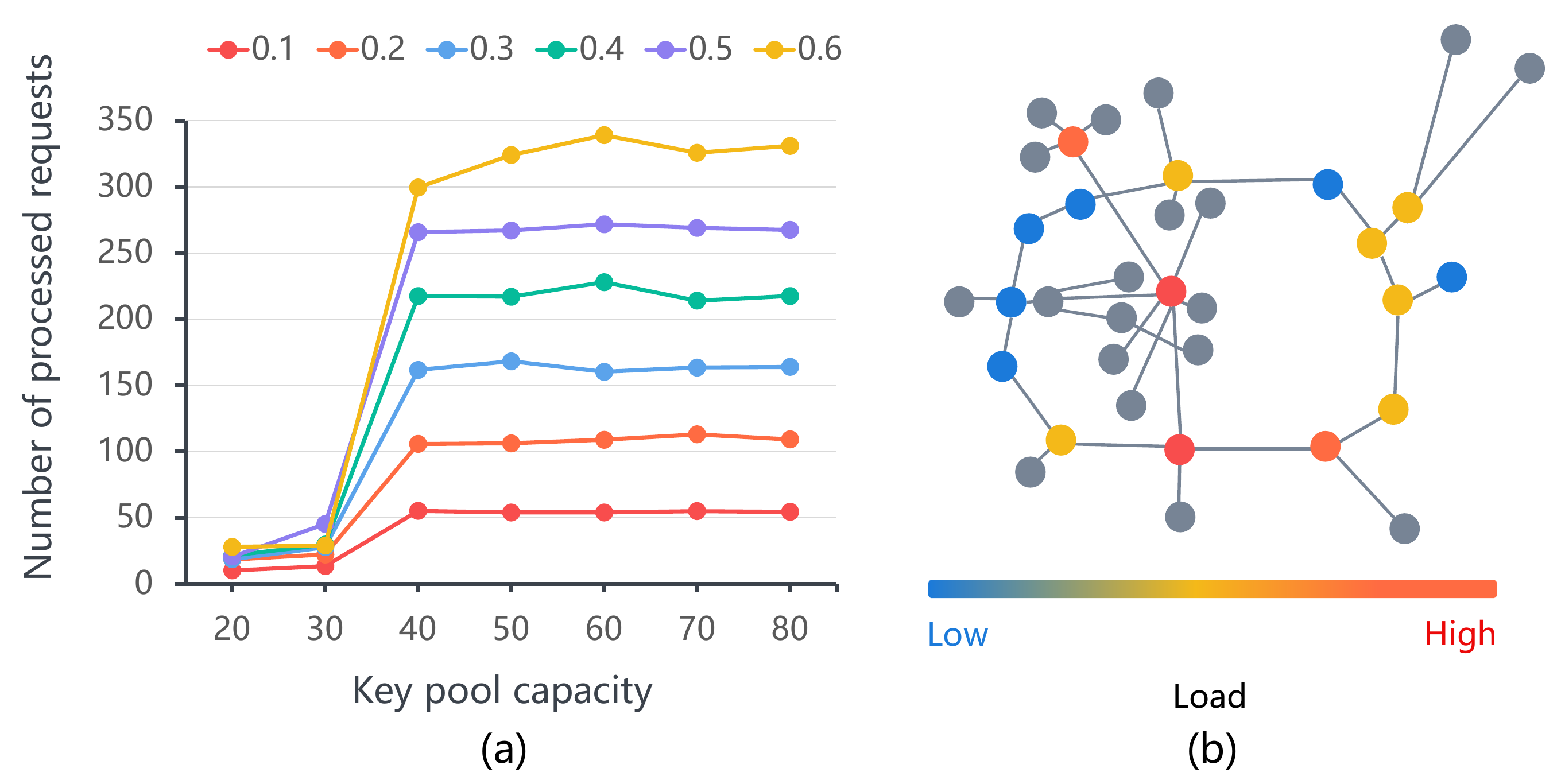}
\caption{The experiment results from the simulation of quantum network architecture with resource management. The left figure shows the relationship between the number of processed requests and the key pool capacity. The right figure showcases the virtualization of the network load at each repeater station. The higher the load, the closer the color is to red.}
\label{fig: protocol_stack_result}
\end{figure}

We use this architecture to examine the capability of request processing with different key pool capacities on the Beijing Quantum Metropolitan Area Network~\cite{chen2021integrated}. In our experiment, we select the receiver and the time at which each endnode creates key requests at random. The amount of each key request is fixed to be $10$ keys, each of length $32$. In the key pool, we set $V_i = V_m/4 = V_r/4$, and the maximum volume (capacity) is chosen from $20$ to $80$ keys with step size $10$; each key is of length $32$. The simulation time is chosen from $0.1$ to $0.6$ seconds with step size $0.1$. The relationship between the average number of processed requests and key pool capacity is depicted in Figure~\ref{fig: protocol_stack_result}. We observe that increasing the key pool capacity increases the number of processed requests and thus the network throughput. However, after a threshold of around $40$ keys, the network throughput barely improves, even if we increase the key pool capacity further. This is because the larger the capacity of the key pool, the more keys it can cache, and thereby the more requests it can process. However, once the key pool capacity reaches a certain point, the capacity of the key pool saturates the demand for keys in the network, leaving the throughput unchanged. This example shows how QNET can be used to assess the performance of a network architecture and quickly determine the best performance settings. Besides, QNET can also be used to monitor network status visually. For example, in Figure~\ref{fig: protocol_stack_result}, the colored nodes are repeater stations, and the closer to red they are, the busier they can be. It is clear that some of them are significantly busier than others and might end up being the bottleneck for the network throughput. This suggests that we might be able to balance the load on the network by giving these nodes more resources.

\subsubsection{CHSH game and its quantum emulation}
\label{sec: chsh game}

\begin{figure}[ht]
\centering
\includegraphics[width=0.4\textwidth]{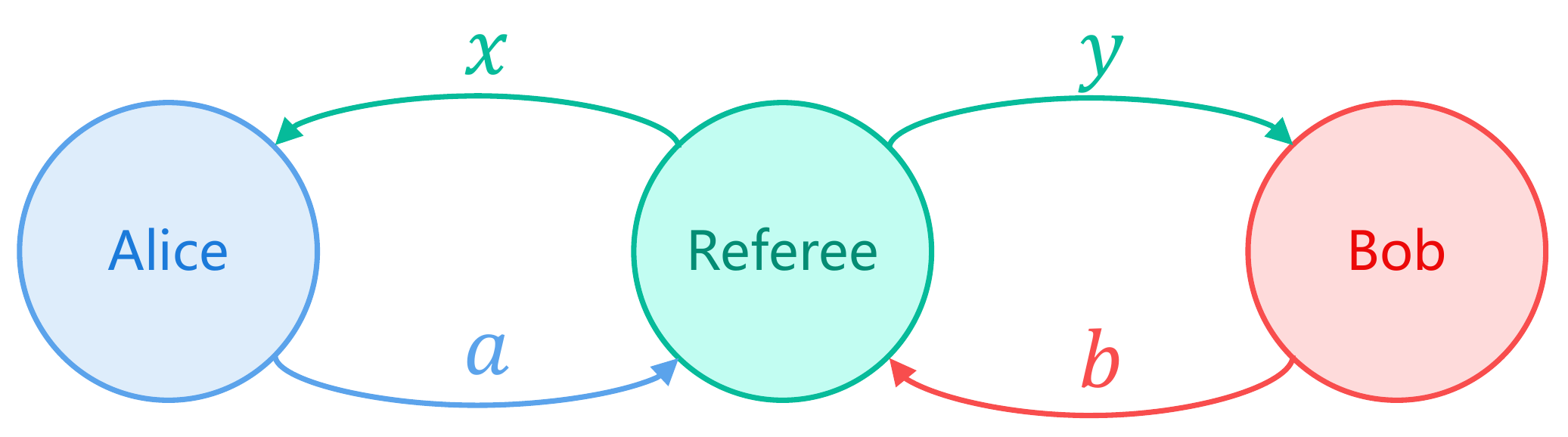}
\caption{An illustration of the CHSH game, which involves one referee and two cooperating players, Alice and Bob. The referee sends questions $x, y \in \{0, 1\}$ to Alice and Bob, respectively, and they need to respond with their answers $a, b \in \{0, 1\}$.}
\label{fig: chsh_game}
\end{figure}

The CHSH game is derived from the CHSH inequality by Clauser, Horne, Shimony, and Holt in 1969~\cite{clauser1969proposed}. This game is commonly used to illustrate quantum advantage. By using quantum entanglement, we can go beyond the limitations of classical information processing. It's worth noting that Aspect, Clauser, and Zeilinger were awarded the Nobel Prize in 2022 for their seminal work on the Bell tests~\cite{nobel-prize-2022}, which can be considered an equivalency of the CHSH game mentioned here. The game features three parties: one referee and two cooperating players, Alice and Bob. The rules are as follows (see Figure~\ref{fig: chsh_game}): 1) The referee chooses two questions $x, y \in \{0, 1\}$ at random and sends one to each player; 2) Alice and Bob must respond to the questions $x$ and $y$ separately and send their answers $a, b \in \{0, 1\}$ to the referee; 3) The referee determines whether the equation $x \cdot y = a \oplus b$ is satisfied. If this is the case, the players win this round; otherwise, they lose. Alice and Bob are allowed to discuss an appropriate strategy before the game begins to increase their chances of victory. However, no communication between them is permitted during the game. It can be easily checked that an optimal classical strategy is provided by Alice and Bob both returning $0$ as their answers regardless of the questions posed to them. In this case, the winning probability, which is also the optimal probability, of this game is $75\%$. The players lose if and only if both questions $x$ and $y$ are equal to $1$.

If Alice and Bob are quantum players and are allowed to share quantum entanglement before the game begins, this winning probability can be increased to $85\%$, far exceeding the classical scenario and exhibiting the quantum advantage. The optimal quantum strategy of the CHSH game is as follows~\cite{montanaro2013survey}: Before the game starts, Alice and Bob need to share an EPR pair $|\Phi\rangle_{AB}=(|00\rangle+|11\rangle) / \sqrt{2}$ which can be created from an entanglement source. Once receiving the questions from the referee, Alice and Bob take measurements on their qubits $A$ and $B$, respectively. Alice measures her qubit $A$ in the $Z$-basis if the question is $x=0$ and in the $X$-basis if the question is $x=1$. She then replies to the referee with her measurement outcome as the answer $a$. Bob measures his qubit $B$ in the $(Z + X)/\sqrt{2}$-basis if the question is $y=0$ and in the $(Z - X)/\sqrt{2}$-basis if the question is $y=1$. Then he replies to the referee with his measurement outcome as the answer $b$.

\begin{figure}[ht]
    \centering
    \includegraphics[width=0.45\textwidth]{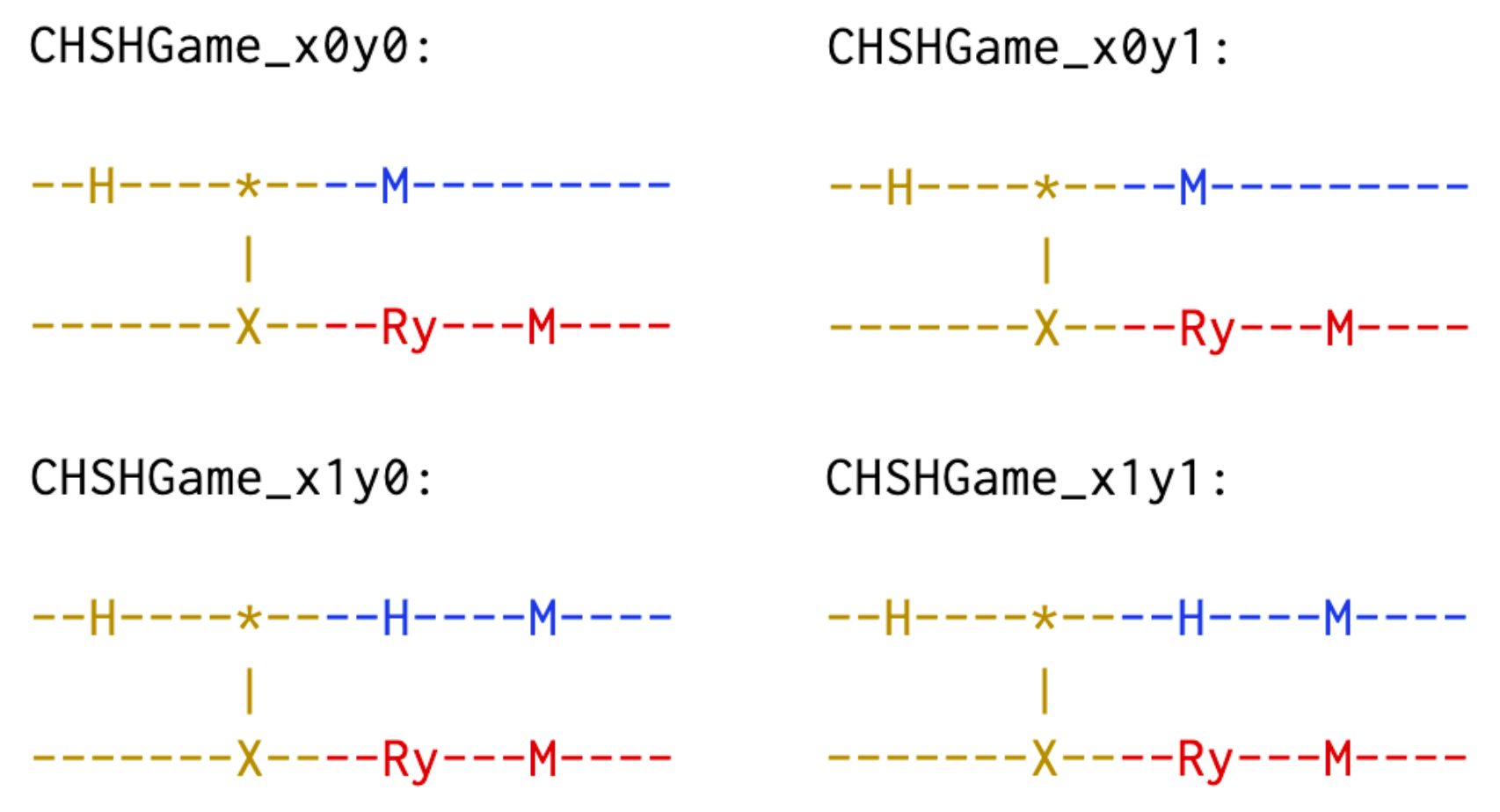}
    \caption{Quantum circuits compiled from the optimal quantum strategy of the CHSH game. Note that the $R_y$ gates in the case of $x=0,y=0$ and $x=1,y=0$ have a rotation angle $-\pi/4$, whereas the other two cases have a rotation angle $\pi/4$.}
    \label{fig: chsh_game circuits}
\end{figure}

Using the protocol compilation in Section~\ref{sec: protocol mapping}, we can compile this three-party protocol into the corresponding quantum circuits~(see Figure~\ref{fig: chsh_game circuits}) and submit them to Baidu's \emph{Qian Shi} (a highly-integrated industry-level superconducting quantum computer~\cite{baidu_quantum}) for evaluation. We observe an average winning probability of around $81.45\%$, surpassing the optimal classical value of $75\%$. More details about this implementation can be found at~\cite{QNET}. We remark that this demonstration differs from the actual long-distance Bell test experiments and cannot avoid the causality loophole. But the point is that we can test network protocols in \emph{quantum} environment through QNET, which can serve as a stepping stone to the next stage, where everything is operated in the actual testbed. 

\begin{figure*}[ht]
    \centering
    \includegraphics[width=\textwidth]{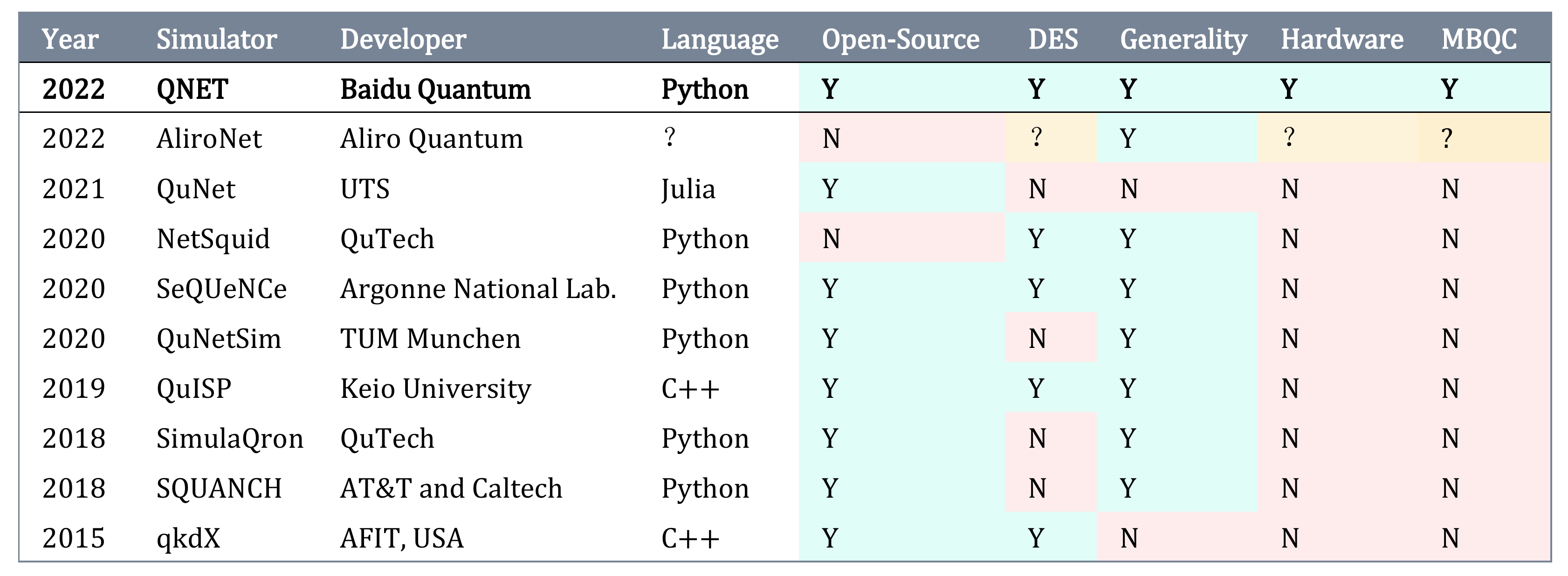}
    \caption{Comparison of different quantum network simulators. "Y" means the feature is supported, "N" means it is not, and a question mark "?" means that we are not sure if it is supported or not.}
    \label{fig: quantum network simulators}
\end{figure*}

\subsection{Comparison with related work}

A comparison of QNET to other quantum network simulators is summarized in Figure~\ref{fig: quantum network simulators}, from which we can draw the following conclusions. First, most simulators are open-source and implemented in Python, but QuISP~\cite{satoh2021quisp} and qkdX~\cite{mailloux2015model} are written in C++ as they are built upon \text{OMNeT++} (a modular, component-based C++ simulation framework that is widely used in classical network simulation). Second, most simulators are designed for general quantum networks, except for QuNet~\cite{leone2021qunet}, which utilizes a method called "quantum cost-vector analysis" to simulate and benchmark entanglement routing in multi-user entanglement networks, and qkdX which is specialized for quantum key distribution. Third, QNET, NetSquid~\cite{netsquid}, SeQUeNCe~\cite{wu2021sequence}, and QuISP are the four general-purpose simulators that utilize the DES framework, which allows for accurate and efficient system tracking. Among these, QuISP makes use of the existing framework from \text{OMNeT++}, whilst the other three simulators build their own standalone simulation engines. There are some distinctions in their DES designs. In particular, \text{SeQUeNCe} schedules events for entities and protocols by calling the method of a global timeline, NetSquid allows event scheduling for all entities and treats protocols as virtual entities, but QNET allows the flexibility of installing an event scheduler to any object on-demand. Finally, while the previous simulators remain at the level of simulation with classical devices, QNET provides the compilation of quantum network protocols and supports their emulation in quantum environments. Moreover, QNET also features the ability to run both quantum circuit and MBQC models.

\section{Conclusions}
\label{sec: conclusions}

In this work, we provide a brief review of the field of quantum networks from both theoretical and experimental perspectives, covering the fundamental ideas, core structures, state-of-the-art developments, and major challenges. Moreover, we introduce a newly developed quantum network toolkit, making it easier for the research and development of this topic. It is the first quantum network toolkit with dual quantum computing engines, making it applicable to the design and simulation of quantum network architectures based on different principles. It also provides the first method for compiling quantum network protocols into quantum circuits so as to interface with quantum hardware devices.

Recalling the history of the internet, we might expect that before we can genuinely build a quantum internet, we first need to build full-fledged quantum computers that operate with millions of qubits. However, there is no cause-and-effect relationship between quantum computing and quantum networking. In particular, the quantum internet is not just about speeding up computations and solving hard problems. It can also be used to improve communication security and provide services that would be impossible in a classical network. Depending on the hardware requirements, quantum internet can have many types of applications at different phases of development~\cite{wehner2018quantum}. It is highly possible that we would first construct a functional quantum internet before we have large-scale quantum computers, despite the fact that the range of applications may be limited. A notable example is quantum key distribution, which requires only single qubits to be created and detected but is sufficient to demonstrate the superiority of the quantum internet over its classical counterpart. There is no doubt that large-scale quantum computers would further enhance the capability of information processing in the quantum internet, but they are just not indispensable for all useful applications.

The future of the quantum internet is promising and significant. However, given the numerous obstacles it faces, its development is still in its early stages, necessitating collaborative efforts from various communities. Future advances in quantum networks include longer-distance quantum communication, larger-scale quantum testbed construction, broader access to quantum services, and the development of more useful applications. In retrospect, the internet's success and diversity were largely due to creative inventors who had access to the technology and wrote programs to make it do fantastic things like remote chatting, online shopping, and cloud computing. In contrast, existing protocols and applications for a quantum internet are primarily developed by quantum scientists. But this is not how successful applications usually come into existence. People must be able to interact hands-on with a quantum internet and play around with it to develop and try out new applications. An open and sustainable quantum ecosystem involving people from different fields will facilitate new innovations and applications. We hope that this work will give interested readers a basic grasp of quantum networks and that QNET as well as its future upgrades can serve as a key platform for further explorations and innovations.

\section*{Acknowledgements}
We would like to thank Yu-Ao Chen, Shusen Liu, Jingbo Wang, Kun Wang, Xin Wang, and Zihe Wang for discussions on the development of QNET. We also thank Jie Lin for suggesting relevant references on single-photon sources and detectors.

\bibliography{Bib}

\onecolumngrid
\appendix

\newpage
\section{Global efforts in quantum networks}
\label{sec: Global efforts in quantum network}

\begin{figure}[H]
\centering
\includegraphics[width=16cm]{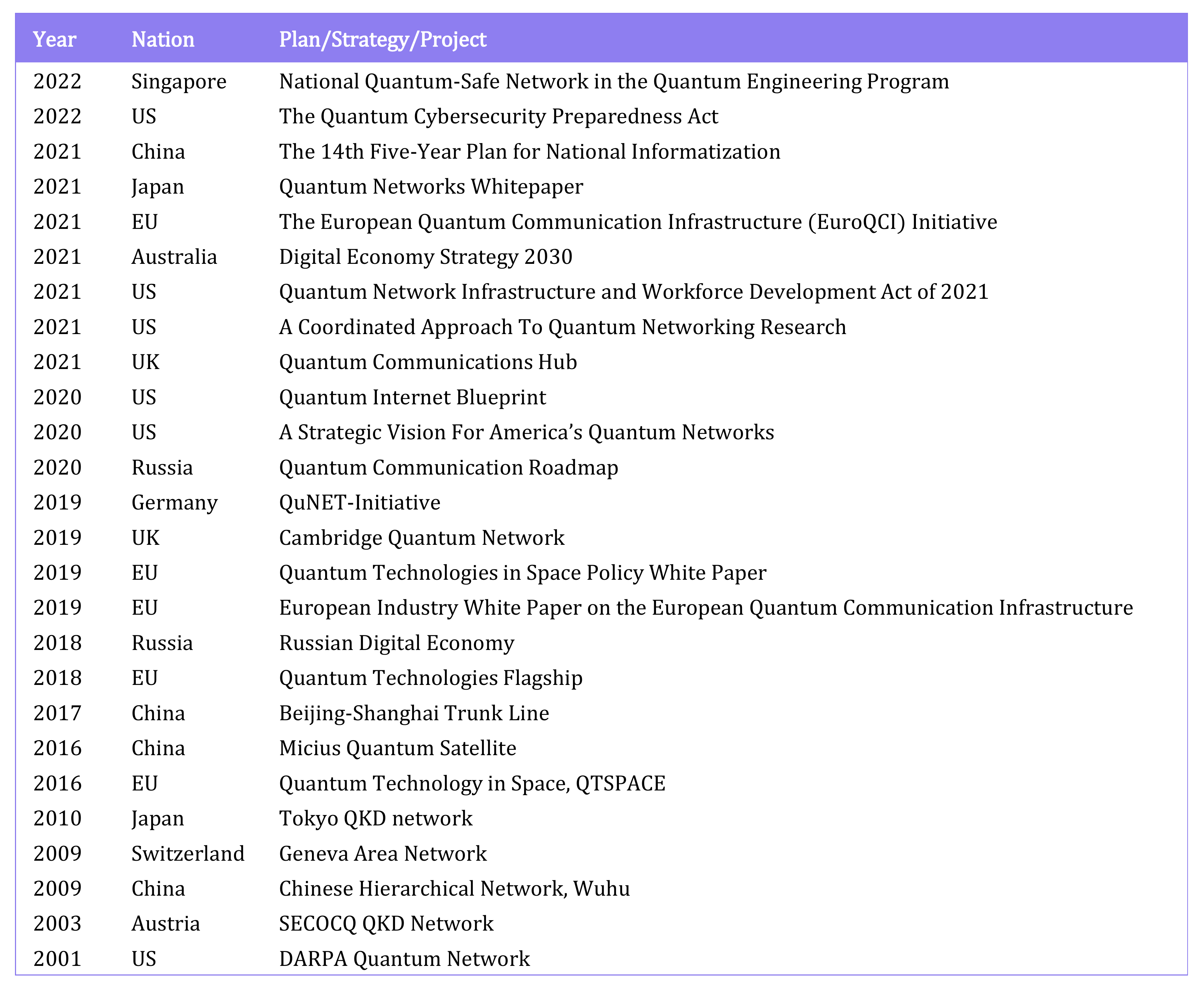}
\caption{A partial list of the national plans/strategies/projects of quantum networks.}
\label{fig: national_strategy}
\end{figure}

\begin{figure}[H]
\centering
\includegraphics[width=16cm]{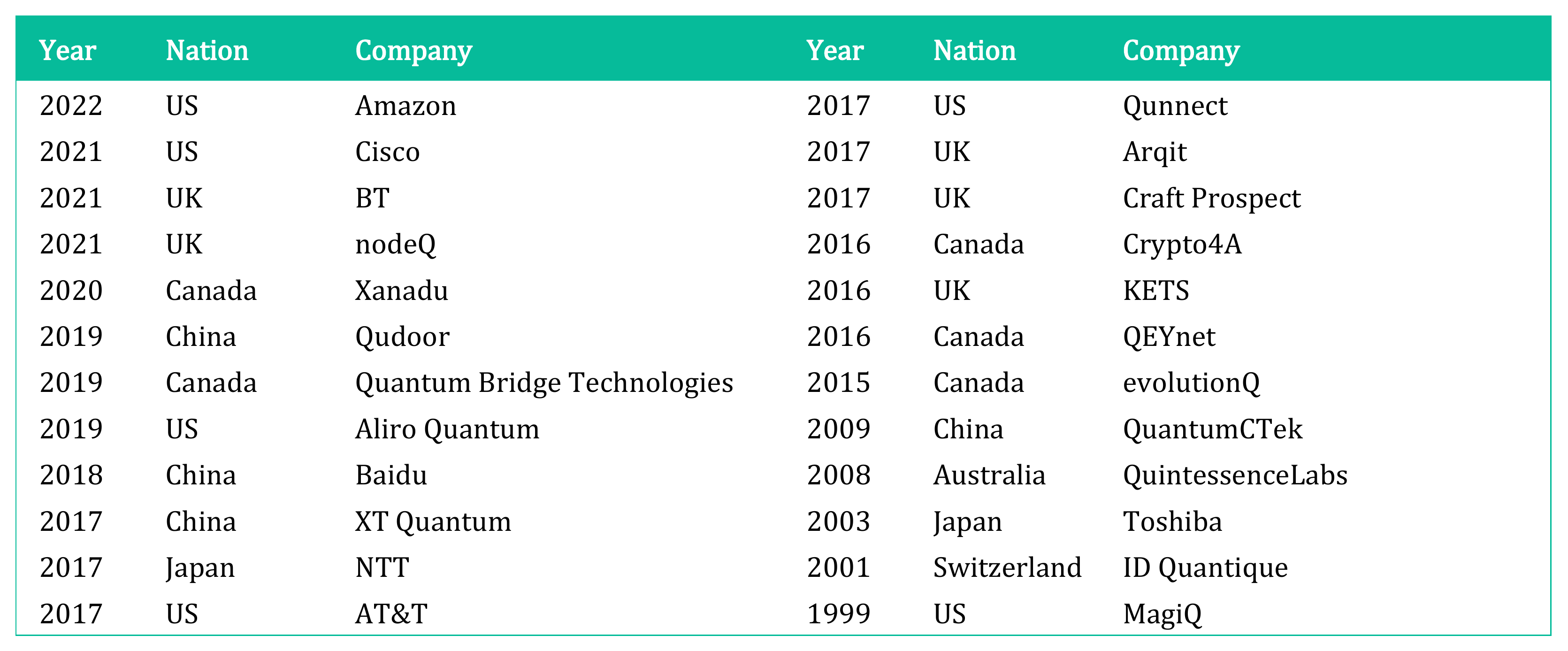}
\caption{A partial list of the industrial participants working on quantum networks.}
\label{fig: company}
\end{figure}

\newpage
\section{UML class diagram of QNET}
\label{sec: UML class diagram}

\begin{figure}[H]
\centering
\includegraphics[width=17cm]{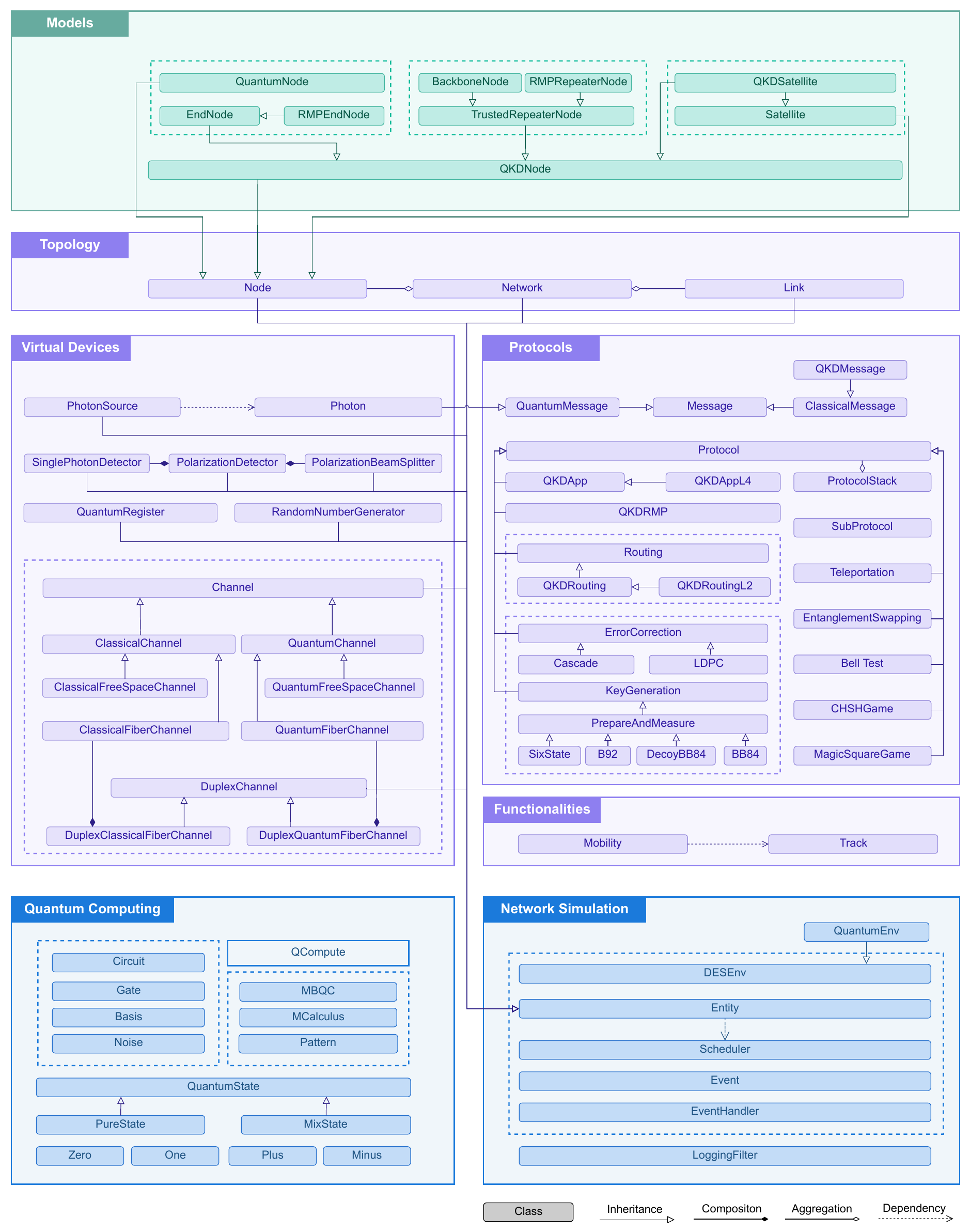}
\caption{A UML class diagram laying out the structure of QNET (based on version 1.2.0). The main classes in each module are highlighted in shaded boxes, and their relationship in terms of inheritance, composition, aggregation and dependency are depicted by different types of arrows.}
\label{fig: ulm_diagram}
\end{figure}

\newpage

\twocolumngrid

\newpage
\section{Compilation of quantum network protocols}
\label{sec: protocol mapping supplements}

In this section, we discuss the compilation techniques used to convert quantum network protocols into quantum circuits in further detail. We start by discussing the instructions that are usually used in quantum circuits. Then we introduce the design of the local quantum register and its instructions. Based on these instructions, we introduce the algorithms that compile local operations as well as qubit transmission to quantum circuit instructions. Putting them together, we can eventually compile a quantum network protocol that involves different parties into one single quantum circuit in general.

\paragraph{Quantum circuit and its instructions}
A quantum circuit usually contains quantum gates and quantum measurements. It is called a \emph{standard quantum circuit} if all measurements are performed at the end of the circuit. According to the execution order, a quantum circuit can be represented as an ordered list of instructions. Each instruction is identified by a triplet [NAME, REG, PAR], indicating the name of the operation, the quantum register it acts on, and the relevant parameters (set to none if no parameters). For example, [rx, 2, pi] means that a $R_x$ rotation gate is applied to qubit 2, and the rotation angle is $\pi$; [cnot, [1, 3], none] means that a controlled-NOT gate is applied to qubits 1 and 3, where qubit 1 is the control qubit and qubit 3 is the target qubit; and [measure, 2, none] means that qubit 2 is measured in the computational basis. More generally, in some quantum circuits, a quantum gate on one qubit can be conditioned on the measurement outcome of the other qubit. Such circuits are called \emph{dynamic quantum circuits}. The classically controlled quantum gate in a dynamic quantum circuit can be represented by a quadruple [NAME, REG, PAR, COND], where the COND indicates that the operation is controlled by the measurement outcome of the qubit to which it points. For example, [x, 2, none, 4] means that a Pauli-$X$ gate is applied to qubit 2 if the measurement outcome of qubit 4 is one; otherwise, the gate is not applied.

\begin{figure}[ht]
\centering
\includegraphics[width=8.2cm]{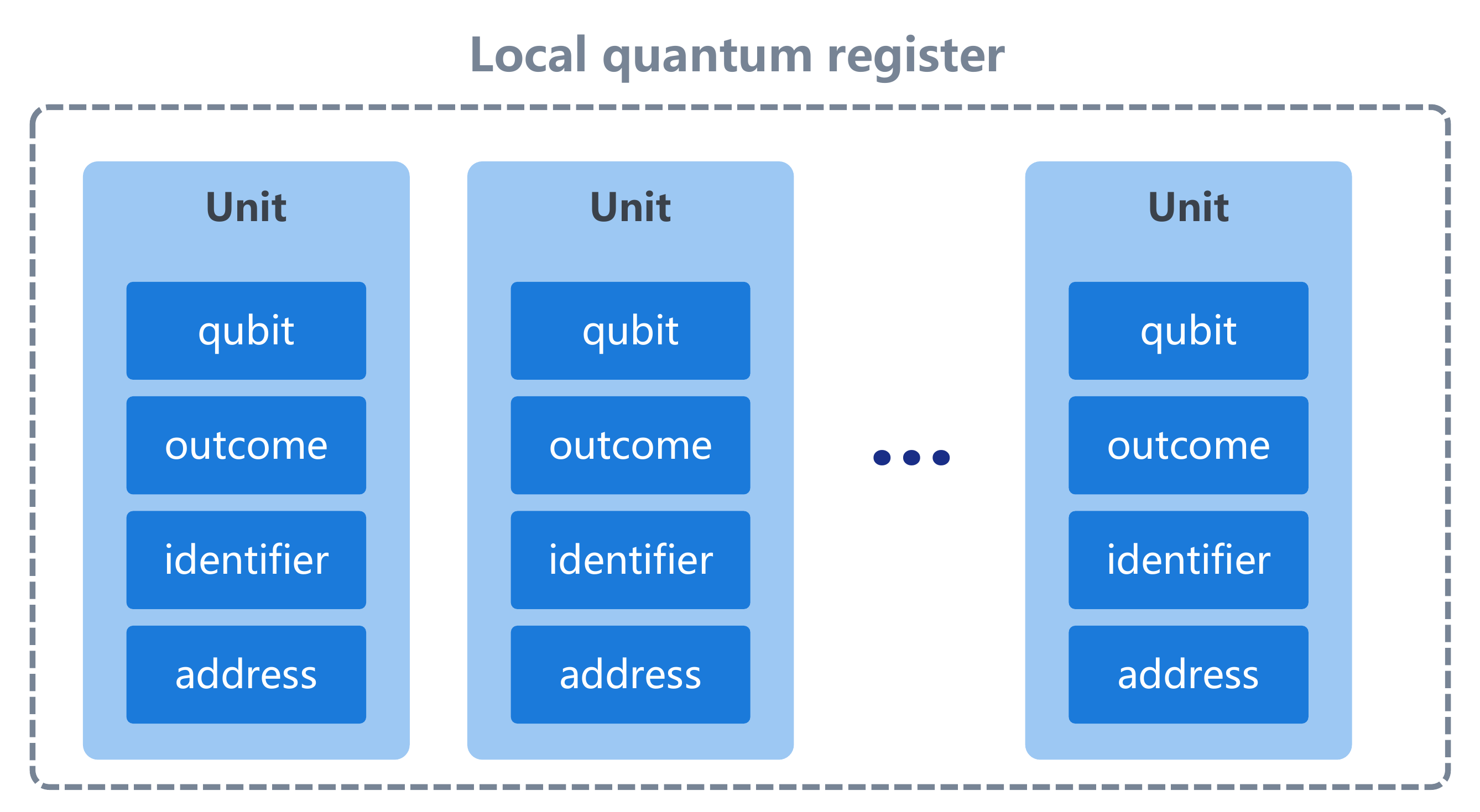}
\caption{The structure of a local quantum register that has several units. Each unit has four attributes, \emph{qubit, outcome, identifier and address}.}
\label{fig: local_quantum_register}
\end{figure}

\paragraph{Local quantum register and its instructions}
Quantum network protocol is different from the quantum circuit. It involves not only local operations at each node but also the transmission of quantum and classical information between different nodes. To formulate the local operations of each node, we propose a design of the local quantum register, which contains multiple units and each unit has four attributes, \emph{qubit, outcome, identifier and address} (see Figure~\ref{fig: local_quantum_register}). The default values for \emph{qubit} and \emph{outcome} are set to be empty, and the default value for the identifier is the node itself. Any instruction on the local register can be identified by a quadruple [NAME, ADDR, PAR, COND], indicating the name of the instruction, the address of the unit to operate (if the instruction is a double-qubit gate, then the ADDR is a list, whose first element is the control qubit and the second element is the target qubit), the relevant parameters of the instruction, and the classically controlled condition of the instruction. For example, the instruction [x, 1, none, none] means that a Pauli-$X$ gate is performed on unit 1 of the local register.
Note that the instructions on the quantum circuit and the local quantum register are both characterized by quadruples. But ADDR and REG have completely different meanings. The former refers to the address of a local register at a network node, whereas the latter refers to the register of a global quantum circuit. The idea of our compilation is trying to build a connection between them.

\paragraph{Compilation of local operations and qubit transmission}
Now we present how instructions for local registers are compiled into instructions for a global quantum circuit. The compilation of single-qubit operation is given in Algorithm~\ref{Alg: single qubit operation mapping}, and the compilation of double-qubit operation is given in Algorithm~\ref{Alg: double qubit operation mapping}.

\begin{algorithm}[H]
    \caption{Compiling single-qubit operation}
    \label{Alg: single qubit operation mapping}
	\KwIn{a local operation [NAME, ADDR, PAR, COND], and the current list of quantum circuit instructions \emph{circuit}} 
	\KwOut{a list of updated circuit instructions}
	\vspace{0.2cm}
	1. \If{the qubit attribute in the local unit of address ADDR is empty}{1.1 get the largest REG index $x$ among all instructions of the input \emph{circuit}\; 1.2 write \emph{$x + 1$} into the \emph{qubit} attribute of the local unit of address ADDR;}
	
	2. get the \emph{qubit} value of the local unit of address ADDR, and record it as REG;
	
	3. append the circuit instruction [NAME, REG, PAR, COND] to the end of \emph{circuit} and return it;
	
	4. \If{NAME is measure}{write the value of REG into the \emph{outcome} attribute of the local unit of address ADDR.}
\end{algorithm}
Note that if the local operation is a quantum measurement, the REG needs to be written into the \emph{outcome} attribute of the local unit of address ADDR. Since the quantum circuit is not yet implemented at this point, the \emph{outcome} here stores the register to be measured rather than an actual measurement outcome.

Similar to the single-qubit operation, we can compile a double-qubit operation as follows.

\begin{algorithm}[H]
    \caption{Compiling double-qubit operation}
    \label{Alg: double qubit operation mapping}
	\KwIn{a local operation [NAME, ADDR, PAR, COND], and the current list of quantum circuit instructions \emph{circuit}}   
	\KwOut{a list of updated circuit instructions}   

    \vspace{0.2cm}
 
    1. set ADDR0 and ADDR1 to be the first and second elements of ADDR, respectively;

    2. get the largest REG index $x$ among all instructions of the input \emph{circuit} and set $f$ to be $x$;
	
    3. \If{the qubit attribute in the local unit of address ADDR0 is empty}{set $f$ to be $f+1$ and then write $f$ into the \emph{qubit} attribute of the local unit of address ADDR0;}
	
    4. \If{the qubit attribute in the local unit of address ADDR1 is empty}{set $f$ to be $f+1$ and then write $f$ into the \emph{qubit} attribute of the local unit of address ADDR1;}
	
    5. get the \emph{qubit} value of the local unit of address ADDR0, and record it as REG0; get the \emph{qubit} value of the local unit of address ADDR1, and record it as REG1;
	
    6. append the circuit instruction [NAME, [REG0, REG1], PAR, COND] to the end of \emph{circuit} and return it.
	
\end{algorithm}

The compilation of qubit transmission is given in Algorithm~\ref{Alg: qubit transmission}. Note that the transmission of a qubit does not change the global quantum circuit, but updates the information of local quantum registers at both the source and destination nodes. The \emph{qubit} stored in the register unit is not really a qubit, but its corresponding register index in the global quantum circuit. By sending this information from source to destination, we effectively transfer the right to manipulate this qubit on the quantum circuit. It is also important that we reset a local unit once the \emph{qubit} is sent so that the no-cloning theorem is not violated. When the destination receives a quantum message, it is stored in the available unit with the smallest address as default, and the \emph{identifier} records the source of this qubit.

\begin{algorithm}[H]
    \caption{Compiling qubit transmission}
    \label{Alg: qubit transmission}
	\KwIn{source node \emph{src}, destination node \emph{dst}, address of local register unit ADDR} 
	\KwOut{updated quantum registers at the source and destination nodes}  

 \vspace{0.2cm}
 
	1. get the \emph{qubit} value of the register unit of address ADDR and record it as \emph{qmsg};
	
	2. reset the unit of address ADDR, i.e., rewrite its \emph{qubit} value as empty and rewrite its \emph{identifier} value as the node itself;
	
	3. send the qubit information \emph{qmsg} to the node \emph{dst}, i.e., find an empty unit with the smallest address at the node \emph{dst}, set the \emph{qubit} value of this unit as the received \emph{qmsg} and set its \emph{identifier} as \emph{src}.
\end{algorithm}

Combining Algorithm \ref{Alg: single qubit operation mapping}, \ref{Alg: double qubit operation mapping} and \ref{Alg: qubit transmission}, we can compile a complete quantum network protocol into one single quantum circuit. This is demonstrated in Algorithm \ref{Alg: protocol to dynamic quantum circuit}.

\begin{algorithm}[H]
    \caption{Compiling quantum network protocol to dynamic quantum circuit}
    \label{Alg: protocol to dynamic quantum circuit}
	\KwIn{a list of instructions of the network protocol}
	\KwOut{a list of instructions of the dynamic circuit} 

 \vspace{0.2cm}
	1. create an empty list of instructions \emph{circuit};
	
	2. \ForEach{instruction from the input}{using Algorithm \ref{Alg: single qubit operation mapping}, \ref{Alg: double qubit operation mapping} and \ref{Alg: qubit transmission} to compile this instruction to a circuit instruction and update the list \emph{circuit};}
	
	3. return the list of circuit instructions \emph{circuit}.
\end{algorithm}

Dynamic quantum circuits provide an accurate and logical description of quantum protocols, but their implementation can still be challenging. Therefore, we will use the principle of deferred measurement to replace a classically controlled quantum gate with a quantum-controlled gate to obtain an equivalent standard quantum circuit. This is presented in the following Algorithm~\ref{Alg: deferred measurement}.

\begin{algorithm}[H]
    \caption{Deferring quantum measurement}
    \label{Alg: deferred measurement}
	\KwIn{a list of circuit instructions \emph{circuit}} 
	\KwOut{a list of updated circuit instructions}    

 \vspace{0.2cm}	
    1. \ForEach{instruction from the input}{
	    \If{COND of this instruction is not empty}{
	        replace this instruction with [CNAME, [COND, REG], PAR], where CNAME is the controlled version of the instruction (e.g., CNAME = cnot if NAME = x);}
	    \If{the current gate is a measurement operation}{
	        add it to the \emph{measured\_gates} list;}
	    }
	    
    2. move all measurement instructions to the end of the list while leaving the remaining items unchanged;
    
    3. return the list of instructions \emph{circuit}.
\end{algorithm}

\end{document}